\PassOptionsToPackage{unicode}{hyperref}
\PassOptionsToPackage{hyphens}{url}
\PassOptionsToPackage{dvipsnames,svgnames,x11names}{xcolor}
\documentclass[
  12pt]{article}
\usepackage{xcolor}
\usepackage{amsmath,amssymb}
\usepackage{iftex}
\ifPDFTeX
  \usepackage[T1]{fontenc}
  \usepackage[utf8]{inputenc}
  \usepackage{textcomp} 
\else 
  \usepackage{unicode-math} 
  \defaultfontfeatures{Scale=MatchLowercase}
  \defaultfontfeatures[\rmfamily]{Ligatures=TeX,Scale=1}
\fi
\usepackage{lmodern}
\ifPDFTeX\else
\fi
\IfFileExists{upquote.sty}{\usepackage{upquote}}{}
\IfFileExists{microtype.sty}{
  \usepackage[]{microtype}
  \UseMicrotypeSet[protrusion]{basicmath} 
}{}
\makeatletter
\@ifundefined{KOMAClassName}{
  \IfFileExists{parskip.sty}{%
    \usepackage{parskip}
  }{
    \setlength{\parindent}{0pt}
    \setlength{\parskip}{6pt plus 2pt minus 1pt}}
}{
  \KOMAoptions{parskip=half}}
\makeatother
\makeatletter
\ifx\paragraph\undefined\else
  \let\oldparagraph\paragraph
  \renewcommand{\paragraph}{
    \@ifstar
      \xxxParagraphStar
      \xxxParagraphNoStar
  }
  \newcommand{\xxxParagraphStar}[1]{\oldparagraph*{#1}\mbox{}}
  \newcommand{\xxxParagraphNoStar}[1]{\oldparagraph{#1}\mbox{}}
\fi
\ifx\subparagraph\undefined\else
  \let\oldsubparagraph\subparagraph
  \renewcommand{\subparagraph}{
    \@ifstar
      \xxxSubParagraphStar
      \xxxSubParagraphNoStar
  }
  \newcommand{\xxxSubParagraphStar}[1]{\oldsubparagraph*{#1}\mbox{}}
  \newcommand{\xxxSubParagraphNoStar}[1]{\oldsubparagraph{#1}\mbox{}}
\fi
\makeatother

\usepackage{longtable,booktabs,array}
\usepackage{calc} 
\usepackage{etoolbox}
\makeatletter
\patchcmd\longtable{\par}{\if@noskipsec\mbox{}\fi\par}{}{}
\makeatother
\IfFileExists{footnotehyper.sty}{\usepackage{footnotehyper}}{\usepackage{footnote}}
\makesavenoteenv{longtable}
\usepackage{graphicx}
\makeatletter
\newsavebox\pandoc@box
\newcommand*\pandocbounded[1]{
  \sbox\pandoc@box{#1}%
  \Gscale@div\@tempa{\textheight}{\dimexpr\ht\pandoc@box+\dp\pandoc@box\relax}%
  \Gscale@div\@tempb{\linewidth}{\wd\pandoc@box}%
  \ifdim\@tempb\p@<\@tempa\p@\let\@tempa\@tempb\fi
  \ifdim\@tempa\p@<\p@\scalebox{\@tempa}{\usebox\pandoc@box}%
  \else\usebox{\pandoc@box}%
  \fi%
}
\def\fps@figure{htbp}
\makeatother

\usepackage[]{natbib}
\usepackage[ruled,vlined,linesnumbered,noend]{algorithm2e}
\usepackage{setspace}
\usepackage{enumitem}
\usepackage{microtype}
\usepackage{booktabs}
\usepackage{multirow}
\usepackage{amsthm}
\SetAlFnt{\small}
\SetAlCapFnt{\small}
\SetAlCapNameFnt{\small}
\setitemize{noitemsep, topsep=0pt, leftmargin=1em}
\newtheoremstyle{break}
  {3pt}
  {3pt}
  {\itshape}
  {}
  {\bfseries}
  {.}
  {\newline}
  {\thmname{#1}\thmnumber{ #2}\thmnote{ (#3)}}
\theoremstyle{break}
\newtheorem{theorem}{Theorem}
\newtheorem{lemma}{Lemma}

\newtheorem{condition}{Condition}

\newtheorem{proposition}{Proposition}
\newtheorem{remark}{Remark}
\newcommand{\suppprop}{\mathrm{SP}}
\usepackage{etoolbox}
\usepackage{graphicx}
\usepackage{subcaption}
\usepackage{pdflscape}
\usepackage{adjustbox}
\AtBeginEnvironment{theorem}{%
  \setlength{\abovedisplayskip}{1pt}
  \setlength{\belowdisplayskip}{1pt}
  \setlength{\abovedisplayshortskip}{2pt}
  \setlength{\belowdisplayshortskip}{2pt}
  \setlength{\jot}{2pt}
}
\numberwithin{equation}{section}
\AtBeginEnvironment{thebibliography}{\singlespacing}
\makeatletter
\@ifpackageloaded{caption}{}{\usepackage{caption}}
\AtBeginDocument{%
\ifdefined\contentsname
  \renewcommand*\contentsname{Table of contents}
\else
  \newcommand\contentsname{Table of contents}
\fi
\ifdefined\listfigurename
  \renewcommand*\listfigurename{List of Figures}
\else
  \newcommand\listfigurename{List of Figures}
\fi
\ifdefined\listtablename
  \renewcommand*\listtablename{List of Tables}
\else
  \newcommand\listtablename{List of Tables}
\fi
\ifdefined\figurename
  \renewcommand*\figurename{Figure}
\else
  \newcommand\figurename{Figure}
\fi
\ifdefined\tablename
  \renewcommand*\tablename{Table}
\else
  \newcommand\tablename{Table}
\fi
}
\@ifpackageloaded{float}{}{\usepackage{float}}
\floatstyle{ruled}
\@ifundefined{c@chapter}{\newfloat{codelisting}{h}{lop}}{\newfloat{codelisting}{h}{lop}[chapter]}
\floatname{codelisting}{Listing}

\makeatother
\makeatletter
\@ifpackageloaded{caption}{}{\usepackage{caption}}
\@ifpackageloaded{subcaption}{}{\usepackage{subcaption}}
\makeatother
\usepackage{bookmark}
\IfFileExists{xurl.sty}{\usepackage{xurl}}{} 
\hypersetup{
  pdftitle={SCAN: Sequentially Detecting Change-points via Adaptive Nonparametric Inference},
  pdfauthor={Ashoka Prabashwara; Patricia Menéndez; Liam Hodgkinson; Stuart Lee},
  pdfkeywords={Long time series, Serial dependence, Distributional
shifts, Bootstrap calibration, Wasserstein distance},
  colorlinks=true,
  linkcolor={blue},
  filecolor={Maroon},
  citecolor={Blue},
  urlcolor={Blue},
  pdfcreator={LaTeX via pandoc}}

\begin{document}

\def\spacingset#1{\renewcommand{\baselinestretch}%
{#1}\small\normalsize} \spacingset{1}


\date{August 28, 2026}
\spacingset{.85}
\title{\bf SCAN: Sequentially Detecting Change-points via Adaptive
Nonparametric Inference}
\author{
Ashoka Prabashwara\\
{\small School of Mathematics and Statistics, University of Melbourne}\\
and\\Patricia Menéndez\\
{\small School of Mathematics and Statistics, University of Melbourne}\\
and\\Liam Hodgkinson\\
{\small School of Mathematics and Statistics, University of Melbourne}\\
and\\Stuart Lee\\
{\small Walter and Eliza Hall Institute, University of Melbourne}\\
}
\maketitle
\spacingset{1}

\bigskip
\bigskip
\begin{abstract}
Modern time series are often long, serially dependent, and
non-stationary. Existing change-point methods either target specific
changes or become computationally intensive when using nonparametric
costs on long series. Many also require thresholds to be carefully
calibrated under serial dependence. We introduce SCAN, an offline method
for detecting multiple distributional change-points in long, serially
dependent univariate time series. SCAN compares adjacent windows using
an integral probability metric, calibrates local discrepancies with a
dependence-aware bootstrap, and refines candidate locations using a
scaled 1-Wasserstein criterion, enabling detection of changes in mean,
variance, and broader distributional structure within a unified
framework. An ensemble over multiple window sizes reduces sensitivity to
window size and threshold specification. We establish consistency of the
estimated number and locations of change-points under exponential
\(\alpha\)-mixing dependence, and show that the localization statistic
reduces to a CUSUM-type statistic under pure mean shifts. In simulations
with up to one million observations, SCAN generally achieves higher
covering and \(F_1\)-scores than competing methods across mean and joint
mean-variance shifts, particularly under serial dependence. On real
data, SCAN identifies labeled activity transitions in sensor data and
interpretable structural changes in hourly Bitcoin prices.
Implementations are available in the Python package \texttt{scan-cpd}
and R package \texttt{scanr}.
\end{abstract}

\noindent%
{\it Keywords:} Long time series, Serial dependence, Distributional
shifts, Bootstrap calibration, Wasserstein distance
\vfill

\newpage
\spacingset{1.9} 

\section{Introduction}\label{sec-intro}

Advances in automated data collection have transformed time series
analysis from a setting involving relatively small to moderately large
datasets to one characterized by massive, high-frequency data. Financial
markets \citep{CHEN2023187}, industrial sensor systems, environmental
monitoring networks \citep{Bitencourt2023}, and healthcare technologies
\citep{1011453531326} now generate millions of observations, requiring
scalable methods capable of extracting information from large and
continuously evolving data. In the context of very long time series,
often comprising millions of observations, the assumption of
stationarity, or even weak stationarity
\citep{f8815cbf-36f9-35f3-9962-055fa42e3de8}, is frequently unrealistic.
For sufficiently long time series, it becomes increasingly unlikely that
a single probability law adequately describes the underlying stochastic
process over the entire observation period. Changes may occur in the
mean, variance, or the distributional law itself, and ignoring such
changes can degrade forecasting accuracy and lead to biased inference.

This has led to extensive research on offline change-point detection
methods for univariate time series, where all observations are available
prior to analysis, which is the case considered in this study. Existing
methods have focused on detecting a change in the mean, with the
cumulative sum (CUSUM) statistic of \citet{Page1954} being one of the
earliest approaches, as well as changes in variance
\citep{Shi2015PanelVolatilityCUSUM}, correlation structure
\citep{Cabrieto2017}, and trend \citep{MaengFryzlewicz2024TrendSegment}.
Similar problems arise in multivariate and high-dimensional settings,
where a common strategy is to reduce the data to one or more univariate
series through projections or other transformations
\citep{WangSamworth2018, Hahn2020, Zhang2024RWPCA, Qin2025}. Therefore,
advances in univariate change-point detection remain relevant beyond the
univariate setting.

For multiple change-points, existing methods can be broadly divided into
exact and approximate approaches. Exact methods formulate detection as a
penalized optimization problem and solve it using dynamic programming
with pruning techniques, including the Segment Neighborhood method of
\citet{Jackson2005OptimalPartition}, the Pruned Exact Linear Time (PELT)
algorithm of \citet{KillickFearnheadEckley2012PELT}, and Functional
Pruning Optimal Partitioning (FPOP) of \citet{Maidstone2017}.
Approximate methods improve computational scalability through recursive
partitioning or local scanning procedures. Examples include Binary
Segmentation \citep{SenSrivastava1975}, Wild Binary Segmentation (WBS)
\citep{Fryzlewicz2014WBS}, Seeded Binary Segmentation (SBS)
\citep{KovacsEtAl2022SBS}, and window-based approaches such as MOSUM
\citep{EichingerKirch2018MOSUM}. While computationally efficient and
effective in many settings, the type of change-point detected by these
methods is largely determined by the associated cost function, which
most commonly targets changes in the mean.

As noted by \citet{TruongOudreVayatis2020RupturesReview}, change-point
methods consist of three main components: a search strategy, a cost or
discrepancy function, and a penalty or threshold used to determine the
number of changes. While extensive research has been devoted to mean
change-point detection in moderately long time series (see
\citet{FearnheadRigaill2020Stat} and \citet{Aminikhanghahi2017} for
reviews), the type of distributional change that a method can detect is
largely determined by its cost function and therefore often needs to be
specified in advance. Nonparametric methods, such as E-divisive
\citep{MattesonJames2014Edivisive} and kernel-based approaches
\citep{arlot2019kernelmultiplechangepointalgorithm}, can detect broader
distributional changes without explicitly targeting a particular
characteristic of the distribution. However, these methods are often
computationally demanding and are typically limited to short or
moderately long series.

In addition to the cost function, change-point methods require a penalty
or threshold to distinguish genuine distributional changes from
stochastic fluctuations. Common choices include
information-criterion-based penalties such as AIC
\citep{Bozdogan1987CAIC}, BIC \citep{Schwarz1978BIC}, and MBIC
\citep{Bogdan2004ModifyingBICQTL}. Yet these approaches typically rely
on a global penalty and are often calibrated under independence. For
dependent time series, serial correlation inflates the long-run
variance, causing such thresholds to underestimate the variability of
the process and potentially leading to excessive false detections.
Although alternative penalty calibration methods have been proposed
\citep{TruongVayatis2017PenaltyLearning, Lavielle2005PenalizedContrasts},
selecting an appropriate penalty remains challenging, particularly in
the presence of dependence. This challenge is amplified in long time
series, where even small threshold misspecifications can lead to
substantial over- or under-segmentation.

Among the search methods discussed above, window-based procedures are
particularly attractive for very long time series because they reduce
the global search problem to a sequence of local comparisons between
adjacent segments, despite their sensitivity to the window size. This
makes them simple to implement and computationally efficient, with
computational cost growing approximately linearly with the series length
\citep{TruongOudreVayatis2020RupturesReview}, a substantial improvement
over the typical quadratic scaling of global nonparametric methods.
Nevertheless, existing window-based methods remain sensitive to the
choice of window size and threshold calibration, particularly under
serial correlation. There remains a need for computationally efficient
change-point detection methods for long time series that can detect
broad distributional changes while accounting for dependence and
requiring minimal tuning.

More broadly, the computational cost of existing change-point methods
varies considerably. Binary segmentation, wild binary segmentation
(WBS), and seeded binary segmentation (SBS) typically scale as
\(O(T\log T)\). PELT is linear under favorable pruning conditions but
has worst-case complexity \(O(T^2)\), while FPOP has the same worst-case
order despite often being fast in practice. Kernel methods such as KCP
can be computationally and memory intensive for long series, as they
evaluate kernel-based segment costs within a global segmentation problem
\citep{arlot2019kernelmultiplechangepointalgorithm, TruongOudreVayatis2020RupturesReview}.

Therefore, viewed collectively, these distinct algorithmic challenges
underscore a frequent trade-off in the literature. While contemporary
frameworks have successfully relaxed specific classical assumptions
piecemeal, existing methodologies rarely address computational
scalability, complex serial dependence, and nonparametric distributional
flexibility simultaneously. Developing a unified approach that satisfies
all three criteria is essential for operating on long time series.

To address these concurrent challenges, we propose SCAN
(\textbf{S}equential Detection of \textbf{C}hange-points via
\textbf{A}daptive \textbf{N}onparametric Inference), a scalable
nonparametric framework for multiple change-point detection in long
univariate time series that combines the efficiency of window-based
methods with the flexibility of nonparametric inference. SCAN performs
local two-sample comparisons between adjacent windows based on an
integral probability metric (IPM, \citet{muller1997integral}), enabling
detection of mean, variance, and general distributional changes within a
unified detection-localization framework. Candidate regions identified
by the scanning procedure are subsequently refined using a newly
proposed localization statistic based on the \(1\)-Wasserstein distance
\citep{Villani2009OptimalTransport}, while detections across multiple
window sizes are aggregated through an ensemble strategy
\citep{Rokach2010} that reduces sensitivity to the choice of a single
window size. We establish theoretical guarantees under exponential
\(\alpha\)-mixing conditions \citep{alphamixing} and further evaluate
the proposed method through simulation studies involving long-memory
processes \citep{grangerlongmemory1980}. Due to its linear computational
complexity with respect to the time series length, SCAN achieves
execution speeds on par with the fastest exact parametric methods for
simple mean shifts, while providing orders-of-magnitude speedups over
competing frameworks when detecting more complex distributional changes.

A second contribution establishes a direct analytical connection between
the proposed localization statistic and the classical CUSUM method. We
show that the \textbf{S}caled 1-\textbf{Wa}sserstein
\textbf{L}ocalization (SWAL) statistic reduces exactly to a CUSUM-type
criterion under a pure mean shift, positioning it as a distributional
generalization of CUSUM. Beyond this special case, the SWAL statistic
retains sensitivity to variance changes, joint mean-variance changes,
and more general distributional shifts, while remaining grounded in a
unified nonparametric framework.

The paper is organized as follows. Section~\ref{sec-psn} defines the
distributional change-point problem for univariate time series.
Section~\ref{sec-meth} introduces SCAN, including the scanning procedure
and ensemble extension. Section~\ref{sec-theory} presents the
theoretical guarantees. Section~\ref{sec-sim-study} reports the
simulation study, and Section~\ref{sec-real-app} applies the method to
real datasets with the conclusion in Section~\ref{sec-conclusion}.
Proofs and additional numerical results are provided in the online
supplementary material.

\section{Model and Data}\label{sec-psn}

Let \(\{X_t\}_{t=1}^T\) be a univariate time series with
\(X_t\in\mathbb R\) that is piecewise strictly stationary
\citep{BrockwellDavis1991TSA} with exponentially \(\alpha\)-mixing
within each stationary segment, with mixing coefficients satisfying
\(\alpha(h)\le e^{-\lambda h}\), where \(\lambda >0\) and \(h \ge 1\).
In this study, we aim to estimate the changes in time series mean,
variance, or distributional properties, as well as their temporal
locations.

For that, we assume there exists an integer \(k \ge 0\) and time points
\(1 \le \tau_1 < \cdots < \tau_k < T\), with \(\tau_0 = 0\) and
\(\tau_{k+1} = T\), such that the process changes at each \(\tau_j\)
either in its mean, variance, or distributional properties. We refer to
times \(\tau_1, \ldots, \tau_k\) as change-points. Then for each segment
\(S_j= (X_{\tau_{j-1}+1}, \ldots, X_{\tau_j})\) with
\(j = 1, \ldots, k+1\), we define the mean, variance and the
distribution of segment \(S_j\) as \(\mu^{(j)}\), \(\sigma^{2(j)}\), and
\(F_{S_j}\) respectively where \(F_{S_l}\) represents the cumulative
distribution function (CDF) of the \(\ell\)th segment for
\(l = 1, \ldots, k+1\).

At each change-point, the distributions of the adjacent stationary
segments differ, so that \(F_{S_j}\neq F_{S_{j+1}}\) for
\(j=1,\ldots,k\). These changes may involve the mean, variance, or other
distributional properties.

To measure the discrepancy between the probability measures associated
with two segments \(S_j\) and \(S_i\), we use Integral Probability
Metrics (IPMs, \citet{muller1997integral}) defined as follows:

Let \(P_{S_j}\) and \(P_{S_i}\) be the probability measures associated
with these segments, and let \(\mathcal G\) be a class of real-valued
measurable functions that are integrable with respect to both
\(P_{S_j}\) and \(P_{S_i}\). The IPM between \(P_{S_j}\) and \(P_{S_i}\)
is defined as \begin{equation*}
d_{\mathcal{G}}(P_{S_j},P_{S_i})
=
\sup_{f\in\mathcal{G}}
\left|
\mathbb{E}_{X\sim P_{S_j}}[f(X)]
-
\mathbb{E}_{Y\sim P_{S_i}}[f(Y)]
\right|.
\end{equation*} Equivalently, if \(F_{S_j}\) and \(F_{S_i}\) denote the
corresponding CDFs, then \begin{equation*}
d_{\mathcal{G}}(P_{S_j},P_{S_i})
=
\sup_{f\in\mathcal{G}}
\left|
\int f(x)\,dF_{S_j}(x)
-
\int f(x)\,dF_{S_i}(x)
\right|.
\end{equation*} Here, \(\int f(x)\,dF(x)\) denotes the
Lebesgue-Stieltjes integral with respect to the probability measure
induced by the CDF \(F\).

\section{Methodology}\label{sec-meth}

SCAN consists of a detection-localization procedure, followed by an
ensemble aggregation step over multiple window sizes via majority voting
\citep{majorityvoting1997} to reduce sensitivity to window size
selection. This separation of testing, localization, and ensemble
stabilization is central to our method and distinguishes it from
approaches that optimize a single global cost or rely on a fixed
penalty.

For window size \(w\), the procedure compares two adjacent
non-overlapping windows of equal size at each candidate split \(t\): the
reference window \(X_{t-w+1:t} = (X_{t-w+1}, \ldots, X_t)\) and the
stride window \(X_{t+1:t+w} = (X_{t+1}, \ldots, X_{t+w})\), with the
equal-size choice motivated by the finite-sample IPM error bound in
Section~\ref{sec-theory}. Throughout, \emph{window} refers to either
local sample \(X_{t-w+1:t}\) or \(X_{t+1:t+w}\); \emph{localization
region} refers to their union \(X_{t-w+1:t+w}\); and \emph{segment}
refers to the underlying stationary regimes between consecutive
change-points, as defined in Section~\ref{sec-psn}.

\begin{figure}[t]
\centering
\includegraphics[width=0.60\textwidth]{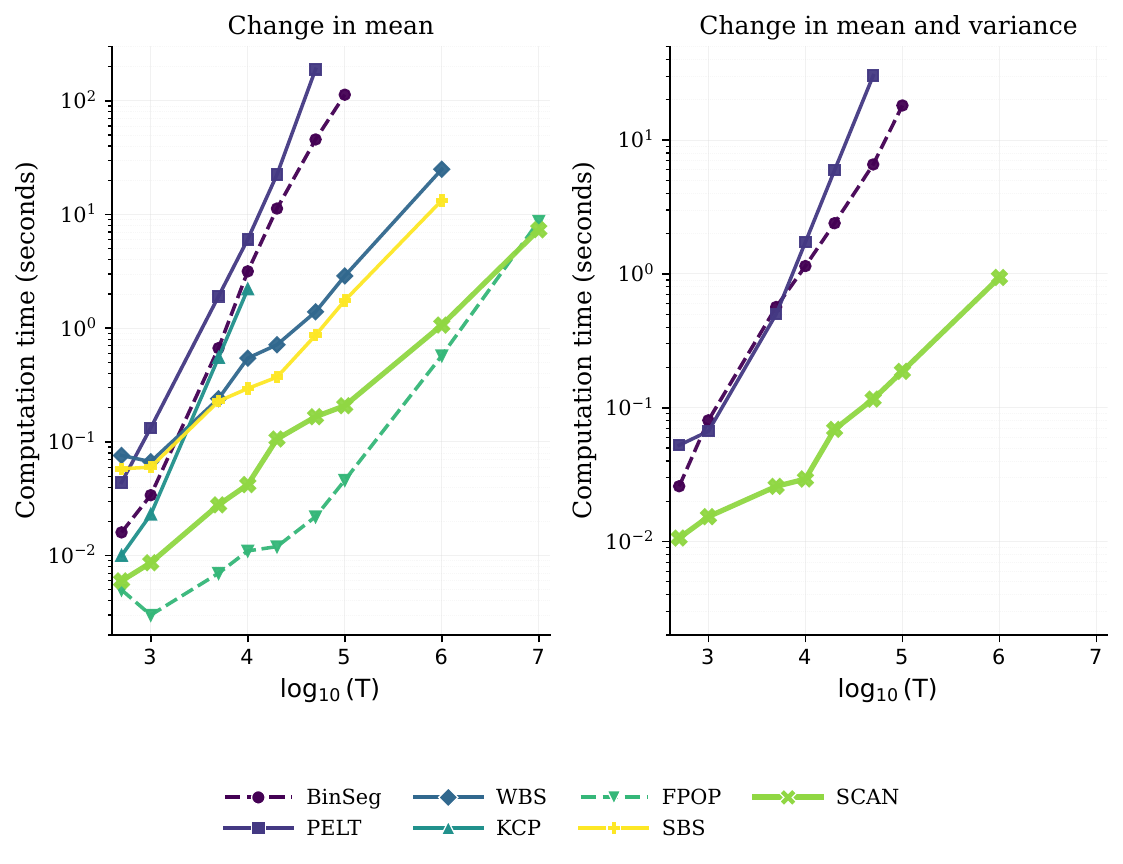}
\caption{Mean computation times over 1,000 simulation runs for competing change-point detection methods applied to white-noise time series with change-points. The left panel shows results for changes in the mean, while the right panel shows results for joint changes in the mean and variance. The horizontal axis represents $\log_{10}(T)$, where $T$ denotes the series length, and the vertical axis represents computation runtime in seconds.}
\label{fig:runtimes}
\end{figure}

By operating on adjacent local windows, SCAN converts the global search
problem into a sequence of local reference-stride comparisons. This
local structure improves computational scalability, as illustrated in
Figure \ref{fig:runtimes}, where SCAN has competitive runtimes for
mean-only change-point methods and shows clear runtime gains for joint
mean-variance changes compared with the main benchmark methods. A
detailed complexity analysis and runtime comparison are provided in the
computational-times subsection of Section~\ref{sec-sim-study}.

\begin{figure}[t]
\centering
\includegraphics[width=\textwidth]{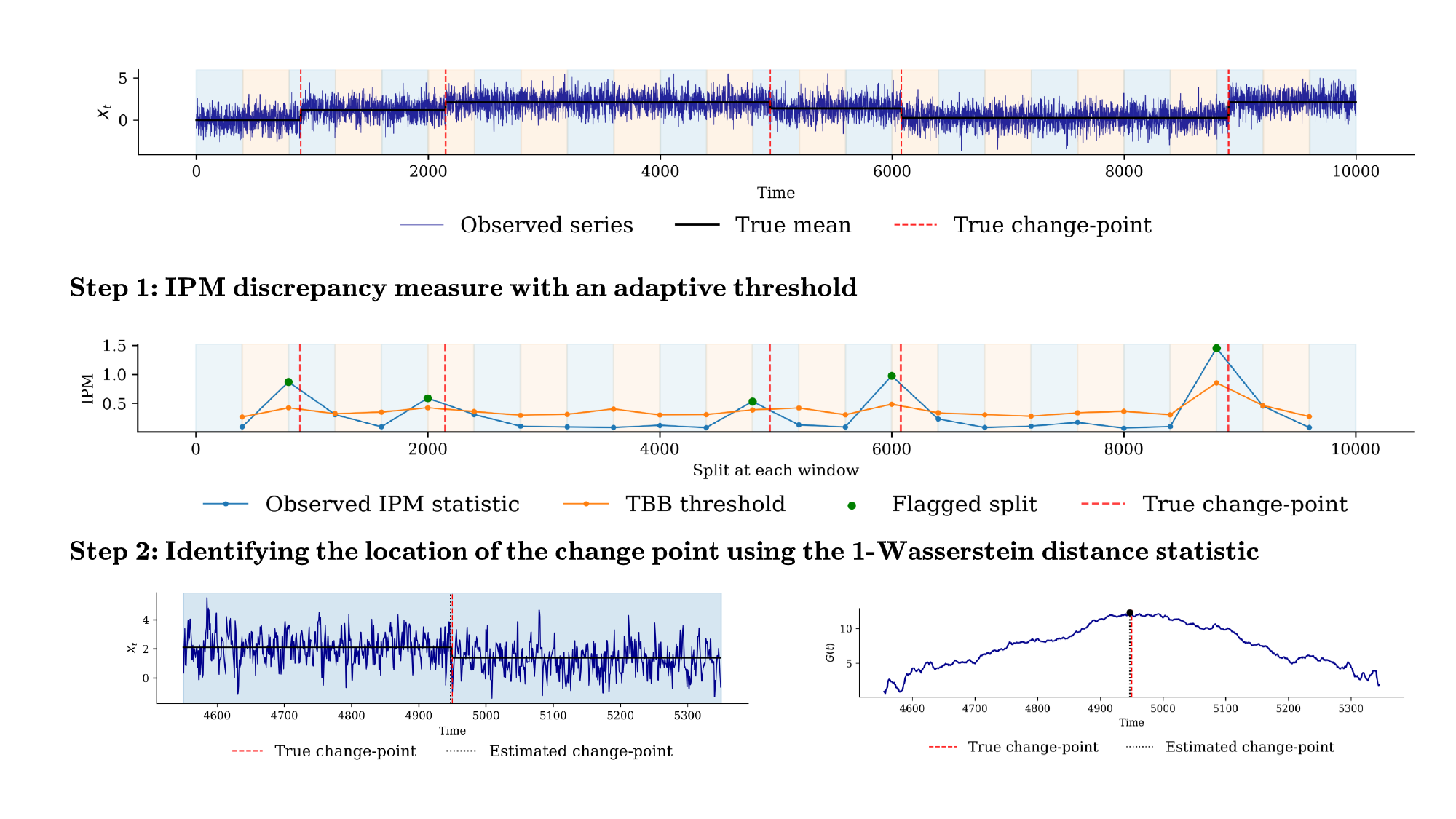}
\caption{
Illustration of the proposed SCAN change-point detection framework. Step 1 compares the local IPM statistic between adjacent windows with a data-driven threshold to flag candidate change-point regions. Step 2 refines each flagged region by maximizing the SWAL statistic over the flagged splits.
}
\label{fig:steps-of-the-method}
\end{figure}

For clarity, the scanning and localization stages are described for a
single candidate split. For a time series of length \(T\) and window
size \(w\), the full procedure applies these steps across all
\(\lfloor T/w \rfloor - 1\) adjacent window pairs, indexed by
\(t_m=mw\). Detailed steps of the proposed \textbf{SCAN framework} are
given below:

\begin{enumerate}

\item \textbf{Window partitioning:} For a candidate split at time $t$, the observed local IPM discrepancy between the empirical distributions of the reference and stride windows is used as the test statistic ${T}_{t,w}=d_{\mathcal{G}}\!\left(\widehat F_{(w),t},\widehat F_{(s),t}\right)$, where $\widehat F_{(w),t}$ and $\widehat F_{(s),t}$ denote the empirical CDFs of $X_{t-w+1:t}$ and $X_{t+1:t+w}$, respectively. The statistic tests $H_{0,t}:F_{(w),t}=F_{(s),t}$ against $H_{1,t}:F_{(w),t}\neq F_{(s),t}$, where $F_{(w),t}$ and $F_{(s),t}$ denote the corresponding population CDFs. Since ${T}_{t,w}$ is a non-negative discrepancy measure, evidence against $H_{0,t}$ occurs only in the upper tail of its null distribution.

\item \textbf{Change-point window selection via a data-driven threshold:}
Let $b_{T,m}$ denote a theoretical or generic threshold for the local IPM test statistic at candidate split $t_m$, with dependence on the window size suppressed. Its empirically calibrated tapered block-bootstrap counterpart is denoted by $\widehat b_{T,m}$. In practice, the local null hypothesis is rejected and the corresponding reference-stride region is flagged whenever
$T_{t_m,w}>\widehat b_{T,m}$. The threshold $\widehat b_{T,m}$ is obtained from a tapered block bootstrap \citep{TaperedBlock2001} approximation of the null distribution of $T_{t_m,w}$; see Section \ref{sec-num-imple} for details.

Rejection of $H_{0,t}$ shows evidence of a distributional change in either the reference or the stride window, but it does not locate the change-point. Localization is therefore treated as a separate inference problem and is applied conditionally within each flagged region. Figure \ref{fig:steps-of-the-method} illustrates both stages: Step 1 flags candidate regions using the data-driven IPM threshold, and Step 2 localizes the exact location of the change-point.

\item \textbf{Change-point detection and localization: } Conditional on rejection of the null hypothesis, and under Condition \ref{cond-1-minimal-sp} in Section \ref{sec-theory}, the detection problem reduces to a single-change-point localization problem. For a rejected candidate split $t$, the corresponding flagged region is defined as the union of the reference window $X_{t-w+1:t}$ and the stride window $X_{t+1:t+w}$. The SWAL statistic is then applied within the localization region $X_{t-w+1:t+w}$ to refine the change-point location. At the population level, the SWAL statistic is defined below: 

For a local sequence $X_{1:n}$ of length $n$ with finite first-order
moments, consider a possible split after observation $k$, which divides the
segment into the two subsamples $X_{1:k}$ and $X_{k+1:n}$. The set of admissible
split locations is $\mathcal{K}_{n} = \{1,\ldots,n-1\}$, so that both subsamples are non-empty. For each $k\in\mathcal{K}_{n}$, the
population scaled 1-Wasserstein statistic is defined as
\begin{equation}
G_n(k)
=
\left\{
\frac{k(n-k)}{n}
\right\}^{1/2}
W_1\left(F_{1:k}, F_{k+1:n}\right),
\label{eq-swal-statistic}
\end{equation}
where $W_1$ denotes the $1$-Wasserstein distance on a compact interval,
and $F_{1:k}$ and $F_{k+1:n}$ denote the population distributions associated with
$X_{1:k}$ and $X_{k+1:n}$, respectively. The population change-point location is
then $\tau_{n}\in\arg\max_{k \in \mathcal{K}_{n}} G_n(k)$.

This statistic operates within a unified nonparametric framework, accommodating changes in the mean, variance, and broader distributional characteristics.

\item \textbf{Ensemble across window sizes: }
Because no single window size is universally optimal and labeled change-points are unavailable in unsupervised settings, standard tuning procedures such as cross-validation are not applicable. We therefore introduce an ensemble extension of SCAN by applying the method over a finite collection of window sizes, $\mathcal{W}=\{w_1,\ldots,w_d\}$. Each SCAN detector associated with a particular window size is referred to as a base detector, and the resulting detections are aggregated through majority voting. This ensemble acts as a stability filter, reducing sensitivity to the choice of any single window size while preserving localization accuracy.

\textbf{Majority-vote aggregation:}
Candidate detections are aggregated over the collection of window sizes
$\mathcal W=\{w_1,\ldots,w_d\}$. For each $w_j\in\mathcal W$, SCAN produces an
estimated change-point set $\widehat{\mathcal R}^{(j)}$, and the pooled set of
candidate detections is $\widehat{\mathcal R}=\bigcup_{j=1}^{d}\widehat{\mathcal R}^{(j)}$.

Let $r_T>0$ denote a tolerance distance used to merge nearby detections that are
plausibly associated with the same underlying change-point. Once $\widehat{\mathcal R}$ is sorted in ascending order, consecutive detections are
assigned to the same cluster if their separation is no larger than $r_T$. Thus,
the pooled set is partitioned into clusters
$\mathcal S=\{\mathcal S_1,\ldots,\mathcal S_L\}$, where $\hat{\tau}_{(m)},\hat{\tau}_{(m+1)}\in\mathcal S_\ell$ whenever $\hat{\tau}_{(m+1)}-\hat{\tau}_{(m)}\leq r_T$.

Each cluster is then evaluated by its support proportion, $$
\suppprop(\mathcal S_\ell)=\frac{1}{d}\sum_{j=1}^{d}\mathbf{1}\left\{\widehat{\mathcal R}^{(j)}\cap \mathcal S_\ell \neq \varnothing\right\},
$$
which measures the fraction of SCAN detectors that contribute at least one
detection to $\mathcal S_\ell$. A cluster is retained when
$\suppprop(\mathcal S_\ell)\ge \nu$, where $\nu\in(0,1]$ is a
fixed voting threshold, treated as a user-specified hyperparameter independent of $T$. For each retained cluster, the vote count at location
$t\in\mathcal S_\ell$ is
$C_{\mathcal S_\ell}(t)=\sum_{j=1}^{d}\mathbf{1}\left\{t\in \widehat{\mathcal R}^{(j)}\right\}$,
and the representative change-point is chosen as
$\widehat{\tau}(\mathcal S_\ell)=\arg\max_{t\in\mathcal S_\ell}C_{\mathcal S_\ell}(t)$.

The final ensemble estimate is therefore $\widehat{\mathcal R}^{\star}=\left\{\widehat{\tau}(\mathcal S_\ell):\suppprop(\mathcal S_\ell)\ge \nu,\ \ell=1,\ldots L\right\}$.
\end{enumerate}

\section{Theoretical Results}\label{sec-theory}

This section establishes the theoretical foundations of the SCAN
change-point detection framework. We begin by introducing the regularity
conditions required for the theoretical analysis, followed by the main
results. The following three conditions are assumed throughout the
theoretical analysis. Conditions \ref{cond-1-minimal-sp} and
\ref{cond-2-min-signal} are standard identifiability requirements
ensuring that change-points are sufficiently separated and that the
signal magnitude at each change-point is bounded away from zero.
Additionally, Condition \ref{condition-bounded-tv} imposes uniform
bounded variation on the function class \(\mathcal{G}\), ensuring that
the theoretical analysis of IPM \(d_{\mathcal{G}}\) can be controlled
via empirical processes theory.

\begin{condition}[Minimal spacing requirement]
\label{cond-1-minimal-sp}
Consecutive elements of the change-point sequence
$0=\tau_0<\tau_1<\cdots<\tau_k<\tau_{k+1}=T$ are separated by at least
$\tau_{\min}>0$, such that $\tau_{j+1} - \tau_j \geq \tau_{\min}$, where $j = 0, \ldots, k$. For the asymptotic theory, $\tau_{\min}$ is allowed to depend on $T$ and satisfies
$\tau_{\min} \to \infty$ such that $\tau_{\min}=o(T)$ as $T \to \infty$.
\end{condition}
\begin{condition}[Minimal Signal Magnitude]
\label{cond-2-min-signal}
There exists a constant $\vartheta_{\mathcal{G}}>0$ such that, for each
$j=1,\ldots,k$, the change is identifiable by the IPM used in the detection stage: $d_{\mathcal{G}}\!\left(F_{S_j}, F_{S_{j+1}}\right) \ge \vartheta_{\mathcal{G}}$.
\end{condition}
\begin{condition}[Uniform bounded variation of the IPM class]
\label{condition-bounded-tv}
Let $\mathcal{G}$ be a class of measurable real-valued functions on $\mathbb{R}$. For any finite partition $Z=\{x_0<x_1<\cdots<x_m\}$ of $\mathbb{R}$, assume there exists a constant $C_{\mathcal{G}}<\infty$ such that $\sup_{f\in\mathcal{G}} \operatorname{TV}_{\mathbb{R}}(f)
=\sup_{f\in\mathcal{G}}\sup_Z\sum_{i=1}^{m}\left|f(x_i)-f(x_{i-1})\right|\le C_{\mathcal{G}}$, where $\operatorname{TV}_{\mathbb R}(f)$ denotes the total variation of $f\in\mathcal G$ on $\mathbb R$.
\end{condition}

Under Condition \ref{condition-bounded-tv}, the functions in
\(\mathcal{G}\) have uniformly bounded total variation. This implies
\begin{equation}
\label{eq-ipm-ks-final}
d_{\mathcal{G}}(F_{(w)},F_{(s)})
\le
C_{\mathcal{G}}
d_{\mathrm{KS}}(F_{(w)},F_{(s)}),
\end{equation} where
\(d_{\mathrm{KS}}(F_{(w)},F_{(s)})=\sup_{x\in\mathbb R}|F_{(w)}(x)-F_{(s)}(x)|\)
is the Kolmogorov-Smirnov distance between \(F_{(w)}\) and \(F_{(s)}\)
\citep{JSSv008i18}. A detailed proof of inequality
\eqref{eq-ipm-ks-final} is provided in the online supplementary
material.

Condition \ref{condition-bounded-tv} is satisfied by several commonly
used IPMs in the one-dimensional setting, including the
Kolmogorov-Smirnov distance and IPMs generated by uniformly
bounded-variation function classes. When the distributions are supported
on a compact interval, the 1-Wasserstein distance satisfies this
condition, since every 1-Lipschitz function has bounded variation on a
compact interval. Similarly, finite collections of polynomial test
functions of fixed order have bounded variation on compact intervals,
thus moment-based discrepancies are also covered in this setting. For
empirical samples, the observed support is finite, so the
bounded-variation condition can be verified on the empirical range.
However, population-level guarantees for distributions with unbounded
support require additional moment, tail, or truncation arguments.

\subsection{Dvoretzky-Kiefer-Wolfowitz Inequality under
Dependence}\label{dvoretzky-kiefer-wolfowitz-inequality-under-dependence}

The bound in inequality \eqref{eq-ipm-ks-final} reduces the problem of
controlling the empirical IPM error to that of controlling the uniform
deviation of the empirical CDF from its population counterpart. Under
independence, this is addressed by the classical
Dvoretzky-Kiefer-Wolfowitz inequality
\citep{DvoretzkyKieferWolfowitz1956}. However, since the observations
within each window are assumed to be exponentially \(\alpha\)-mixing
rather than independent, a dependent analogue of this inequality is
required. Lemma \ref{lem:lemma1} provides this result.

\begin{lemma}[DKW-type inequality under exponentially $\alpha$-mixing]
\label{lem:lemma1}
Let $\{X_t\}_{t\ge1}$ be a strictly stationary univariate time series
with continuous marginal CDF $F$. Suppose its strong-mixing coefficients satisfy
$\alpha(h)\le \exp(-\lambda h)$ with $h\ge1$ for some $\lambda>0$. For a sample  $X_1,\ldots,X_T$, let $\widehat F_T(x) = \frac{1}{T}\sum_{i=1}^T\mathbf 1\{X_i\le x\},$ and define $Y_i(x)=\mathbf 1\{X_i\le x\}-F(x)$ and let $
\sigma^2=\sup_{x\in\mathbb R}\sup_{i\ge1}\left[\operatorname{Var}\{Y_i(x)\}+2\sum_{j>i}\left|\operatorname{Cov}\{Y_i(x),Y_j(x)\}\right|\right]$. Then $\sigma^2<\infty$, and there exists a constant $C>0$, depending only
on $\lambda$, such that, for every $T\ge 2$ and every $\varepsilon>0$,
$$
\Pr\left(
\sup_{x\in\mathbb R}
|\widehat F_T(x)-F(x)|>\varepsilon
\right)
\le
\frac{6}{\varepsilon}
\exp\left[
-\frac{CT\varepsilon^2}
{4\left\{
\sigma^2+T^{-1}
+(\varepsilon/2)(\log T)^2
\right\}}
\right],
$$
\end{lemma}

\subsection{Finite-Sample IPM Error Bounds and Optimal Window
Size}\label{finite-sample-ipm-error-bounds-and-optimal-window-size}

Lemma \ref{lem:lemma1} controls the uniform deviation of a single
empirical CDF under exponential \(\alpha\)-mixing. Theorem
\ref{theo:AWEBU} extends this to bound the absolute error of the
empirical IPM computed between each pair of reference and stride
windows, providing the finite-sample guarantee required to control
Type-I and Type-II errors of the local hypothesis test at each candidate
split.

\begin{theorem}[Absolute IPM Error Bound under $\alpha$-mixing and Condition \ref{condition-bounded-tv}]
\label{theo:AWEBU}
Let $\{X_t\}_{t=1}^T\subset\mathbb R$ be a univariate time series, and
consider reference and stride windows of integer sizes $w,s\ge2$.
Assume that the observations within each window are strictly stationary
and exponentially $\alpha$-mixing, with mixing coefficients satisfying
$\alpha(h)\le e^{-\lambda h}$ for some $\lambda>0$ and all $h\ge1$.
Assume that the marginal CDF of each window is continuous.

Let $F_{(w)}$ and $\widehat F_{(w)}$ (respectively, $F_{(s)}$ and
$\widehat F_{(s)}$) denote the population and empirical distribution
functions of the reference and stride windows, and assume that the IPM
function class $\mathcal G$ satisfies Condition
\ref{condition-bounded-tv} with $0<C_{\mathcal G}<\infty$. Then, for every $\varepsilon>0$,
$$
\begin{aligned}
&\Pr\left(
\left|
d_{\mathcal G}\left(\widehat F_{(w)},\widehat F_{(s)}\right)
-
d_{\mathcal G}\left(F_{(w)},F_{(s)}\right)
\right|>\varepsilon
\right)\\
&\qquad\le
\frac{12C_{\mathcal G}}{\varepsilon}
\left[
\exp\{-w\varepsilon^2a_w(\varepsilon)\}
+
\exp\{-s\varepsilon^2a_s(\varepsilon)\}
\right],
\end{aligned}
$$
where, for $n\in\{w,s\}$, $a_n(\varepsilon)=\frac{C}{16C_{\mathcal G}^2\left\{\sigma^2+n^{-1}+\frac{\varepsilon}{4C_{\mathcal G}}(\log n)^2\right\}}$, $C>0$ is the constant in Lemma \ref{lem:lemma1}, and $\sigma^2<\infty$ is a common upper bound for the covariance-variance quantities in that lemma for the reference and stride windows.
\end{theorem}
\begin{remark}[Windows containing a change-point]
\label{rem:contaminated-window}
For equal window sizes satisfying $2w_T<\tau_{\min}$, each combined reference-stride region contains at most one change-point, and applying Lemma~\ref{lem:lemma1} separately to its stationary subsegments yields an analogous absolute IPM error bound with the same convergence rate and adjusted constants; see the online supplementary material for details.
\end{remark}

Consequently, under the conditions of Theorem \ref{theo:AWEBU}, with
\(C_{\mathcal G}\) and \(\sigma^2\) uniformly bounded, the absolute IPM
error has rate \(O_p\!\left(\sqrt{\log(w)/w}+\sqrt{\log(s)/s}\right)\)
as \(\min\{w,s\}\to\infty\). The two terms are the respective
contributions of the reference and stride windows. For a fixed local
length \(N=w+s\), the corresponding conservative error bound is
minimized by balancing the windows, \(w=s=N/2\) and justifies SCAN's use
of adjacent equal-size windows. Formal corollary statements and proofs
for above results are provided in the online supplementary material.

\subsection{Consistency of the SWAL
Statistic}\label{consistency-of-the-swal-statistic}

We establish consistency of the proposed SWAL estimator
\(\widehat{\tau}_n\) by applying Theorem \ref{theo:AWEBU} to the
1-Wasserstein distance as the IPM under the compact-interval setting. By
uniformly controlling the difference between the empirical and
population SWAL statistics in Equation \eqref{eq-swal-statistic}, we
then show that the population criterion admits a unique maximizer,
yielding consistency of the empirical estimator.

\begin{theorem}[Consistency of $\widehat{\tau}_n$]
\label{theo:consistcp}
Fix $\eta\in(0,1/2)$ and set $\mathcal K_n=\{k:\lceil\eta n\rceil\le k\le\lfloor(1-\eta)n\rfloor\}$, $G_n(k)=\left\{k(n-k)/n\right\}^{1/2}W_1\left(F_{1:k},F_{k+1:n}\right)$, and $\tau_n\in\arg\max_{k\in\mathcal K_n}G_n(k)$. Let $\widehat G_n(k)$ be the empirical counterpart, let $\widehat\tau_n\in\mathcal K_n$ be its maximizer, and set $\bar G_n(k)=n^{-1/2}G_n(k)$ and $\bar{\widehat G}_n(k)=n^{-1/2}\widehat G_n(k)$.

Suppose \textbf{(i)} $\sup_{k\in\mathcal K_n}|\bar{\widehat G}_n(k)-\bar G_n(k)|\xrightarrow{p}0$; \textbf{(ii)} $\tau_n$ is the unique maximizer of $G_n$ over $\mathcal K_n$ and, for every $\varepsilon>0$, there exists $c_\varepsilon>0$, independent of $n$, such that $\bar G_n(\tau_n)-\sup_{\substack{k\in\mathcal K_n\\|k-\tau_n|\ge\varepsilon n}}\bar G_n(k)\ge c_\varepsilon$ whenever the comparison set is nonempty; and \textbf{(iii)} $\widehat G_n(\widehat\tau_n)\ge\widehat G_n(\tau_n)-o_p(1)$. Then $|\widehat\tau_n-\tau_n|/n\xrightarrow{p}0$ as $n\to\infty$.
\end{theorem}

Lemmas 1.1-1.3 in the online supplementary material establish that the
conditions assumed in Theorem \ref{theo:consistcp} are satisfied for the
proposed SWAL estimator. Beyond consistency, the population SWAL
criterion admits a natural connection to classical mean-change detection
as shown in Proposition \ref{prop:1}.

\begin{proposition}[Similarity with the CUSUM statistic]\label{prop:1}
Let $\{X_t\}_{t=1}^{n}$ follow the single change in mean model with
$X_t=\mu_1+Z_t$ for $t\le\tau_n$ and $X_t=\mu_2+Z_t$ for $t>\tau_n$,
where the $Z_t$ share a mean-zero distribution with finite first moment,
and let $G_n(k)$ denote the population SWAL criterion in Equation \eqref{eq-swal-statistic}. Then,
for every $k\in\{1,\dots,n-1\}$,
\[
G_n(k)=\left\{\frac{k(n-k)}{n}\right\}^{1/2}
W_1\left(F_{1:k},F_{k+1:n}\right)
=\left\{\frac{k(n-k)}{n}\right\}^{1/2}
\left|\mu_{1:k}-\mu_{k+1:n}\right|.\]
which coincides with the CUSUM statistic.
\end{proposition}

Under a pure mean shift with all other distributional properties fixed,
the proposed \(1\)-Wasserstein localization statistic reduces to the
classical CUSUM statistic (a proof is provided in the online
supplementary material). Beyond this special case, the statistic remains
sensitive to broader distributional shifts, including changes in
variance and joint mean-variance shifts. This makes it suitable as a
nonparametric localization cost for offline change-point methods that
traditionally employ CUSUM-type criteria. The online supplementary
material further illustrates numerically that the SWAL statistic is
maximized near the true change-point under changes in mean, variance,
joint mean-variance, and more general distributional structure.

\subsection{Consistency of the Multiple Change-Point Detection
Framework}\label{consistency-of-the-multiple-change-point-detection-framework}

The estimator scans the candidate split points
\(\{t_m=mw_T:m=1,\dots,M\}\), rejecting split \(t_m\) whenever
\(T_{t_m}>b_{T,m}\), and returns a single change-point estimate from
each flagged region via the SWAL localization step. The proof that the
resulting estimator is consistent for both the number and the locations
of the change-points combines two arguments. First, the concentration
bound of Theorem \ref{theo:AWEBU}, applied with a threshold of order
\(\sqrt{\log T/w_T}\) and a sufficiently large fixed multiplier,
controls the Type-I and Type-II error probabilities uniformly over the
grid; under Conditions \ref{cond-1-minimal-sp} and
\ref{cond-2-min-signal}, the grid-coverage and error-control lemmas in
the supplement show that both error probabilities vanish asymptotically,
so that, with probability tending to one, no change-point-free split is
flagged and no true change-point is missed. Second, on this event the
condition \(w_T=o(\tau_{\min})\) forces the rejected splits to cluster
into exactly \(k\) maximal runs, each isolating a single change-point;
hence \(\widehat k=k\), and Theorem \ref{theo:consistcp} establishes
consistency of the estimated location within each run. Detailed proofs
for the lemmas are provided in the online supplementary material.

\begin{theorem}[Consistency of the Consolidated SCAN Estimator]
\label{theo:consistNcp}
Let $\widehat k$ and $\widehat\tau_1<\dots<\widehat\tau_{\widehat k}$ be the output of the estimated-change-point SCAN algorithm (Algorithm~\ref{alg:SCAN_full}), run with window size $w_T=\lfloor T^{\beta}\rfloor$, where $\beta\in(0,1)$ and $w_T=o(\tau_{\min})$; $M$ two-sample tests; and thresholds $b_{T,m}=A_{T,m}\sqrt{\log T/w_T}$. Suppose that $k$ is fixed as $T\to\infty$, that
Conditions \ref{cond-1-minimal-sp} and \ref{cond-2-min-signal} hold, that the local IPM
bound of Theorem~\ref{theo:AWEBU} applies with $c>0$; and that the conditions of
Theorem \ref{theo:consistcp} hold with rate $\rho_T=o(1)$. Assume further that there exist fixed constants $A_*,A^*>0$ and $T_0<\infty$ such that, for all $T\ge T_0$ and all $m$, $0<A_*\le A_{T,m}\le A^*<\infty$ and $cA_*^2\ge 1-\tfrac{\beta}{2}$.
Then
$$
\Pr\left(\widehat k=k\ \text{ and }\
\max_{1\le j\le k}\left|\widehat\tau_j-\tau_j\right|\le T\rho_T\right)\to 1,
$$
so the framework is consistent in both the number and the relative locations of the
change-points.
\end{theorem}

Note that the consistency result relies on deterministic thresholds of
order \(\sqrt{\log T/w_T}\). In practice, we use tapered block-bootstrap
thresholds that adapt to the local dependence structure, as explained in
detail in Section~\ref{sec-num-imple}. Establishing uniform bootstrap
validity over a growing collection of local tests would require a
separate analysis of dependent empirical processes and extreme bootstrap
quantiles and is beyond the scope of this work.

For a fixed finite ensemble, if every base SCAN detector satisfies
Theorem \ref{theo:consistNcp} with rate \(\rho_T=o(1)\) and the merging
radius satisfies \(2T\rho_T\le r_T<\tau_{\min}/2\), then, for any fixed
voting threshold \(\nu\in(0,1]\), Ensemble-SCAN consistently estimates
both the number and relative locations of the change-points; a detailed
explantion is provided in the online supplementary material.

\section{Numerical Implementation}\label{sec-num-imple}

The theoretical results are stated for abstract admissible window sizes,
a fixed finite ensemble size, a fixed voting threshold \((\nu)\), and a
tolerance distance \((r_T)\) satisfying the required separation
conditions. In practice, defaults must be chosen to balance statistical
performance and computational cost. We set the localization radius to
\(r_T = \min(\mathcal{W})\), since using the smallest window size
provides the finest localization resolution.

\subsection{Adaptive Threshold}\label{adaptive-threshold}

\textbf{Null distribution of test statistic via tapered block bootstrap}

The generic threshold \(b_{T,m}\) introduced in Section \ref{sec-meth}
is implemented locally at candidate split \(t=t_m\) and window size
\(w\) using a tapered block bootstrap. We denote the resulting
empirically calibrated threshold by \(\widehat b_{T,m}\), with
dependence on the fixed window size suppressed. Under the null
hypothesis \(H_{0,t_m}\), we assume both the stride and reference
windows come from identical distributions. We therefore pool
observations from both windows into the local sequence
\(Z_{t,1:2w}=\left(X_{t-w+1},\ldots,X_t,X_{t+1},\ldots,X_{t+w}\right),\)
with the goal of estimating the distribution of the test statistic
\({T}_{t,w}\) under \(H_{0,t}\).

To do so, we generate the collection of \(B\) tapered block bootstrap
replicates \(\left\{Z_{t,1:2w}^{*(b)}\right\}_{b=1}^{B}\) from the
pooled local sequence \(Z_{t,1:2w}\). Compared to the classical
bootstrap \citep{efron1994introduction}, block bootstrap methods are
better suited to dependent data, as they preserve the serial dependence
structure by resampling contiguous blocks of observations. Among these,
the tapered block bootstrap is adopted here, as it additionally produces
smoother transitions between adjacent resampled blocks than the standard
block bootstrap. The block length is set to \(\ell\asymp(2w)^{1/3}\),
corresponding to the length \((2w)\) of the pooled local region. While
the theoretical calibration is stated for the ideal bootstrap
distribution, equivalently assuming that \(B\to\infty\), the numerical
implementation approximates this distribution using a finite number
\(B\) of tapered block bootstrap replications. Then for each replicate
\(b = 1, \ldots, B\), we partition \(Z_{t,1:2w}^{*(b)}\) into a
bootstrap reference segment and a bootstrap stride segment of lengths
\(w\) and \(s\), respectively, and compute the corresponding bootstrap
test statistic \({T}_{t,w}^{*(b)}
=
d_{\mathcal{G}}\left(\widehat{F}_{(w),t}^{(b)},\, \widehat{F}_{(s),t}^{(b)}\right),\)
where \(\widehat{F}_{(w),t}^{(b)}\) and \(\widehat{F}_{(s),t}^{(b)}\)
denote the empirical distribution functions of the bootstrap reference
and stride segments at candidate split \(t=t_m\). The data-driven
threshold is then defined as the empirical \((1-\alpha')\)-quantile of
the bootstrap distribution: \(\widehat b_{T,m}
=\inf\left\{q \in \mathbb{R} :\frac{1}{B}\sum_{b=1}^{B}\mathbf{1}\left({T}_{t,w}^{*(b)} \le q\right)\ge 1-\alpha'\right\}.\)

\textbf{FWER control in the multiple testing framework}

For a fixed window size, the scan performs \(M=\lfloor T/w\rfloor-1\)
local tests. We use the Bonferroni-adjusted level \(\alpha'=\alpha/M\),
which controls the family-wise error rate without requiring independence
among the local statistics. Less conservative procedures such as
Holm-Bonferroni \citep{Holm-Bonferroni} and Hochberg \citep{10HOCHBERG}
require global ordering across the many local tests and are not used.
Pseudocode appears as Algorithm S1 in the online supplement.

\subsection{SCAN}\label{scan}

We summarize the full SCAN framework in Algorithm \ref{alg:SCAN_full}.
For a fixed window size \(w\), SCAN examines the candidate splits
\(t_m=mw\): at each split it compares the non-overlapping reference and
stride windows through the local IPM discrepancy \(T_{t_m}\) and rejects
the local null \(H_{0,t_m}\) when \(T_{t_m}\) exceeds the
bootstrap-calibrated threshold \(\widehat b_{T,m}\). In the
implementation, this threshold is the tapered-block-bootstrap quantile
calibrated under the local null, with a Bonferroni adjustment across the
\(M\) local tests that controls the family-wise error rate. This step
returns the set of rejected splits. Because the grid spacing equals
\(w\), a single change-point can fall within the windows of two adjacent
splits and trigger both; we show in Theorem \ref{theo:consistNcp} that
it triggers at most two, and that these are necessarily consecutive. We
therefore group consecutive rejected splits into a single flagged region
before localizing, which prevents one change-point from being reported
twice. Within each flagged region, we then estimate the change-point
location as the point that maximizes the SWAL statistic, yielding an
accurate estimate of the true change-point.

\begin{algorithm}[!htbp]
\DontPrintSemicolon
\caption{SCAN}
\label{alg:SCAN_full}

\KwIn{Time series $\{X_t\}_{t=1}^T$; window size $w$; bootstrap replications $B$; significance level $\alpha$.}
\KwOut{$\mathcal{R}$: list of detected change-points.}

Set $M \leftarrow \lfloor T/w\rfloor-1$, $\alpha'\leftarrow\alpha/M$,
$\widehat{\mathcal J}\leftarrow\emptyset$, and $\mathcal R\leftarrow\emptyset$.\;

\For{$m=1,\ldots,M$}{
    Define the reference and stride windows as
    $X_m^{(w)}\leftarrow X_{(m-1)w+1:mw}$ and
    $X_m^{(s)}\leftarrow X_{mw+1:(m+1)w}$\;
    Compute the data-driven threshold
    $\widehat b_{T,m}\leftarrow\textsc{AdaptiveThreshold}(X_m^{(w)},X_m^{(s)},B,\alpha')$
    and the IPM test statistic
    $T_m\leftarrow d_{\mathcal G}(X_m^{(w)},X_m^{(s)})$\;
    \If{$T_m>\widehat b_{T,m}$}{
        Flag split $m$ as rejected:
        $\widehat{\mathcal J}\leftarrow\widehat{\mathcal J}\cup\{m\}$\;
    }
}
Merge adjacent rejected splits forming $k$ localization-regions $\widehat{\mathcal C}_1,\ldots,\widehat{\mathcal C}_{\widehat k}$;
\For{$\ell=1,\ldots,\widehat k$}{
    Apply SWAL to the consolidated region
    $X_\ell^{(\mathrm{loc})}\leftarrow X_{(\underline m-1)w+1:\,(\overline m+1)w}$,
    with $\underline m=\min\widehat{\mathcal C}_\ell$,
    $\overline m=\max\widehat{\mathcal C}_\ell$, to obtain the change-point
    estimate $\widehat\tau_\ell$, and set
    $\mathcal R\leftarrow\mathcal R\cup\{\widehat\tau_\ell\}$\;
}

\Return{$\mathcal R$}\;
\end{algorithm}

\subsection{Ensemble-SCAN}\label{ensemble-scan}

Let \(\mathcal W=\{w_1,\ldots,w_d\}\) denote the set of window sizes.
For each \(w_j\in\mathcal W\), we apply SCAN to obtain an estimated
change-point set \(\widehat{\mathcal R}^{(j)}
=\{\widehat\tau_1^{(j)},\ldots,\widehat\tau_{\widehat k_j}^{(j)}\}\).
The ensemble then combines detections across window sizes. Genuine
change-points are expected to receive support from several SCAN
detectors, whereas spurious detections are less likely to persist across
scales. Thus, the ensemble acts as a stability filter, reducing
sensitivity to the choice of a single window size.

\begin{figure}[t]
\centering
\includegraphics[width=\textwidth]{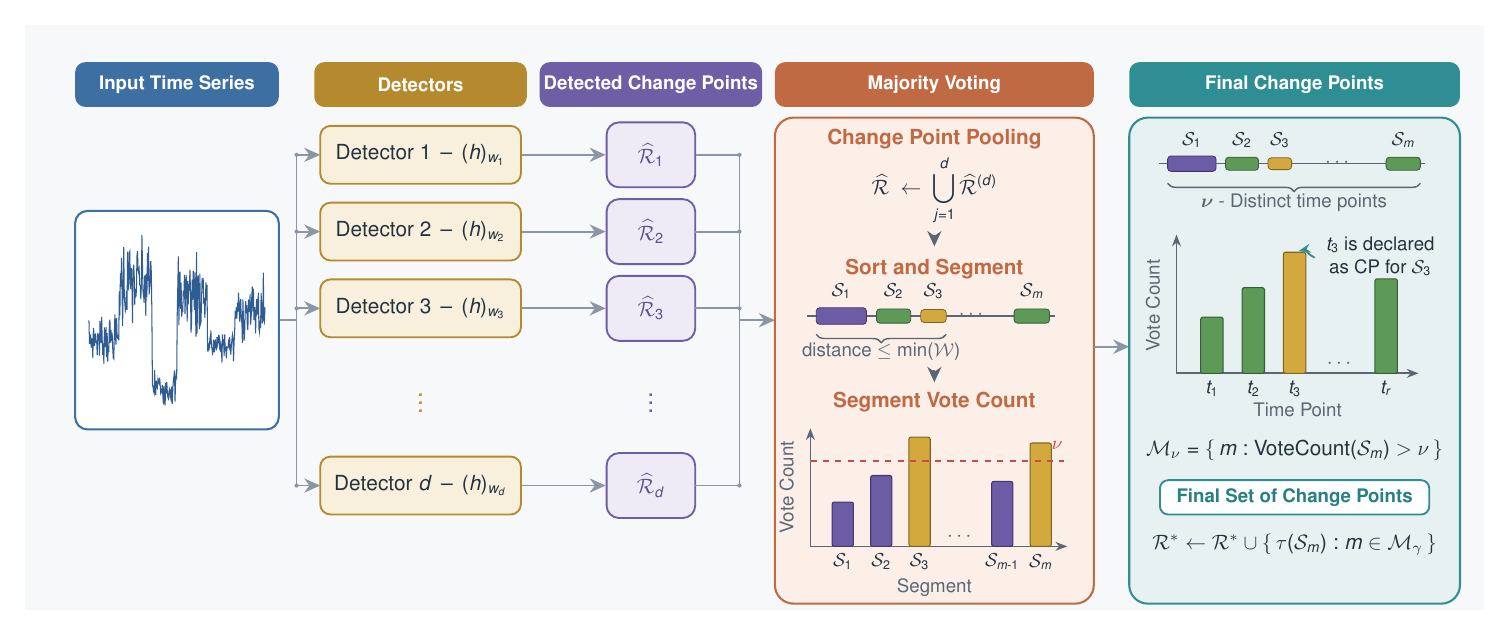}
\caption{Overview of the Ensemble-SCAN change-point detection framework. SCAN is applied across a collection of admissible window sizes. The resulting detections are pooled, grouped into local candidate clusters, and retained when their support proportion exceeds the majority-voting threshold.}
\label{fig:illustration-ensemble}
\end{figure}

Nearby detections are clustered, clusters supported by at least a
proportion \(\nu\) of the base detectors are retained, and each retained
cluster is represented by its most-supported location; full definitions
and pseudocode appear as Algorithm S2 in the online supplement. An
overview appears in Figure \ref{fig:illustration-ensemble}. In the
remainder of the paper, Ensemble-SCAN is referred to simply as SCAN
unless stated otherwise. Furthermore, practical guidance on selecting
the window-size range and voting threshold, along with default parameter
settings, is provided in the online supplementary material.

\section{Simulation Studies}\label{sec-sim-study}

\subsection{Data Generation}\label{data-generation}

We consider changes in mean and joint changes in mean and variance for
series lengths
\(T\in\{500,\allowbreak 1{,}000,\allowbreak 5{,}000,\allowbreak 10{,}000,\allowbreak 20{,}000,\allowbreak 50{,}000,\allowbreak 100{,}000,\allowbreak 1{,}000{,}000\}\),
with the number of change-points increasing with \(T\) while satisfying
the minimum spacing between consecutive changes. This extends beyond the
fixed-\(k\) regime used in our theoritical framework and provides a more
demanding empirical assessment of SCAN. Details of the change-point
generation are given in the online supplement.

We simulate these change-points under several dependence structures
ranging from independence to long-range dependence, as summarized in
Table \ref{tbl-dependence-structures}. Mean shifts range from 2 to 6
(both positive and negative), while variance changes are generated on
the log scale over \([0,\log(10)/2]\), following
\citet{KillickFearnheadEckley2012PELT}. We conduct 1,000 replications
for each scenario. The long-range dependence setting additionally
evaluates SCAN outside the dependence assumptions used for the
theoretical guarantees.

\begin{table}

\caption{\label{tbl-dependence-structures}Dependence structures for univariate simulations.}

\centering{

\centering
\captionsetup{skip=3pt}

\small
\setlength{\tabcolsep}{6pt}
\renewcommand{\arraystretch}{0.95}
\begin{tabular}{@{}lll@{}}
\toprule
Dependence structure & Model & Parameters \\
\midrule
Independent noise & AR(1) (white noise) & $\rho=0,\;\varepsilon_t\sim\mathcal N(0,1)$ \\
Short-term dependence & AR(1) & $\rho\sim\mathrm{Unif}[0,1],\;\varepsilon_t\sim\mathcal N(0,1)$ \\
& AR(1) & $\rho=0.7,\;\varepsilon_t\sim\mathcal N(0,1)$ \\
& ARMA(1,1) & $\rho=0.5,\;\theta=0.3,\;\varepsilon_t\sim\mathcal N(0,1)$ \\
Long-memory dependence & ARFIMA(1,$d$,1) & $d=0.35,\;\varepsilon_t\sim\mathcal N(0,1)$ \\
\bottomrule
\end{tabular}

}

\end{table}%

\textbf{Performance evaluation}

We use the covering metric \citep{VanDenBurgWilliams2020} and
\(F_1\)-score
\citep{TruongOudreVayatis2020RupturesReview, Aminikhanghahi2017} as
primary evaluation measures. The covering metric assesses agreement
between the segmentations induced by the true and estimated
change-points, while the \(F_1\)-score balances true positives, false
positives, and false negatives. For \(F_1\) computation, an estimate
\(\hat{\tau}\) is counted as a true positive if
\(|\hat{\tau}-\tau| \le \xi\), allowing for small localization errors.
Throughout the simulation study, we set \(\xi=10\).

\textbf{Comparison methods and parameter settings}

We compare SCAN with Binary Segmentation, PELT, FPOP, WBS, SBS, and KCP.
For joint mean-variance changes, methods restricted to mean-change
detection are omitted, while Binary Segmentation and PELT are
implemented using a Gaussian log-likelihood cost; KCP is excluded for
because of its substantial computational cost for long series
\((T\>10{,}000)\). Methods requiring more than 2 minutes for a given
series length are also excluded. Competing methods use their default
implementations with BIC-type penalties, with full implementation
details given in the online supplement. For SCAN, we use \(B=400\)
tapered block-bootstrap replications, ensemble size \(d=7\), voting
threshold \(\nu=0.5\), and minimum window size \(w_{\min}=30\), with
window sizes sampled uniformly from
\([w_{\min},\lfloor T^{2/3}\rfloor]\). These settings are fixed across
all simulation scenarios.

\subsection{Simulation Results}\label{simulation-results}

\textbf{Change in Mean}

\begin{figure}[t]
\centering
\includegraphics[width=\textwidth]{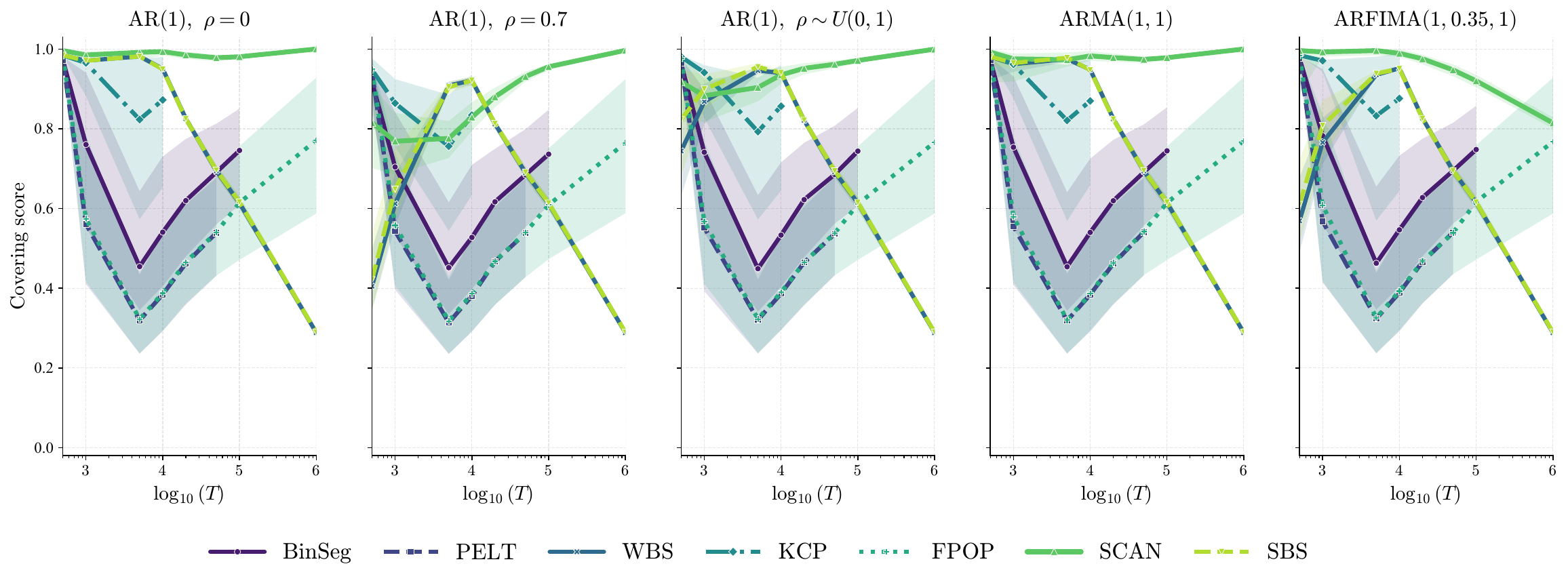}
\caption{Covering metric for change-in-mean detection across the series lengths and dependence structures considered in the simulation study. Lines and shaded bands show the median and interquartile range over 1{,}000 runs. Under the 2-minute limit, KCP is excluded for $T>10^4$, and PELT and Binary Segmentation for $T=10^6$.}
\label{fig:covering-univariate}
\end{figure}

Figure \ref{fig:covering-univariate} shows that SCAN is competitive for
short series and delivers its strongest relative performance for
moderately long to very long series across the dependence structures
considered. Similar patterns are observed for the \(F_1\)-score; see
online supplementary material. In settings where the theoretical
assumptions are satisfied, the covering metric remains close to 1 and,
in several cases, improves as the series length increases, providing
empirical support for the consistency of the proposed method. In most
longer-series settings, SCAN generally achieves higher detection
accuracy than competing change-point detection methods based on fixed
penalties.

Figure \ref{fig:covering-univariate} reports the median covering metric
over 1000 simulation runs, with the shaded bands representing the first
and third quartiles. For SCAN, these bands remain concentrated near the
median, indicating stable performance across the repeated simulations.
At shorter series lengths, several competing methods match or exceed
SCAN in some settings, whereas for longer series the competing methods
typically exhibit lower median performance together with more dispersed
bands, suggesting greater sensitivity to sampling variability and
dependence structure. Considered together, these results indicate that
SCAN's main empirical advantage is most pronounced in the long-series
regime targeted by the proposed method.

Under the ARFIMA setting, SCAN performs less favorably than under
short-range dependence and is more sensitive to the ensemble voting
threshold. This is expected because the bootstrap calibration assumes
exponentially \(\alpha\)-mixing dependence, whereas ARFIMA processes
exhibit long memory. The resulting inflation of the local IPM statistic
increases false detections, particularly at the default threshold
\(\nu=0.5\), and reduces both the covering score and the \(F_1\)-score.
As shown in Figure S1 of the online supplementary material, increasing
\(\nu\) mitigates this effect and yields more stable performance for
large \(T\). When the dependence structure is unknown, the proposed
scree-plot criterion provides an unsupervised method for selecting
\(\nu\) without requiring specification of the memory parameter.

\textbf{Change in Mean and Variance}

\begin{figure}[t]
\centering
\includegraphics[width=\textwidth]{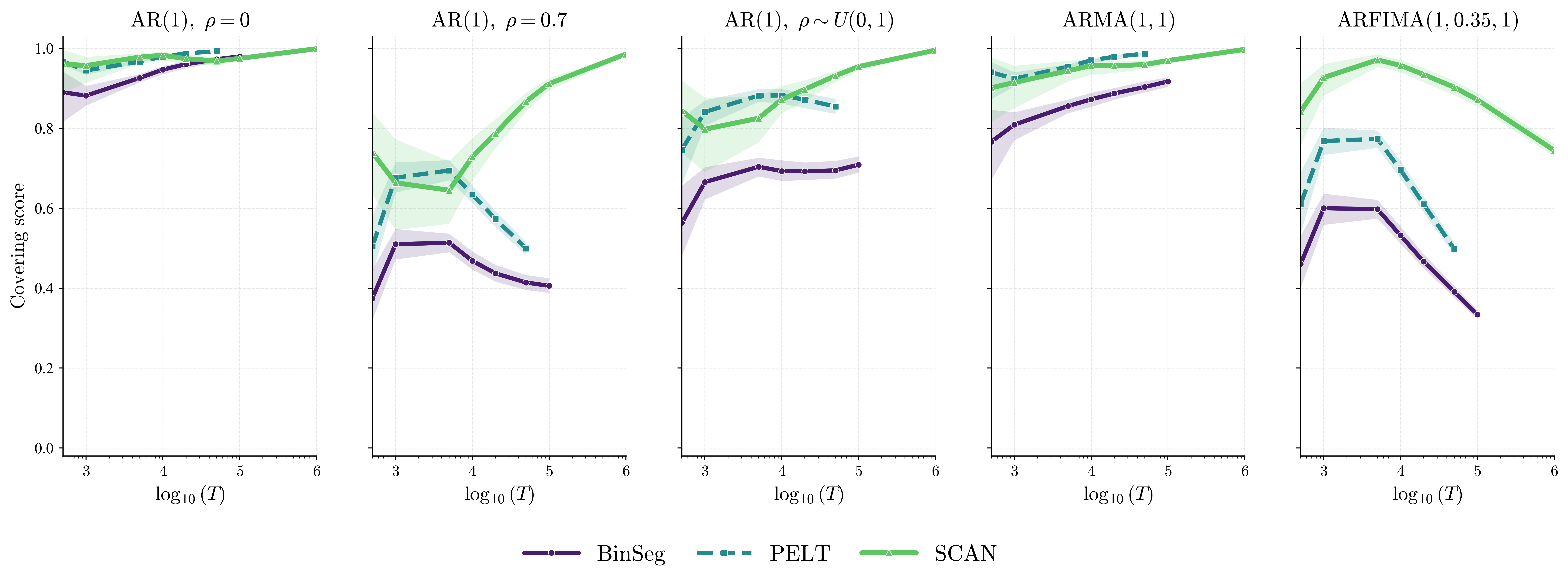}
\caption{Covering metric for joint mean-and-variance change detection, using the settings and display conventions of Figure \ref{fig:covering-univariate}. SBS, WBS, and FPOP are excluded because their implementations detect mean changes only; KCP is omitted because of its computational cost. PELT and Binary Segmentation are excluded for $T\ge10^5$ under the 2-minute limit.}
\label{fig:covering-univariate-meanvar}
\end{figure}

The results in Figure \ref{fig:covering-univariate-meanvar} show that
SCAN delivers its strongest relative covering performance for moderately
long to very long series across the dependence structures considered.
Similar patterns are observed for the \(F_1\)-score; see online
supplementary material. In settings where the theoretical assumptions
are satisfied, the covering metric for SCAN remains close to 1 and, in
several cases, improves as the series length increases, providing
empirical support for the consistency of the proposed method. For short
series, the results are more mixed, with some competing methods
performing similarly or better in selected settings. In most
longer-series settings, SCAN generally achieves higher detection and
localization accuracy than competing change-point detection methods
based on a BIC penalty.

In contrast, the performance of BinSeg and PELT deteriorates under
serial dependence. Both methods are implemented using a Gaussian
log-likelihood cost, so their segmentation criterion becomes
misspecified when the data are correlated. In particular, serial
dependence inflates the long-run variance, which can distort the cost
function and reduce the effectiveness of penalty terms such as BIC.
Consequently, their covering performance declines more noticeably in
dependent settings.

\begin{figure}[!htbp]
\centering
\captionsetup[subfigure]{skip=1pt}
\begin{subfigure}[t]{0.49\textwidth}
\centering
\includegraphics[width=\linewidth]{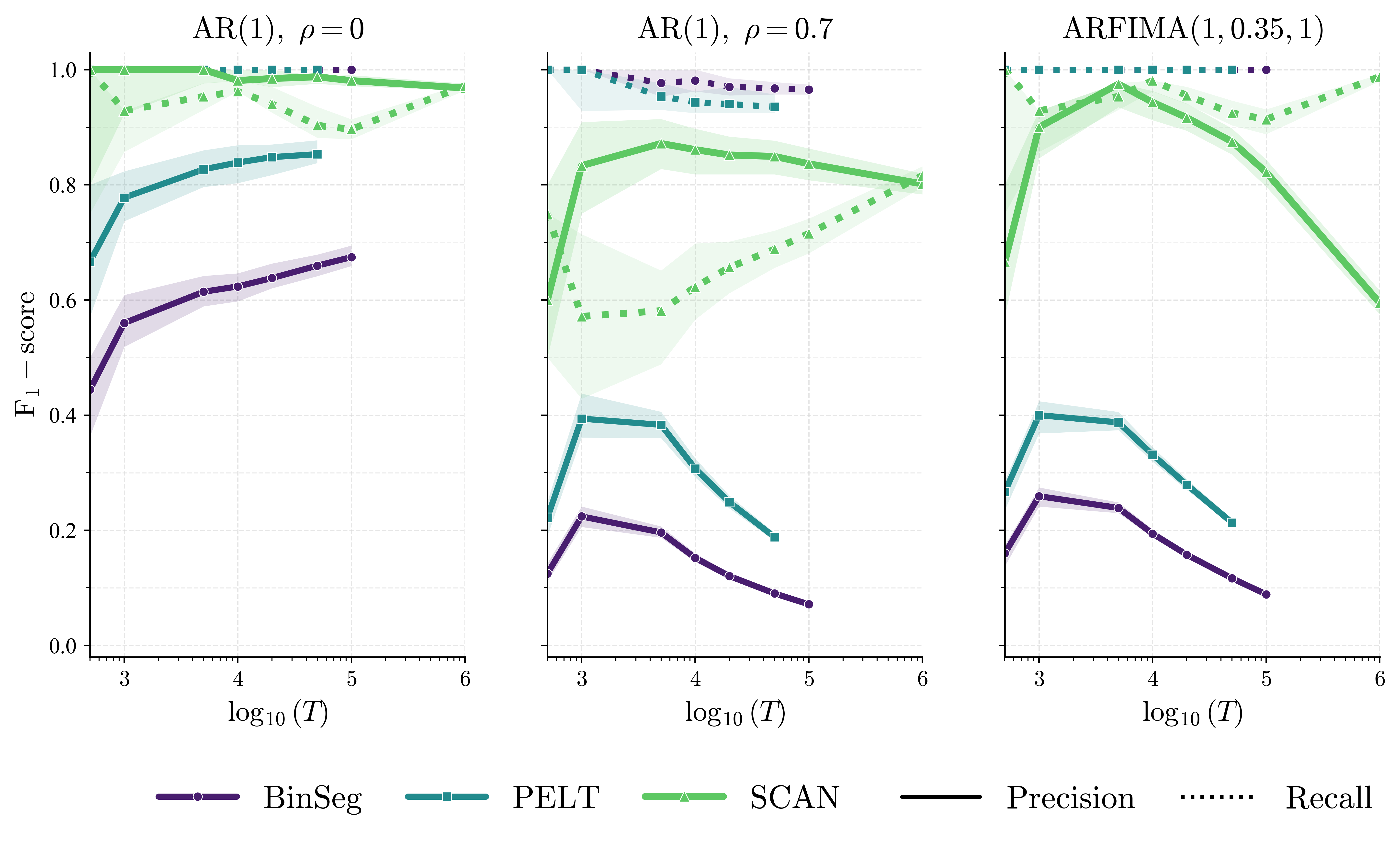}
\caption{}
\label{fig:precision-recall-panel}
\end{subfigure}\hfill
\begin{subfigure}[t]{0.49\textwidth}
\centering
\includegraphics[width=\linewidth]{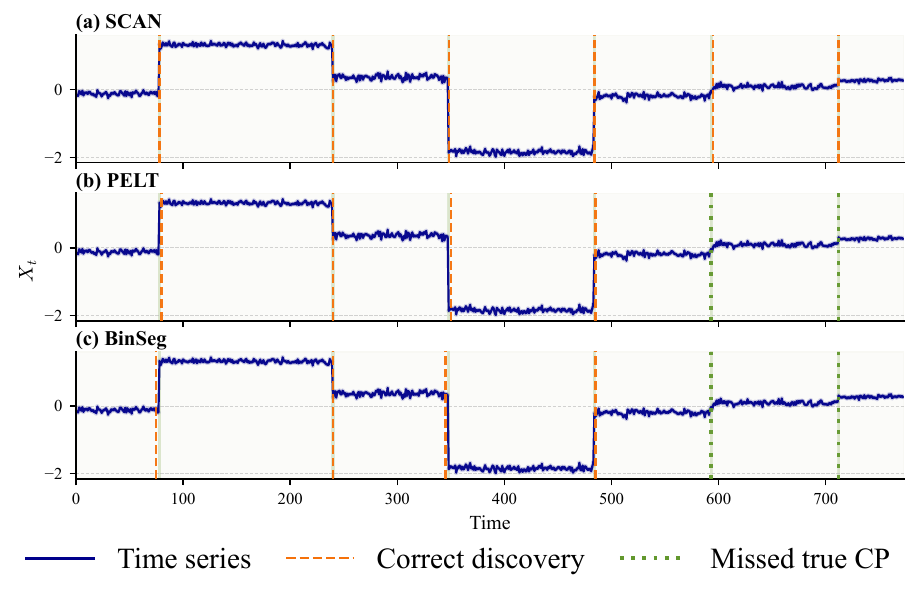}
\caption{}
\label{fig:benchmark-panel}
\end{subfigure}
\caption{Simulation diagnostics and benchmark comparison: (a) decomposition of precision and recall scores across series lengths under independence, serial correlation (AR(1), $\rho = 0.7$), and long-range dependence. The solid and dotted lines denote precision and recall, respectively, while the shaded regions represent the IQR; (b) benchmark comparison under varying shifts in mean and variance.}
\label{fig:simulation-diagnostics-benchmark}
\end{figure}

A complementary breakdown in terms of false positives and false
negatives is provided in Figure \ref{fig:precision-recall-panel}. This
shows that the precision of BinSeg and PELT is lower under dependence
than in the independent case, indicating a greater tendency to
over-segment through false positive detections. Figure
\ref{fig:precision-recall-panel} further highlights the consistency of
SCAN, which maintains a better balance between precision and recall.
Although the precision of SCAN also deteriorates under long-range
dependence, where its assumptions are violated, it remains competitive
with methods based on Gaussian cost functions.

Figure \ref{fig:benchmark-panel} compares the performance of SCAN, PELT,
and Binary Segmentation on a simulated series with heteroscedastic mean
and variance shifts (realization of a simulated series). SCAN
successfully detects all true change-points, including those
corresponding to relatively small shifts. In contrast, PELT and Binary
Segmentation with a BIC penalty miss several of the more subtle changes,
indicating that SCAN is more sensitive to both large and small shifts in
the mean and variance, thereby providing more accurate localization.

\textbf{Computational Times}

In the univariate setting, computing the empirical 1-Wasserstein
distance within a window of size \(w\) requires sorting and has cost
\(O(w\log w)\). Since SCAN performs \(O(T/w)\) local tests for a fixed
window size \(w\), the bootstrap-calibrated scan has complexity
\(O(BT\log w)\) for fixed \(w\), where \(B\) is the number of bootstrap
replications. For an ensemble over \(d\) window sizes, with \(B\) and
\(d\) treated as fixed hyperparameters, the resulting complexity is
approximately linear in \(T\) up to logarithmic factors.

Figure \ref{fig:runtimes} compares method runtimes (in seconds) from the
simulation study. SCAN remains computationally efficient across all
series lengths, rising from 0.006 seconds at \(T = 500\) to 1.065
seconds at \(T = 10^6\). A least-squares regression of \(\log_{10}\)
runtime on \(\log_{10} T\) yields empirical gradients of 0.71 (SCAN),
0.72 (SBS), 0.78 (WBS), 0.78 (FPOP), 1.75 (Binary Segmentation), 1.78
(PELT), and 1.84 (KCP). SCAN thus combines a small computational
constant with one of the lowest observed growth rates: at \(T = 10^6\)
it is approximately 23.5\(\times\) faster than WBS and 12.5\(\times\)
faster than SBS, and at \(T = 10^7\) it remains faster than FPOP.
Although SCAN incurs a higher initial cost than FPOP owing to the
tapered block bootstrap used for threshold calibration, its runtime
grows more slowly with \(T\). All simulations were run on a 2.4 GHz AMD
EPYC 7763 processor with 16 cores and 16 GB of RAM. Further details are
in the online supplementary material.

\section{Real Data Application}\label{sec-real-app}

We illustrate the performance of our proposed change-point detection
method using two real-world datasets. The first is the Human Activity
Sensing Consortium (HASC) dataset \citep{KawaguchiEtAl2011HASC}, which
contains high-frequency motion-sensor recordings collected during
sequences of human activities such as walking, jogging, standing, and
walking upstairs or downstairs. Activity transitions induce abrupt
changes in the distribution of the sensor signal, particularly through
changes in variance, making this dataset useful for evaluating
variance-change detection. The second is the Bitcoin (BTC-USD)
closing-price series \citep{zielinski_bitcoin} from January 2017 to
April 2026, which is used to evaluate detection of changes in mean,
variance, and trend.

\textbf{HASC Dataset}

\begin{figure}[!t]
\centering
\captionsetup[subfigure]{skip=1pt}
\begin{subfigure}{\textwidth}
\centering
\includegraphics[width=\linewidth]{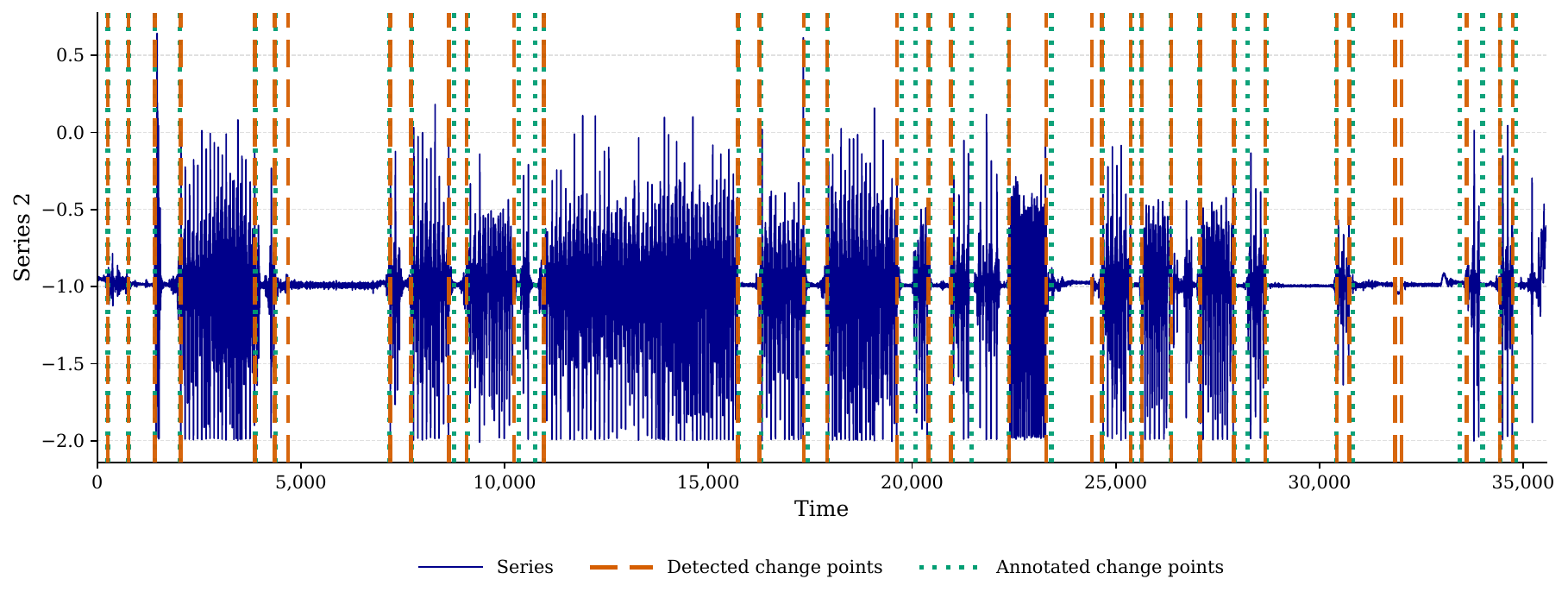}
\caption{}
\label{fig:hasc-with-detectedcps}
\end{subfigure}

\begin{subfigure}{\textwidth}
\centering
\includegraphics[width=\linewidth]{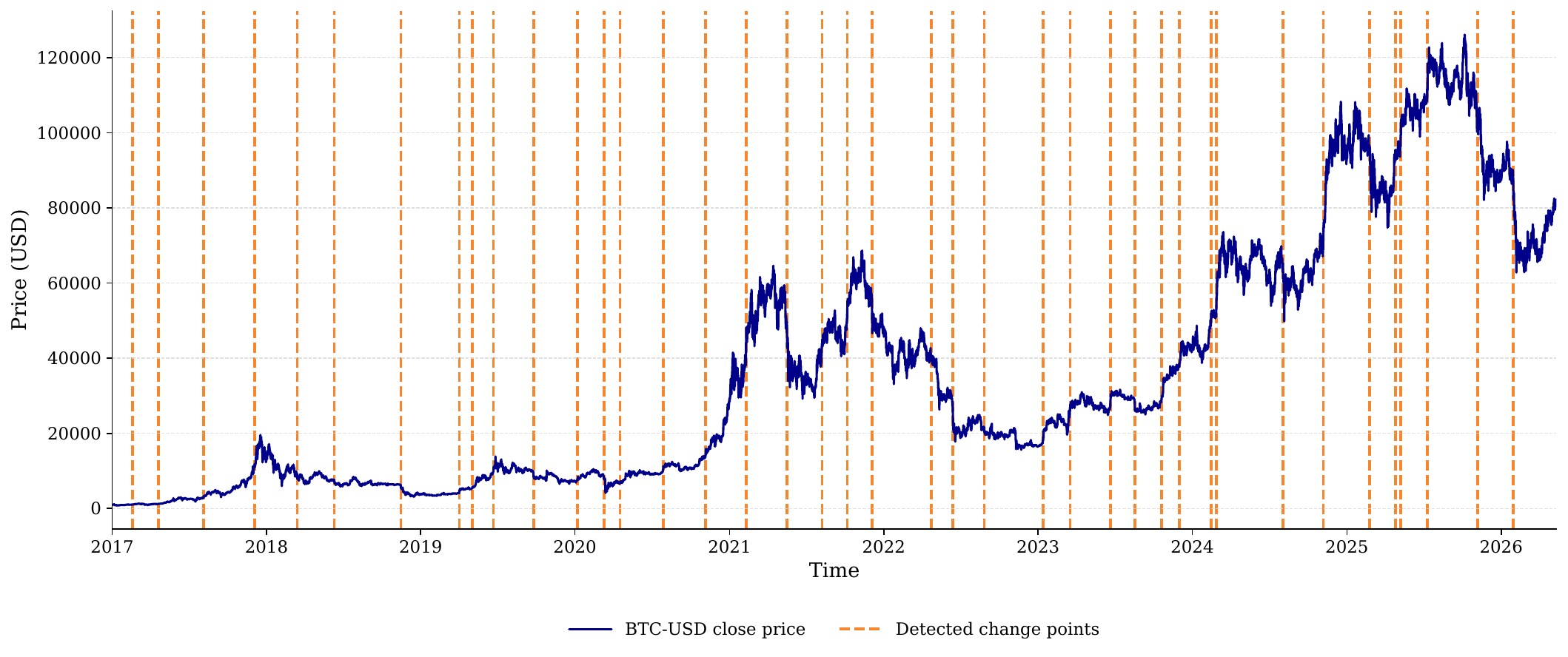}
\caption{}
\label{fig:btc-with-detectedcps}
\end{subfigure}
\caption{Real-data applications of SCAN: (a) Series 2 from HASC 2010 (individual 671-a), with dashed detected change-points and dotted annotated activity transitions; (b) BTC-USD hourly closing prices with detected mean and variance change-points.}
\label{fig:real-data-applications}
\end{figure}

For the HASC application, we use Series 2 from individual 671-a,
consisting of approximately 40,000 observations and 38 annotated
activity transitions. Figure \ref{fig:hasc-with-detectedcps} provides a
visual benchmark between the annotated change-points and those detected
by SCAN. The annotated change-points are mapped to the nearest
observation index for plotting.

Using a tolerance of \(\xi=120\) observations for matching detected and
annotated change-points, SCAN selects 37 change-points with an
\(F_1\)-score of 0.80 and a covering metric of 0.8360. Most detected
change-points align closely with the annotated activity transitions,
indicating that SCAN captures the main distributional changes in the
HASC signal. Further parameter settings for this application are
provided in the online supplementary material.

\textbf{Bitcoin Price}

Bitcoin provides a challenging setting for change-point detection
because it trades continuously and responds rapidly to changes in market
sentiment, regulation, institutional activity, and macroeconomic
conditions. We apply SCAN to the hourly BTC-USD closing price series
\citep{zielinski_bitcoin}, comprising 81,079 observations over 3,378
days. With no natural window scale specified a priori, we use the
ensemble formulation in Section~\ref{sec-theory}, sampling window sizes
uniformly up to \(w=\lfloor T^{2/3}\rfloor\).

SCAN detects change-points on 40 distinct days, many of which coincide
with major market events. Examples include the March 2020 COVID-19 crash
\cite{cawrey2020market}, the 9 February 2021 price surge following
Tesla's Bitcoin purchase announcement \cite{white2021bitcoin}, and the 6
November 2024 rally during the U.S. election period amid expectations of
more favorable cryptocurrency regulation \cite{westbrook2024bitcoin}.
The final detected change-point, in January 2026, coincides with a sharp
market decline associated with broad risk-off sentiment and liquidations
\cite{westbrook2026bitcoin}. Overall, 33 of the 40 detected
change-points coincide with dates linked to relevant Reuters reports.
Details of the labeled events in Figure \ref{fig:btc-with-detectedcps}
are provided in the online supplementary material.

This application illustrates that SCAN can extract interpretable
structural changes from a long, noisy, and highly non-stationary
financial time series. The resulting segmentation reflects both abrupt
market stress events and periods of rapid repricing, providing a
practical summary of how Bitcoin price dynamics evolve over time.

\section{Conclusion}\label{sec-conclusion}

We introduced SCAN, a scalable nonparametric framework for multiple
change-point detection in long univariate time series that combines
IPM-based local two-sample tests with tapered block-bootstrap
calibration, avoiding parametric cost functions and fixed global
penalties that can be difficult to calibrate under serial dependence and
distributional shifts. Under pure mean shifts, the proposed SWAL
statistic reduces to a CUSUM-type statistic, connecting SCAN to
classical methodology while retaining sensitivity to more general
distributional changes. Simulations with up to one million observations
show that the empirical gains of SCAN are most pronounced for moderately
long to very long series, where it generally achieves higher covering
and \(F_1\)-scores than Binary Segmentation, PELT, FPOP, WBS, SBS, and
KCP for mean and joint mean-variance shifts, particularly under serial
dependence where fixed-penalty methods tend to over-segment. We further
establish consistency of the estimated number and locations of
change-points under a regularity condition on the IPM class and
exponential \(\alpha\)-mixing dependence. Extending these guarantees to
long-memory processes, and the framework to multivariate and
high-dimensional settings, are natural directions for future work.

\textbf{Disclosure Statement}

The authors report there are no competing interests to declare.

\textbf{Data Availability Statement}

The HASC and BTC-USD datasets are publicly available. Replication code
and processed data will be released before publication.

\textbf{Declaration of generative AI}

While preparing this work, the authors used ChatGPT-5 to proofread and
enhance its language. After using this tool, the authors reviewed and
edited the content as needed.

\phantomsection\label{online-supplementary-material}
\bigskip

\begin{center}

{\large\bf ONLINE SUPPLEMENTARY MATERIAL}

\end{center}

The material contains lemmas, proofs computational details, additional
results from the paper. SCAN is implemented in the Python package
scan-cpd and R package scanr.

\bibliography{bibliography}

\end{document}


\maketitle

\section{Proofs of Theoretical
Results}\label{proofs-of-theoretical-results}

\subsection{The IPM Discrepancy Is Controlled by the Kolmogorov-Smirnov
Distance}\label{the-ipm-discrepancy-is-controlled-by-the-kolmogorov-smirnov-distance}

\begin{proof}
Let $H(x)=F_{(w)}(x)-F_{(s)}(x)$, for $x\in\mathbb R$. Since \(F_{(w)}\) and \(F_{(s)}\) are distribution functions, $\lim_{x\to-\infty}H(x)=\lim_{x\to\infty}H(x)=0$.
For \(f\in\mathcal G\), let \(df\) denote its finite signed
Lebesgue-Stieltjes measure under right continuity. For a function \(G\) of bounded variation, define  \(G(x-)=\lim_{y\uparrow x}G(y)\) for its left limit, as it exists
at every \(x\in\mathbb R\) and equals \(G(x)\) at continuity points.
For finite \(a<b\), the Stieltjes integration-by-parts formula gives
\[
\int_{(a,b]} f(x)\,dH(x)
+
\int_{(a,b]} H(x-)\,df(x)
=
f(b)H(b)-f(a)H(a),
\]
where \(H(x-)=F_{(w)}(x-)-F_{(s)}(x-)\).

Under the Condition 3 from the main text, the functions in $\mathcal{G}$ have uniformly bounded total variation. Therefore, applying the integration-by-parts formula for discrete distributions the IPM representation yields, 
\[
\int_{\mathbb R}
f(x)\,d\{F_{(w)}-F_{(s)}\}(x)
=
-
\int_{\mathbb R}
\{F_{(w)}(x-)-F_{(s)}(x-)\}\,df(x).
\]
Consequently,
\begin{align}
d_{\mathcal G}(F_{(w)},F_{(s)})
&=
\sup_{f\in\mathcal G}
\left|
\int_{\mathbb R}
f(x)\,d\{F_{(w)}-F_{(s)}\}(x)
\right|
\nonumber\\
&=
\sup_{f\in\mathcal G}
\left|
\int_{\mathbb R}
\{F_{(w)}(x-)-F_{(s)}(x-)\}\,df(x)
\right|
\nonumber\\
&\le
\sup_{f\in\mathcal G}
\left[
\sup_{x\in\mathbb R}
\left|
F_{(w)}(x-)-F_{(s)}(x-)
\right|
\int_{\mathbb R}|df(x)|
\right].
\label{eq-ipm-ks-bound}
\end{align}
For every \(x\in\mathbb R\),
\[
\left|F_{(w)}(x-)-F_{(s)}(x-)\right|
=
\lim_{y\uparrow x}
\left|F_{(w)}(y)-F_{(s)}(y)\right|
\le
\sup_{y\in\mathbb R}
\left|F_{(w)}(y)-F_{(s)}(y)\right|,
\]
because the left limit is the limit of the corresponding difference along
a sequence \(y\uparrow x\). Moreover, by Condition 3, $\int_{\mathbb R}|df(x)|=\operatorname{TV}_{\mathbb R}(f)\le C_{\mathcal G}$ uniformly over \(f\in\mathcal G\). Hence,
\[
d_{\mathcal G}(F_{(w)},F_{(s)})
\le
C_{\mathcal G}
\sup_{x\in\mathbb R}
\left|F_{(w)}(x)-F_{(s)}(x)\right|.
\]
Since $d_{\mathrm{KS}}(F_{(w)},F_{(s)})=\sup_{x\in\mathbb R}|F_{(w)}(x)-F_{(s)}(x)|$ is the Kolmogorov-Smirnov distance \cite{JSSv008i18} between $F_{(w)}$ and $F_{(s)}$, \eqref{eq-ipm-ks-bound} implies \begin{equation}
\label{eq-ipm-ks-final}
d_{\mathcal{G}}(F_{(w)},F_{(s)})
\le
C_{\mathcal{G}}
d_{\mathrm{KS}}(F_{(w)},F_{(s)}).
\end{equation}
\end{proof}

\subsection{Proof of Lemma 1}\label{proof-of-lemma-1}

\emph{Outline}: We apply Theorem 2 of
\cite{merlevede2012bernsteininequalitymoderatedeviations} to obtain a
Bernstein-type concentration inequality for centered \(\alpha\)-mixing
sequences, deriving a pointwise bound on \(|\widehat F_T(x)-F(x)|\). A
discretization argument based on a finite grid subsequently extends the
pointwise bound to a uniform bound over \(x\in\mathbb R\), yielding a
DKW-type inequality under \(\alpha\)-mixing dependence.

\begin{proof}
Let $\{X_t\}_{t\ge 1}$ satisfy the assumptions of the main-text DKW-type inequality under exponentially $\alpha$-mixing. For the observed sample $X_1,\ldots,X_T$, define $Y_i(x)=\mathbf{1}\{X_i\le x\}-F(x)$, so that $|Y_i(x)|\le 1$ and $\mathbb{E}[Y_i(x)]=0$. Define $S_T(x)=\sum_{i=1}^T Y_i(x)$ by
\begin{equation}
S_T(x)=\sum_{i=1}^T\bigl(\mathbf{1}\{X_i\le x\}-F(x)\bigr)=T\bigl(\widehat F_T(x)-F(x)\bigr).
\label{eq:lem1-sn-fn}
\end{equation}

For each fixed $x\in\mathbb R$, $Y_i(x)$ is a measurable transformation of
$X_i$. Since the $\sigma$-algebra generated by $Y_i$ is contained in the $\sigma$-algebra generated by $X_i$, 
$$
\alpha_Y(h)\le \alpha_X(h)\le e^{-\lambda h}.
$$
Moreover,
$$
\mathbb E\{Y_i(x)\}=0,
\qquad
|Y_i(x)|\le 1.
$$
Thus, for every fixed $x$, the sequence $\{Y_i(x)\}_{i\ge 1}$ satisfies
the centering, uniform boundedness, and exponential strong-mixing conditions
of Theorem 2 of \cite{merlevede2012bernsteininequalitymoderatedeviations}.
Applying the probability bound in that theorem with $M=1$ gives
\begin{equation}
\Pr\bigl(|S_T(x)|>t\bigr)
\le
\exp\left(
-\frac{Ct^2}{T\sigma^2 + 1 + t(\log T)^2}
\right),
\label{eq:lem1-bernstein}
\end{equation}
where $C>0$ depends only on $\lambda$ and
$$
\sigma^2
=
\sup_{x\in\mathbb R}\sup_{i\ge 1}
\left[
\operatorname{Var}\{Y_i(x)\}
+
2\sum_{j>i}
\left|\operatorname{Cov}\{Y_i(x),Y_j(x)\}\right|
\right].
$$
This variance bound is finite. Indeed, the strong-mixing covariance inequality \cite[eqn. (1.4)]{Bradley2005}
and $|Y_i(x)|\le 1$ imply
$$
\left|
\operatorname{Cov}\{Y_i(x),Y_{i+h}(x)\}
\right|
\le
4\alpha_X(h)
\le
4e^{-\lambda h}.
$$
Also,
$$
\operatorname{Var}\{Y_i(x)\}
=
F(x)\{1-F(x)\}
\le
\frac14.
$$
Consequently,
$$
\sigma^2
\le
\frac14+8\sum_{h=1}^{\infty}e^{-\lambda h}
=
\frac14+\frac{8e^{-\lambda}}{1-e^{-\lambda}}
<\infty.
$$
Using \eqref{eq:lem1-sn-fn} with $t=T\varepsilon$, we obtain the pointwise bound
\begin{equation}
\Pr\bigl(|\widehat F_T(x)-F(x)|>\varepsilon\bigr)
\le
\exp\left(
-\frac{C T\varepsilon^2}{\sigma^2+T^{-1}+\varepsilon(\log T)^2}
\right),
\label{eq:lem1-pointwise}
\end{equation}

For the uniform bound, it is enough to consider deviations in $(0,1]$, since
$\sup_x|\widehat F_T(x)-F(x)|\le 1$ makes the claimed probability zero for
deviations greater than one. We first fix an auxiliary
$\varepsilon\in(0,1)$ and choose an integer $k$ such that
$
\varepsilon^{-1}\le k\le \varepsilon^{-1}+1.
$
Since $F$ is continuous, choose $x_j\in\mathbb R$ satisfying
$F(x_j)=j/k$ for $j=1,\ldots,k-1$, and set
$x_0=-\infty$ and $x_k=\infty$, with
$F(x_0)=\widehat F_T(x_0)=0$ and
$F(x_k)=\widehat F_T(x_k)=1$.
For every $x\in\mathbb{R}$ there exists $j\in\{0,\ldots,k-1\}$ such that $x_j\le x\le x_{j+1}$. Since both $F$ and $\widehat F_T$ are nondecreasing because they are CDFs,
$$
F(x_j)\le F(x)\le F(x_{j+1}),
\qquad
\widehat F_T(x_j)\le \widehat F_T(x)\le \widehat F_T(x_{j+1}),
$$
and, by grid construction,
\begin{equation}
F(x_{j+1})-F(x_j)=\frac{1}{k}\le \varepsilon.
\label{eq:lem1-grid-gap}
\end{equation}

Define the event
$$
A
=
\left\{
\sup_{x\in\mathbb{R}}|\widehat F_T(x)-F(x)|>2\varepsilon
\right\},
$$
and write $A\subset A^+\cup A^-$, where
$$
A^+
=
\left\{
\sup_{x\in\mathbb{R}}\bigl(\widehat F_T(x)-F(x)\bigr)>2\varepsilon
\right\},
\qquad
A^-
=
\left\{
\sup_{x\in\mathbb{R}}\bigl(F(x)-\widehat F_T(x)\bigr)>2\varepsilon
\right\}.
$$

On the event $A^+$, there exists some $x\in\mathbb{R}$ such that
$
\widehat F_T(x)-F(x)>2\varepsilon.
$
Choose $j$ such that $x_j\le x\le x_{j+1}$. The monotonicity of the CDFs,
together with \eqref{eq:lem1-grid-gap}, gives
$$
\widehat F_T(x)-F(x)
\le
\bigl(\widehat F_T(x_{j+1})-F(x_{j+1})\bigr)+\varepsilon.
$$
It follows that $\widehat F_T(x_{j+1})-F(x_{j+1})>\varepsilon$, and hence
\begin{equation}
A^+
\subset
\bigcup_{j=1}^{k}
\left\{
|\widehat F_T(x_j)-F(x_j)|>\varepsilon
\right\}.
\label{eq:lem1-Aplus}
\end{equation}

Similarly, on the event $A^-$, there exists some $x\in\mathbb{R}$ such that
$
F(x)-\widehat F_T(x)>2\varepsilon.
$
Choose $j$ such that $x_j\le x\le x_{j+1}$. By monotonicity and
\eqref{eq:lem1-grid-gap},
$$
F(x)-\widehat F_T(x)
\le
F(x)-\widehat F_T(x_j)
\le
F(x_j)-\widehat F_T(x_j)+\frac{1}{k}
\le
F(x_j)-\widehat F_T(x_j)+\varepsilon,
$$
and so
\begin{equation}
A^-
\subset
\bigcup_{j=0}^{k-1}
\left\{
|\widehat F_T(x_j)-F(x_j)|>\varepsilon
\right\}.
\label{eq:lem1-Aminus}
\end{equation}

Combining \eqref{eq:lem1-Aplus}, \eqref{eq:lem1-Aminus}, and the union bound gives
\begin{align}
\Pr\left(\sup_{x\in\mathbb{R}}|\widehat F_T(x)-F(x)|>2\varepsilon\right)
&\le
\sum_{j=0}^{k}\Pr\bigl(|\widehat F_T(x_j)-F(x_j)|>\varepsilon\bigr) \notag\\
&\le
(k+1)\exp\left(
-\frac{C T\varepsilon^2}{\sigma^2+T^{-1}+\varepsilon(\log T)^2}
\right).
\label{eq:lem1-union}
\end{align}
Since $k+1\le 3/\varepsilon$ for $\varepsilon\in(0,1)$, substituting in \eqref{eq:lem1-union} gives
\begin{equation}
\Pr\left(\sup_{x\in\mathbb{R}}|\widehat F_T(x)-F(x)|>2\varepsilon\right)
\le
\frac{3}{\varepsilon}
\exp\left(
-\frac{C T\varepsilon^2}{\sigma^2+T^{-1}+\varepsilon(\log T)^2}
\right).
\end{equation}

Replacing $\varepsilon$ in the preceding inequality by $\varepsilon/2$ gives
for $0<\varepsilon\le 1$,
$$
\Pr\left(\sup_{x\in\mathbb{R}}|\widehat F_T(x)-F(x)|>\varepsilon\right)
\le
\frac{6}{\varepsilon}
\exp\left(
-\frac{C T\varepsilon^2}{4\left\{\sigma^2+T^{-1}+(\varepsilon/2)(\log T)^2\right\}}\right).
$$
For $\varepsilon>1$, the result is immediate because
$\sup_x|\widehat F_T(x)-F(x)|\le 1$.

\end{proof}

\subsection{Proof of Theorem 1}\label{proof-of-theorem-1}

\emph{Outline}: The proof uses the uniform bounded-variation condition
to control the IPM discrepancy measure \(d_{\mathcal{G}}\) using the
Kolmogorov-Smirnov distance. Applying the triangle inequality on
\(d_{\mathcal{G}}\) yields an upper bound on the absolute error
\(\left|d_{\mathcal{G}}(\widehat F_{(w)},\widehat F_{(s)})-d_{\mathcal{G}}(F_{(w)},F_{(s)})\right|\),
in terms of the two one-sample errors
\(d_{\mathcal{G}}(\widehat F_{(w)},F_{(w)})\) and
\(d_{\mathcal{G}}(\widehat F_{(s)},F_{(s)})\). Finally, we bound each
one-sample IPM error by applying the main-text DKW-type inequality under
exponentially \(\alpha\)-mixing to derive the stated finite-sample
probability bound.

\begin{proof}
Consider the time series $\{X_t\}_{t=1}^T\subset\mathbb{R}$ defined in the main text, with a split at time $t$ and corresponding reference and stride windows of sizes $w$ and $s$, respectively. Under the uniform bounded-variation condition, the IPM discrepancy between the CDFs of these windows satisfies

\begin{equation}
d_{\mathcal{G}}\left(F_{(w)},F_{(s)}\right)
\le
C_{\mathcal{G}}
\sup_{x\in\mathbb{R}}
|F_{(w)}(x)-F_{(s)}(x)|
=
C_{\mathcal{G}} \, d_{\mathrm{KS}}(F_{(w)},F_{(s)})
\label{eq:thm1-w1-sup}
\end{equation}

Applying the triangle inequality,
\begin{align}
d_{\mathcal{G}}\bigl(\widehat F_{(w)},\widehat F_{(s)}\bigr)
&\le
d_{\mathcal{G}}\bigl(\widehat F_{(w)},F_{(w)}\bigr)
+
d_{\mathcal{G}}\bigl(F_{(w)},F_{(s)}\bigr)
+
d_{\mathcal{G}}\bigl(F_{(s)},\widehat F_{(s)}\bigr),
\label{eq:thm1-tri-1}\\
d_{\mathcal{G}}\bigl(F_{(w)},F_{(s)}\bigr)
&\le
d_{\mathcal{G}}\bigl(F_{(w)},\widehat F_{(w)}\bigr)
+
d_{\mathcal{G}}\bigl(\widehat F_{(w)},\widehat F_{(s)}\bigr)
+
d_{\mathcal{G}}\bigl(\widehat F_{(s)},F_{(s)}\bigr).
\label{eq:thm1-tri-2}
\end{align}
Combining \eqref{eq:thm1-tri-1} and \eqref{eq:thm1-tri-2} yields
\[
\left|
d_{\mathcal{G}}\bigl(\widehat F_{(w)},\widehat F_{(s)}\bigr)
-
d_{\mathcal{G}}\bigl(F_{(w)},F_{(s)}\bigr)
\right|
\le
d_{\mathcal{G}}\bigl(\widehat F_{(w)},F_{(w)}\bigr)
+
d_{\mathcal{G}}\bigl(\widehat F_{(s)},F_{(s)}\bigr).
\]
Therefore,
\begin{align}
&\Pr\left(
\left|
d_{\mathcal{G}}\bigl(\widehat F_{(w)},\widehat F_{(s)}\bigr)
-
d_{\mathcal{G}}\bigl(F_{(w)},F_{(s)}\bigr)
\right|>\varepsilon
\right) \notag\\
&\qquad\le
\Pr\left(
d_{\mathcal{G}}\bigl(\widehat F_{(w)},F_{(w)}\bigr)
+
d_{\mathcal{G}}\bigl(\widehat F_{(s)},F_{(s)}\bigr)>\varepsilon
\right) \notag\\
&\qquad\le
\Pr\left(
d_{\mathcal{G}}\bigl(\widehat F_{(w)},F_{(w)}\bigr)>\frac{\varepsilon}{2}
\right)
+
\Pr\left(
d_{\mathcal{G}}\bigl(\widehat F_{(s)},F_{(s)}\bigr)>\frac{\varepsilon}{2}
\right).
\label{eq:thm1-union}
\end{align}

By substituting \eqref{eq:thm1-w1-sup} into \eqref{eq:thm1-union},
\[
d_{\mathcal{G}}\bigl(\widehat F_{(w)},F_{(w)}\bigr)>\frac{\varepsilon}{2}
\quad\Longrightarrow\quad
\sup_x\left|\widehat F_{(w)}(x)-F_{(w)}(x)\right|>\frac{\varepsilon}{2C_{\mathcal{G}}},
\]
and similarly for the stride window. Hence, applying the main-text DKW-type inequality under exponentially $\alpha$-mixing with sample sizes $w$ and $s$ gives
\begin{align}
&\Pr\left(
\left|
d_{\mathcal{G}}\bigl(\widehat F_{(w)},\widehat F_{(s)}\bigr)
-
d_{\mathcal{G}}\bigl(F_{(w)},F_{(s)}\bigr)
\right|>\varepsilon
\right) \notag\\
&\qquad\le
\frac{12C_{\mathcal{G}}}{\varepsilon}
\exp\left(
-\frac{Cw(\varepsilon/2C_{\mathcal{G}})^2}
{4\left\{\sigma^2+w^{-1}+(\varepsilon/4C_{\mathcal{G}})(\log w)^2\right\}}
\right)
+\\
&\qquad\quad\frac{12C_{\mathcal{G}}}{\varepsilon}
\exp\left(
-\frac{Cs(\varepsilon/2C_{\mathcal{G}})^2}
{4\left\{\sigma^2+s^{-1}+(\varepsilon/4C_{\mathcal{G}})(\log s)^2\right\}}
\right).
\label{eq:thm1-main}
\end{align}
Equivalently,
\begin{equation}
\Pr\left(
\left|
d_{\mathcal{G}}\bigl(\widehat F_{(w)},\widehat F_{(s)}\bigr)
-
d_{\mathcal{G}}\bigl(F_{(w)},F_{(s)}\bigr)
\right|>\varepsilon
\right)
\le
\frac{12C_{\mathcal{G}}}{\varepsilon}
\left\{
e^{-w\varepsilon^2 a_w(\varepsilon)}
+
e^{-s\varepsilon^2 a_s(\varepsilon)}
\right\},
\label{eq:thm1-main-aw}
\end{equation}
where
\[
a_w(\varepsilon)
=
\frac{C}{
16C_{\mathcal{G}}^2
\left\{
\sigma^2+w^{-1}
+
(\varepsilon/(4C_{\mathcal{G}}))(\log w)^2
\right\}}, 
\qquad
a_s(\varepsilon)
=
\frac{C}{
16C_{\mathcal{G}}^2
\left\{
\sigma^2+s^{-1}
+
(\varepsilon/(4C_{\mathcal{G}}))(\log s)^2
\right\}}.
\]
\end{proof}
\begin{remark}
\label{rem:contaminated-window}
Lemma 1 is stated for a strictly stationary time series. However, if a window of length $w$ contains a change point, the observations within that window are piecewise stationary. The window can therefore be decomposed into stationary subsegments, to each of which Lemma 1 can be applied. Suppose $X_1,\ldots,X_r \sim F_A$ and $X_{r+1},\ldots,X_w \sim F_B$. Then the population and empirical CDFs of the window satisfy

$$
F_{(w)}
=
\frac{r}{w}F_A
+
\frac{w-r}{w}F_B,
\qquad
\widehat F_{(w)}
=
\frac{r}{w}\widehat F_{A,r}
+
\frac{w-r}{w}\widehat F_{B,w-r}.
$$

Hence,

$$
\sup_x
\left|
\widehat F_{(w)}(x)-F_{(w)}(x)
\right|
\leq
\frac{r}{w}
\sup_x
\left|
\widehat F_{A,r}(x)-F_A(x)
\right|
+
\frac{w-r}{w}
\sup_x
\left|
\widehat F_{B,w-r}(x)-F_B(x)
\right|.
$$

Therefore, by a union bound,
$$\begin{aligned}
\Pr\left(
\sup_x
\left|
\widehat F_{(w)}(x)-F_{(w)}(x)
\right|>\epsilon
\right)
&\leq
\Pr\left(
\sup_x
\left|
\widehat F_{A,r}(x)-F_A(x)
\right|
>
\frac{w\epsilon}{2r}
\right)
\\
&\quad+
\Pr\left(
\sup_x
\left|
\widehat F_{B,w-r}(x)-F_B(x)
\right|
>
\frac{w\epsilon}{2(w-r)}
\right).
\end{aligned}$$

Each term involves a stationary sub-segment and can therefore be controlled using Lemma 1, yielding the same convergence order, but with adjusted sample constants. For a strictly stationary window, Lemma 1 gives, $\Pr\left(\sup_x \left|\widehat F(x)-F(x)\right|>\epsilon\right)\leq C\exp(-cw\epsilon^2)$. Decomposing into two stationary sub-segments with allocations $\epsilon/2$ to each yields $\Pr\left(\sup_x \left|\widehat F_{(w)}(x)-F_{(w)}(x)\right|>\epsilon\right)\leq 2C\exp\left(-\frac{c}{4}w\epsilon^2\right)$. Thus, the exponential convergence order in $w$ is preserved, although the finite-sample constants need not remain the same.

More generally, if a window contains multiple change-points, the same argument extends by expressing the empirical and population CDFs as weighted mixtures over the resulting stationary subsegments and applying a union bound across them.

\end{remark}

\subsection{Proof of Corollary 1}\label{proof-of-corollary-1}

\begin{corollary}[Convergence rate for the absolute IPM error]\label{cor:cor1}
Under the conditions of the main-text absolute IPM error bound, suppose that $C_{\mathcal G}$ and
$\sigma^2$ are bounded uniformly as the window sizes grow. Then, as
$\min\{w,s\}\to\infty$, the absolute IPM error satisfies
$$
\left|\,d_{\mathcal{G}}\left(\widehat{F}_{(w)},\,\widehat{F}_{(s)}\right)-d_{\mathcal{G}}\left(F_{(w)},\,F_{(s)}\right)\right| = O_{p}\left(\sqrt{\frac{\log w}{w}}+\sqrt{\frac{\log s}{s}}\right).
$$
\end{corollary}

\emph{Outline}: Let
\(Z_{w,s}=\left|d_{\mathcal{G}}(\widehat F_{(w)},\widehat F_{(s)})-d_{\mathcal{G}}(F_{(w)},F_{(s)})\right|\),
and \(r_{w,s}=\sqrt{\frac{\log w}{w}}+\sqrt{\frac{\log s}{s}}\). Using
the main-text absolute IPM error bound, we show that for
\(\varepsilon = M r_{w,s}\), with fixed \(M\ge1\) chosen sufficiently
large, we get \(\Pr\left(Z_{w,s} > M r_{w,s}\right) \longrightarrow 0\)
as \(\min\{w,s\}\to\infty\). Thus,
\(Z_{w,s}=O_p\left(\sqrt{\frac{\log w}{w}}+\sqrt{\frac{\log s}{s}}\right)\).

\begin{proof}
Let $Z_{w,s}=\left|d_{\mathcal{G}}\bigl(\widehat F_{(w)},\widehat F_{(s)}\bigr)-d_{\mathcal{G}}\bigl(F_{(w)},F_{(s)}\bigr)\right|$ and suppose $r_{w,s}=\sqrt{\frac{\log w}{w}}+\sqrt{\frac{\log s}{s}}.$
We show that $Z_{w,s}=O_p(r_{w,s})$ as $\min\{w,s\}\to\infty$. This is the convergence rate of the absolute IPM error under the uniform bounded-variation condition.
By the main-text absolute IPM error bound, for every $\varepsilon>0$,
\begin{equation}
\label{final:eq:tm1}
\Pr(Z_{w,s}>\varepsilon)
\le
\frac{12C_{\mathcal{G}}}{\varepsilon}
\left\{
\exp\{-w\varepsilon^2a_w(\varepsilon)\}
+
\exp\{-s\varepsilon^2a_s(\varepsilon)\}
\right\},
\end{equation}
where
$$
a_w(\varepsilon)
=
\frac{C}{
16C_{\mathcal{G}}^2
\left\{
\sigma^2+w^{-1}
+
\frac{\varepsilon}{4C_{\mathcal{G}}}(\log w)^2
\right\}},
$$
and $a_s(\varepsilon)$ is defined analogously with $s$ replacing $w$.
Consider a fixed $M\ge1$ and $\varepsilon=M r_{w,s}$. Since $r_{w,s}\to 0$ as $\min\{w,s\}\to\infty$, we have $\varepsilon\to0$. For $n\in\{w,s\}$, define $r_n=\sqrt{\frac{\log n}{n}}$.
Since $r_{w,s}\ge r_n$, we have $\varepsilon=M r_{w,s}\ge M r_n.$
For either $n\in\{w,s\}$, consider
$$
D_n(\varepsilon)
=
\sigma^2+n^{-1}
+
\frac{\varepsilon}{4C_{\mathcal{G}}}(\log n)^2.
$$
For all sufficiently large $n$, there exist constants $A,B>0$, independent
of $w$ and $s$, such that
$$
D_n(\varepsilon)\le A+B\varepsilon(\log n)^2.
$$
Therefore, for some constant $c_0>0$ depending on $C_{\mathcal{G}}$ and $C$, but independent of $n$, 
$$
n\varepsilon^2a_n(\varepsilon)
=
\frac{Cn\varepsilon^2}{16C_{\mathcal{G}}^2D_n(\varepsilon)}
\ge
c_0
\min\left\{
n\varepsilon^2,
\frac{n\varepsilon}{(\log n)^2}
\right\}.
$$
Since $\varepsilon\ge M\sqrt{\log n/n}$, we have $n\varepsilon^2 \ge M^2\log n$.
Also,
$$
\frac{n\varepsilon}{(\log n)^2}
\ge
M\frac{\sqrt{n\log n}}{(\log n)^2}
=
M\frac{\sqrt n}{(\log n)^{3/2}}.
$$
It follows that, for all sufficiently large $n$,
$
\frac{n\varepsilon}{(\log n)^2}
\ge
M\log n.
$
Hence, for fixed $M\ge1$ and all sufficiently large $n$,
\begin{equation}
\label{final:min}
n\varepsilon^2a_n(\varepsilon)
\ge
c_1M\log n,
\end{equation}
for some constant $c_1>0$.
Applying \eqref{final:min} to \eqref{final:eq:tm1} gives
$$
\exp\{-w\varepsilon^2a_w(\varepsilon)\}
\le
w^{-c_1M},
\qquad
\exp\{-s\varepsilon^2a_s(\varepsilon)\}
\le
s^{-c_1M}.
$$

Therefore,
\begin{equation}
\label{final-eq}
\Pr(Z_{w,s}>M r_{w,s})
\le
\frac{12C_{\mathcal{G}}}{M r_{w,s}}
\left\{
w^{-c_1M}
+
s^{-c_1M}
\right\}.
\end{equation}

Since $r_{w,s}\ge \sqrt{\log w/w}$ and $r_{w,s}\ge \sqrt{\log s/s}$, combining with \eqref{final-eq} we obtain,
$$
\frac{1}{r_{w,s}}w^{-c_1M}
\le
\frac{\sqrt w}{\sqrt{\log w}}w^{-c_1M}
=
\frac{w^{1/2-c_1M}}{\sqrt{\log w}},
$$
and similarly,
$$
\frac{1}{r_{w,s}}s^{-c_1M}
\le
\frac{s^{1/2-c_1M}}{\sqrt{\log s}}.
$$

Choosing a sufficiently large $M$ such that $c_1M>1/2$, both terms converge to zero as $\min\{w,s\}\to\infty$. Therefore,
$
\Pr(Z_{w,s}>M r_{w,s})\to 0.
$

Finally,
$$
\left|
d_{\mathcal{G}}\bigl(\widehat F_{(w)},\widehat F_{(s)}\bigr)
-
d_{\mathcal{G}}\bigl(F_{(w)},F_{(s)}\bigr)
\right|
=
O_p\left(
\sqrt{\frac{\log w}{w}}
+
\sqrt{\frac{\log s}{s}}
\right).
$$
\end{proof}

\subsection{Proof of Corollary 2}\label{proof-of-corollary-2}

\emph{Outline:} We apply the main-text absolute IPM error bound to the
IPM estimation error \(Z_{w,s}\). Writing
\(h_\varepsilon(n)=n\varepsilon^2a_n(\varepsilon)\), the two exponential
terms in the bound become \(\exp\{-h_\varepsilon(w)\}\) and
\(\exp\{-h_\varepsilon(s)\}\). We show that \(h_\varepsilon(n)\) is
increasing for \(n>7\), so the overall bound is controlled by the
smaller window size, \(\min\{w,s\}\). Subject to the constraint
\(N=w+s\), this minimum window size is maximized when \(w=s=N/2\), which
minimizes the resulting error bound.

\begin{corollary}[Optimal equal allocation of local window sizes]
Under the conditions of the main-text absolute IPM error bound, suppose $w,s>\exp(2)$ and $N=w+s$ is fixed. The upper bound for
$$
\left|\,d_{\mathcal{G}}\left(\widehat F_{(w)},\,\widehat F_{(s)}\right)-d_{\mathcal{G}}\left(F_{(w)},\,F_{(s)}\right)\right|
$$
is minimized when $w=s=N/2$.
\end{corollary}
\begin{proof}
Let
\[
Z_{w,s}
=
\left|
d_{\mathcal{G}}\left(\widehat F_{(w)},\widehat F_{(s)}\right)
-
d_{\mathcal{G}}\left(F_{(w)},F_{(s)}\right)
\right|.
\]
By the main-text absolute IPM error bound, for any fixed \(\varepsilon>0\),
\begin{equation}
\label{cor2:eq1}
\Pr(Z_{w,s}>\varepsilon)
\le
\frac{12C_{\mathcal G}}{\varepsilon}
\left[
\exp\{-w\varepsilon^2a_w(\varepsilon)\}
+
\exp\{-s\varepsilon^2a_s(\varepsilon)\}
\right].
\end{equation}
where
\[
a_n(\varepsilon)
=
\frac{C}
{
16C_{\mathcal{G}}^2
\left\{
\sigma^2+n^{-1}
+
\frac{\varepsilon}{4C_{\mathcal{G}}}(\log n)^2
\right\}
}.
\]
Let $h_\varepsilon(n)=n\varepsilon^2 a_n(\varepsilon)$. Once it is
shown that \(h_\varepsilon(n)\) is increasing in \(n\), we have
\[
\exp\{-h_\varepsilon(w)\}
+
\exp\{-h_\varepsilon(s)\}
\le
2\exp\{-h_\varepsilon(\min\{w,s\})\}.
\]

Thus the bound in \eqref{cor2:eq1} admits a conservative worst-window
bound controlled by $\min\{w,s\}$.
For a fixed and even local segment length $N=w+s$, this conservative
bound is minimized by maximizing $\min\{w,s\}$, which occurs at
$w=s=N/2$.
It remains to prove that \(h_\varepsilon(n)\) is increasing in \(n\).
Since \(C\varepsilon^2/(16C_{\mathcal{G}}^2)\) is positive and does not depend on \(n\),
this is equivalent to showing
\[
\phi(n)
=
\frac{n}
{
\sigma^2+n^{-1}
+
\frac{\varepsilon}{4C_{\mathcal{G}}}(\log n)^2
}
\]
is strictly increasing. For notational simplicity, we consider $D_n(\varepsilon)=\sigma^2+n^{-1}+\frac{\varepsilon}{4C_{\mathcal{G}}}(\log n)^2$.
Therefore,
\[
\phi(n)
=
\frac{n}{D_n(\varepsilon)}.
\]
Differentiating $\phi(n)$ with respect to $n$ yields
\[
\phi'(n)
=
\frac{
\sigma^2+2n^{-1}
+
\frac{\varepsilon}{4C_{\mathcal{G}}}
\left\{(\log n)^2-2\log n\right\}
}
{
\left\{
\sigma^2+n^{-1}
+
\frac{\varepsilon}{4C_{\mathcal{G}}}(\log n)^2
\right\}^2
}.
\]
Therefore, if \(n>e^2\),
\[
(\log n)^2-2\log n
=
\log n(\log n-2)>0.
\]
Because $n$ is an integer and $e^2\approx 7.389$, the condition $n>7$ implies $n\ge 8>e^2$. Hence $\phi'(n)>0$, so $h_\varepsilon(n)$ is strictly increasing for $n>7$.
\end{proof}

\subsection{Supporting Lemmas for Proving Theorem
2}\label{supporting-lemmas-for-proving-theorem-2}

The Lemma \ref{lem:uniform-convergence-G} states that, the scaled
empirical SWAL criterion converges uniformly in probability to its
population counterpart over all candidate split points away from the
boundaries.

\begin{lemma}
\label{lem:uniform-convergence-G}
Let $\{X_t\}_{t=1}^{n}$ be a time series and let
$$
\mathcal K_n(\eta)
=
\left\{
k:\lceil\eta n\rceil
\le k\le
\lfloor(1-\eta)n\rfloor
\right\},
\qquad
\eta\in(0,1/2),
$$
where $\eta$ is fixed as $n\to\infty$.
Suppose that the main-text absolute IPM error bound, together with its
extension in Remark 1 in the main text for subsamples containing
a single change-point, holds uniformly over
$k\in\mathcal K_n(\eta)$, with common constants $C_0$, $C_{\mathcal G}$,
and $C$, and a uniform upper bound on $\sigma^2$. Suppose also that the
population criterion $G_n(k)$ is the SWAL criterion. Define
$$
\bar G_n(k)=n^{-1/2}G_n(k),
\qquad
\bar{\widehat G}_n(k)=n^{-1/2}\widehat G_n(k).
$$
Then, as $n\to\infty$,
$$
\sup_{k\in\mathcal K_n(\eta)}
\left|
\bar{\widehat G}_n(k)-\bar G_n(k)
\right|
\xrightarrow{p}0.
$$
\end{lemma}
\begin{proof}
We prove uniform convergence of the scaled empirical SWAL criterion over the
trimmed index set \(\mathcal K_n(\eta)\).
For \(k\in\mathcal K_n(\eta)\), define
\[
R_n(k)
=
\left|
d_{\mathcal{G}}\left(\widehat F_{1:k},\widehat F_{k+1:n}\right)
-
d_{\mathcal{G}}\left(F_{1:k},F_{k+1:n}\right)
\right|.
\]
The \(1\)-Wasserstein distance on a common fixed compact interval satisfies
the uniform bounded-variation condition. For a candidate split \(k\), a
subsample lying within one stationary segment is covered directly by the
main-text absolute IPM error bound, while a subsample containing the single
change-point is covered by Remark 1. Hence, using the common uniform prefactor \(C_0\), for every
\(k\in\mathcal K_n(\eta)\) and \(\varepsilon>0\),
\begin{equation}
\label{eq:lemma:loc-consis-1}
\Pr\{R_n(k)>\varepsilon\}
\le
\frac{C_0}{\varepsilon}
\left[
\exp\{-k\varepsilon^2a_k(\varepsilon)\}
+
\exp\{-(n-k)\varepsilon^2a_{n-k}(\varepsilon)\}
\right],
\end{equation}
where \(a_\ell(\varepsilon)\) is defined as in the main-text absolute IPM error bound with window size
\(\ell\).
Let \(r_n=\sqrt{\log n/n}\), fix \(M>0\), and set
\(\varepsilon_n=Mr_n\). For every \(k\in\mathcal K_n(\eta)\) and
\(\ell\in\{k,n-k\}\),
\[
\eta n\le \ell\le n
\quad\text{and}\quad
\varepsilon_n(\log \ell)^2
\le
M\frac{(\log n)^{5/2}}{\sqrt n}
\longrightarrow0.
\]
Together with \(\ell^{-1}\le(\eta n)^{-1}\) and the uniform bounds in the
lemma, the definition of \(a_\ell\) therefore yields a constant \(c_0>0\),
independent of \(n\), \(k\), and \(\ell\), such that
\[
\inf_{k\in\mathcal K_n(\eta)}
\min\{a_k(\varepsilon_n),a_{n-k}(\varepsilon_n)\}
\ge c_0
\]
for all sufficiently large \(n\). Consequently, by
\eqref{eq:lemma:loc-consis-1}, the union bound, and
\(|\mathcal K_n(\eta)|\le n\), for any \(M>0\),
\begin{equation}
\begin{aligned}
\label{eq:lemma:loc-consis-2}
\Pr\left(
\sup_{k\in\mathcal K_n(\eta)}R_n(k)>M r_n
\right)
&\le
\sum_{k\in\mathcal K_n(\eta)}
\Pr\{R_n(k)>M r_n\} \\
&\le
\frac{C_0|\mathcal K_n(\eta)|}{M r_n}
\left[
\exp\{-c_0\eta n M^2 r_n^2\}
+
\exp\{-c_0\eta n M^2 r_n^2\}
\right].
\end{aligned}
\end{equation}
Since \(n r_n^2=\log n\), \eqref{eq:lemma:loc-consis-2} gives
\[
\Pr\left(
\sup_{k\in\mathcal K_n(\eta)}R_n(k)>M r_n
\right)
\le
\frac{2C_0}{M\sqrt{\log n}}
n^{3/2-c_0\eta M^2}.
\]
Choosing \(M\) sufficiently large makes the exponent
\(3/2-c_0\eta M^2\) negative. Hence
\[
\sup_{k\in\mathcal K_n(\eta)}R_n(k)
=
O_p\left(\sqrt{\frac{\log n}{n}}\right).
\]
By the definition of the scaled SWAL criterion,
\begin{equation}
\label{eq:lemma:loc-consis-3}
\left|
\bar{\widehat G}_n(k)-\bar G_n(k)
\right|
=
\frac{\sqrt{k(n-k)}}{n} R_n(k),
\end{equation}
and \(\sqrt{k(n-k)}/n\le 1/2\). Therefore,
\[
\sup_{k\in\mathcal K_n(\eta)}
\left|
\bar{\widehat G}_n(k)-\bar G_n(k)
\right|
\le
\frac{1}{2}
\sup_{k\in\mathcal K_n(\eta)}R_n(k)  = O_p\left(\sqrt{\frac{\log n}{n}}\right).
\]
Since \(\log n/n\to0\), the right-hand side is \(o_p(1)\), and hence
\[
\sup_{k\in\mathcal K_n(\eta)}
|\bar{\widehat G}_n(k)-\bar G_n(k)|\xrightarrow{p}0.
\]
\end{proof}

Lemma \ref{lem:wasserstein-mixture-linearity} below, shows that the
1-Wasserstein distance between two mixtures of the same distributions
\(F_0\) and \(F_1\) is proportional to the difference in their mixing
weights.

\begin{lemma}
\label{lem:wasserstein-mixture-linearity}
Let $F_0$ and $F_1$ be probability measures on $\mathbb{R}$ with finite first moments, and define $\Delta := W_1(F_0,F_1)$. For every $\lambda\in[0,1]$,
\[
W_1\bigl(F_0,(1-\lambda)F_0+\lambda F_1\bigr)
=
\lambda \Delta .
\]
More generally, for every $a,b\in[0,1]$, if
\[
M_a=(1-a)F_0+ aF_1,
\qquad
M_b=(1-b)F_0+ bF_1,
\]
then
\[
W_1(M_a,M_b)=|a-b|\Delta.
\]
\end{lemma}
\begin{proof}
By the Kantorovich-Rubinstein dual representation of the $1$-Wasserstein
distance, for two probability measures $P$ and $Q$ with finite first moments,
\[
W_1(P,Q)
=
\sup_{f\in \mathcal{F}_1}
\left|
\int f\,dP-\int f\,dQ
\right|,
\qquad
\mathcal{F}_1
=
\{f:\|f\|_{\mathrm{Lip}}\le 1\}.
\]
For every $f\in\mathcal F_1$, linearity of integration gives
\[
\begin{aligned}
\int f\,dM_a-\int f\,dM_b
&=
(1-a)\int f\,dF_0+a\int f\,dF_1
-
(1-b)\int f\,dF_0-b\int f\,dF_1\\
&=
(a-b)
\left(
\int f\,dF_1-\int f\,dF_0
\right).
\end{aligned}
\]
Taking the supremum over $f\in\mathcal F_1$ and pulling out the scalar,
\[
W_1(M_a,M_b)
=
|a-b|
\sup_{f\in\mathcal F_1}
\left|
\int f\,dF_1-\int f\,dF_0
\right|
=
|a-b|\,W_1(F_0,F_1)
=
|a-b|\Delta.
\]
The first identity follows by taking $a=0$, $b=\lambda$, since then
$M_0=F_0$ and $M_\lambda=(1-\lambda)F_0+\lambda F_1$, giving
$W_1\bigl(F_0,(1-\lambda)F_0+\lambda F_1\bigr)=\lambda\Delta$.
\end{proof}

Lemma \ref{lem:population-separation} shows that, in the presence of a
single change-point, the population SWAL criterion is uniquely maximized
at the true change-point and remains uniformly separated from all
sufficiently distant candidate locations.

\begin{lemma}
\label{lem:population-separation}
Let \(G_n(k)\) be the population SWAL criterion over
\(\mathcal K_n(\eta)=\{k:\lceil \eta n\rceil \le k \le \lfloor (1 - \eta)n\rfloor\}\),
and suppose there is a single change-point \(\tau_n\in\mathcal K_n(\eta)\)
with marginal distributions \(F_u=F_0\) for \(u\le\tau_n\), \(F_u=F_1\) for
\(u>\tau_n\), where $F_0$ and $F_1$ are probability measures with finite
first moments and $0<\Delta=W_1(F_0,F_1)<\infty$. Then the following holds:
\begin{enumerate}
\item[(i)] \emph{(Unique maximizer)} \(G_n(k) < G_n(\tau_n)\) for all
\(k\in\mathcal K_n(\eta)\), \(k\ne\tau_n\).
\item[(ii)] \emph{(Uniform separation)} For every
\(\varepsilon\in(0,1-2\eta)\) there exists \(c_\varepsilon>0\), depending only
on \((\varepsilon,\eta,\Delta)\), such that, whenever the comparison set is
nonempty,
\[
 \bar G_n(\tau_n) - \sup_{k\in\mathcal K_n(\eta):\,|k-\tau_n|\ge\varepsilon n} \bar G_n(k)
\;\ge\; c_\varepsilon.
\]
\end{enumerate}
\end{lemma}
\begin{proof}
For \(k\le\tau_n\), \(F_{1:k}=F_0\) and
\(F_{k+1:n}=\frac{\tau_n-k}{n-k}F_0+\frac{n-\tau_n}{n-k}F_1\); for
\(k>\tau_n\), \(F_{1:k}=\frac{\tau_n}{k}F_0+\frac{k-\tau_n}{k}F_1\) and
\(F_{k+1:n}=F_1\). By Lemma \ref{lem:wasserstein-mixture-linearity},
\(W_1(F_0,(1-\lambda)F_0+\lambda F_1)=\lambda\Delta\) for
\(\lambda\in[0,1]\), so
\begin{equation}
G_n(k)=
\begin{cases}
\dfrac{(n-\tau_n)\Delta}{\sqrt n}\sqrt{\dfrac{k}{n-k}}, & k\le\tau_n,\\[2ex]
\dfrac{\tau_n\Delta}{\sqrt n}\sqrt{\dfrac{n-k}{k}}, & k>\tau_n.
\end{cases}
\label{eq:Gn-closed-form}
\end{equation}

\emph{Proof of (i).}
Since \(k\mapsto\sqrt{k/(n-k)}\) is strictly increasing and
\(k\mapsto\sqrt{(n-k)/k}\) strictly decreasing, \eqref{eq:Gn-closed-form}
shows \(G_n\) is strictly increasing on \(\{k\le\tau_n\}\) and strictly
decreasing on \(\{k>\tau_n\}\); both branches agree at \(k=\tau_n\), so
\(\tau_n\) is the unique maximizer.

\emph{Proof of (ii).}
With \(\rho_n=\tau_n/n\) and \(q=k/n\), dividing
\eqref{eq:Gn-closed-form} by $\sqrt n$ gives
\(\bar G_n(k)=H(\rho_n,q)\), where
\[
H(\rho,q)=
\begin{cases}
(1-\rho)\Delta\sqrt{q/(1-q)}, & q\le\rho,\\
\rho\Delta\sqrt{(1-q)/q}, & q>\rho,
\end{cases}
\]
is jointly continuous on \([\eta,1-\eta]^2\) and, for each fixed \(\rho\),
strictly increasing in \(q\) on \(q\le\rho\) and strictly decreasing on
\(q>\rho\). The set
\(\mathcal A_\varepsilon=\{(\rho,q)\in[\eta,1-\eta]^2:|q-\rho|\ge\varepsilon\}\)
is compact and nonempty for \(\varepsilon<1-2\eta\), and
\(D(\rho,q)=H(\rho,\rho)-H(\rho,q)\) is continuous and strictly positive on
it, so \(c_\varepsilon:=\min_{\mathcal A_\varepsilon}D>0\). For any \(n\) and
\(k\in\mathcal K_n(\eta)\) with \(|k-\tau_n|\ge\varepsilon n\), the pair
\((\rho_n,k/n)\) lies in \(\mathcal A_\varepsilon\), hence
\(\bar G_n(\tau_n)-\bar G_n(k)=D(\rho_n,k/n)\ge c_\varepsilon\). Taking the supremum
over such \(k\) gives (ii).
\end{proof}

\subsection{Proof of Theorem 2}\label{proof-of-theorem-2}

\begin{proof}
Scaling the criterion by $n^{-1/2}$ does not change the location of its
maximizer. Define
$$
U_n
=
\sup_{k\in\mathcal K_n(\eta)}
\left|
\bar{\widehat G}_n(k)-\bar G_n(k)
\right|.
$$
By Lemma \ref{lem:uniform-convergence-G},
$
U_n\xrightarrow{p}0,
$
and hence $U_n=o_p(1)$.
Condition (iii), after division by $\sqrt n$, implies that there exists
a sequence $\delta_n=o_p(1)$ such that
$$
\bar{\widehat G}_n(\widehat\tau_n)
\ge
\bar{\widehat G}_n(\tau_n)-\delta_n.
$$
Fix $\varepsilon>0$. If the set
$
\left\{
k\in\mathcal K_n(\eta):
|k-\tau_n|\ge\varepsilon n
\right\}
$
is empty, then
$
\Pr\left(
|\widehat\tau_n-\tau_n|\ge\varepsilon n
\right)=0,
$
and the result is immediate. Otherwise, Lemma
\ref{lem:population-separation} gives a constant $c_\varepsilon>0$,
independent of $n$, such that
$$
\bar G_n(\tau_n)
-
\sup_{\substack{k\in\mathcal K_n(\eta)\\
|k-\tau_n|\ge\varepsilon n}}
\bar G_n(k)
\ge
c_\varepsilon.
$$

On the event
$\left\{|\widehat\tau_n-\tau_n|\ge\varepsilon n\right\},$
the uniform separation inequality implies
$
\bar G_n(\tau_n)-\bar G_n(\widehat\tau_n)
\ge
c_\varepsilon.
$
On the other hand,
$$
\begin{aligned}
\bar G_n(\tau_n)-\bar G_n(\widehat\tau_n)
&=
\bar G_n(\tau_n)
-
\bar{\widehat G}_n(\tau_n)\\
&\quad+
\bar{\widehat G}_n(\tau_n)
-
\bar{\widehat G}_n(\widehat\tau_n)\\
&\quad+
\bar{\widehat G}_n(\widehat\tau_n)
-
\bar G_n(\widehat\tau_n)\\
&\le
\left|
\bar G_n(\tau_n)
-
\bar{\widehat G}_n(\tau_n)
\right|\\
&\quad+
\left[
\bar{\widehat G}_n(\tau_n)
-
\bar{\widehat G}_n(\widehat\tau_n)
\right]\\
&\quad+
\left|
\bar{\widehat G}_n(\widehat\tau_n)
-
\bar G_n(\widehat\tau_n)
\right|\\
&\le
2U_n+\delta_n.
\end{aligned}
$$
The last inequality follows from the definition of $U_n$ and the
approximate-maximization condition.

Consequently,
$$
\left\{
|\widehat\tau_n-\tau_n|\ge\varepsilon n
\right\}
\subseteq
\left\{
2U_n+\delta_n\ge c_\varepsilon
\right\}.
$$
Therefore,
$$
\begin{aligned}
\Pr\left(
\frac{|\widehat\tau_n-\tau_n|}{n}\ge\varepsilon
\right)
&=
\Pr\left(
|\widehat\tau_n-\tau_n|\ge\varepsilon n
\right)\\
&\le
\Pr\left(
2U_n+\delta_n\ge c_\varepsilon
\right)\\
&\longrightarrow0,
\end{aligned}
$$
because $U_n=o_p(1)$, $\delta_n=o_p(1)$, and
$c_\varepsilon>0$ is independent of $n$.
Since this holds for every $\varepsilon>0$,
$$
\frac{|\widehat\tau_n-\tau_n|}{n}
\xrightarrow{p}0.
$$
\end{proof}

\subsection{Proof of Proposition 1}\label{proof-of-proposition-1}

\begin{proposition}[Population equivalence with CUSUM statistic]
Under a pure mean-shift model with a single change-point and common mean-zero noise distribution with finite first moments, for every \(k\in\{1,\ldots,n-1\}\),
$$
G_n(k)
=
\left\{\frac{k(n-k)}{n}\right\}^{1/2}
W_1\left(F_{1:k},F_{k+1:n}\right)
=
\left\{\frac{k(n-k)}{n}\right\}^{1/2}
\left|\mu_{1:k}-\mu_{k+1:n}\right|.
$$
\end{proposition}

\emph{Outline:} Consider the pure location-shift model, \[X_t =
\begin{cases}
\mu_1+Z_t, & t\le \tau_n,\\
\mu_2+Z_t, & t>\tau_n,
\end{cases}\] where \(\mu_1\) and \(\mu_2\) are the pre- and post-change
means, and \(Z_t\) has a common mean-zero distribution \(F_0\). For a
candidate split \(k\ne\tau_n\), one or both candidate segments are
mixtures of the pre- and post-change distributions. We express these
mixtures in terms of the proportions of post-change observations on
either side of \(k\) and apply the linearity of the 1-Wasserstein
distance along the resulting mixture path to show that
\(W_1(F_{1:k},F_{k+1:n})=|\mu_{1:k}-\mu_{k+1:n}|\).

\begin{proof}
Let $\{X_t\}_{t=1}^{n}$ follow the pure location-shift model
\[
X_t =
\begin{cases}
\mu_1+Z_t, & t\le \tau_n,\\
\mu_2+Z_t, & t>\tau_n,
\end{cases}
\]
where $Z_t$ has a common mean-zero distribution $F_0$ and a finite first moment.
Let $F_{\mu_j}$ denote the distribution of $\mu_j+Z_t$, for $j=1,2$.
We first record two elementary consequences of the location-shift structure.
For any location parameter $\mu$,
\[
F_\mu^{-1}(u)=\mu+F_0^{-1}(u),\qquad 0<u<1.
\]
Therefore,
\[
W_1(F_{\mu_1},F_{\mu_2})
=
\int_0^1
\left|F_{\mu_1}^{-1}(u)-F_{\mu_2}^{-1}(u)\right|\,du
=
|\mu_1-\mu_2|.
\]
Next, for any $a,b\in[0,1]$, define
\[
M_a=(1-a)F_{\mu_1}+aF_{\mu_2},
\qquad
M_b=(1-b)F_{\mu_1}+bF_{\mu_2}.
\]
By Lemma \ref{lem:wasserstein-mixture-linearity}, the mixture identity can be written as
\[
W_1(M_a,M_b)=|a-b|\,|\mu_2-\mu_1|.
\]

Now fix a candidate split $k\in\{1,\ldots,n-1\}$. Let
\[
\pi_L(k)=\frac{1}{k}\sum_{t=1}^{k}\mathbf{1}\{t>\tau_n\},
\qquad
\pi_R(k)=\frac{1}{n-k}\sum_{t=k+1}^{n}\mathbf{1}\{t>\tau_n\}
\]
be the proportions of post-change observations in the left and right candidate
segments. The corresponding segment distributions are therefore
\[
F_{1:k}
=
\{1-\pi_L(k)\}F_{\mu_1}+\pi_L(k)F_{\mu_2},
\qquad
F_{k+1:n}
=
\{1-\pi_R(k)\}F_{\mu_1}+\pi_R(k)F_{\mu_2}.
\]
Hence, applying the mixture identity above with
$a=\pi_L(k)$ and $b=\pi_R(k)$,
\[
W_1(F_{1:k},F_{k+1:n})
=
|\pi_R(k)-\pi_L(k)|\,|\mu_2-\mu_1|.
\]

The corresponding population means are
\[
\mu_{1:k}
=
\{1-\pi_L(k)\}\mu_1+\pi_L(k)\mu_2,
\qquad
\mu_{k+1:n}
=
\{1-\pi_R(k)\}\mu_1+\pi_R(k)\mu_2.
\]
Therefore,
\[
|\mu_{1:k}-\mu_{k+1:n}|
=
|\pi_R(k)-\pi_L(k)|\,|\mu_2-\mu_1|.
\]
Combining the last two displays gives
\[
W_1(F_{1:k},F_{k+1:n})
=
|\mu_{1:k}-\mu_{k+1:n}|.
\]
Substituting this into the SWAL statistic yields
\[
G_n(k)
=
\sqrt{\frac{k(n-k)}{n}}\,
W_1(F_{1:k},F_{k+1:n})
=
\sqrt{\frac{k(n-k)}{n}}\,
|\mu_{1:k}-\mu_{k+1:n}|,
\]
which is the population CUSUM statistic for a pure mean shift.
\end{proof}

\subsection{Consistency of the Change-Point
Detector}\label{consistency-of-the-change-point-detector}

Lemma \ref{lem:typeI} shows the consistency of global Type-I error under
the analytical threshold.

\begin{lemma}
\label{lem:typeI}
Let \(w_T=\lfloor T^\beta\rfloor\), \(\beta\in(0,1)\), with candidate splits \(t_m=mw_T\),
\(m=1,\ldots,M_T:=\lfloor T/w_T\rfloor-1\), test statistics \(T_{t_m}\),
population counterpart \(D_{T,m}\), and null set
\(\mathcal M_0:=\{m:D_{T,m}=0\}\). Suppose 
\(
b_{T,m}:=A_{T,m}\sqrt{\frac{\log T}{w_T}}\),
where the threshold multipliers \(A_{T,m}\) are deterministic constants
(possibly varying with \(T\) and \(m\)) and there exist fixed constants
\(A_*\) and \(A^*\) satisfying
\[
0 < A_*\le A_{T,m}\le A^*<\infty
\]
uniformly in \(T\) and \(m\). Suppose the main-text absolute IPM error
bound holds uniformly over \(\mathcal M_0\). In addition, suppose there
exist constants \(c>0\) and \(T_0\) such that
\[
\inf_{m\in\mathcal M_0}a_{w_T}(b_{T,m})\ge c
\]
for every \(T\ge T_0\), where the infimum over an empty set is defined to
be \(+\infty\), and assume that \(cA_*^2\ge1-\beta/2\). Then
\(\Pr\bigl(\mathcal E^{I}_{T,w_T}\bigr)\to0\) as \(T\to\infty\).
\end{lemma}
\begin{proof}
Define the global Type-I error event by
\[
\mathcal E^I_{T,w_T}
:=
\bigcup_{m\in\mathcal M_0}\{T_{t_m}>b_{T,m}\}.
\]
For every \(m\in\mathcal M_0\), \(D_{T,m}=0\), so the main-text absolute IPM error bound gives
$$
\Pr(T_{t_m} > u) \le \frac{24C_{\mathcal{G}}}{u}\exp\{-w_T u^2 a_{w_T}(u)\}.
$$
Because \(A_{T,m}\le A^*\) uniformly in \(T\) and \(m\), and
\(w_T\asymp T^\beta\),
\[
\sup_m b_{T,m}(\log w_T)^2
\le
A^*\sqrt{\frac{\log T}{w_T}}(\log w_T)^2
\longrightarrow0.
\]
Since
\[
w_Tb_{T,m}^2=A_{T,m}^2\log T,
\]
and, by assumption,
\[
\inf_{m\in\mathcal M_0}a_{w_T}(b_{T,m})\ge c
\]
for all sufficiently large \(T\), we have
\(a_{w_T}(b_{T,m})\ge c\) uniformly over
\(m\in\mathcal M_0\). If \(\mathcal M_0=\varnothing\), then
\(\mathcal E^I_{T,w_T}=\varnothing\) and the result is immediate. Otherwise,
evaluating at $u = b_{T,m}$ and applying the union bound over $\mathcal{M}_0$,
$$
\Pr\left(\mathcal{E}^{I}_{T,w_T}\right) 
\le \sum_{m \in \mathcal{M}_0} \frac{C}{b_{T,m}} T^{-cA_{T,m}^2}
\le M_T \cdot \frac{C}{b_{T,\min}} T^{-cA_*^2},
$$
where $b_{T,\min} = \min_{m \in \mathcal{M}_0} b_{T,m}$. Since
$b_{T,\min}\ge A_*\sqrt{\log T/w_T}$,
$w_T\asymp T^\beta$, $M_T=O(T^{1-\beta})$, and
$cA_*^2\ge 1-\beta/2$, it follows that
\[
\Pr\!\left(\mathcal E^I_{T,w_T}\right)
\le
\frac{C}{A_*\sqrt{\log T}}
T^{1-\beta/2-cA_*^2}
\le
\frac{C}{A_*\sqrt{\log T}}
\longrightarrow 0.
\]
\end{proof}

In Lemma \ref{lem:grid-coverage}, we show that, for sufficiently large
samples, every true change-point is covered by an isolated candidate
split whose discrepancy retains at least half of the true signal
strength and whose associated region contains no other change-points,
thereby ensuring that localization step is well defined.

\begin{lemma}
\label{lem:grid-coverage}
Let the candidate splits be \(t_m=mw_T\) with reference and stride windows of
common length \(w_T\), and suppose \(w_T=o(\tau_{\min})\). By the minimal-spacing condition,
$\tau_j\ge\tau_{\min}$, and $T-\tau_j\ge\tau_{\min}$. Since $w_T=o(\tau_{\min})$, every change-point is eventually more than
$w_T$ away from the boundaries. Hence its nearest grid point belongs to the admissible candidate set $m=1,\ldots,M_T$.
Then, for all sufficiently large \(T\), each change-point \(\tau_j\) admits a split
\(t_{m(j)}\) with \(|t_{m(j)}-\tau_j|\le w_T/2\) whose combined region
\((t_{m(j)}-w_T,\,t_{m(j)}+w_T]\) contains no other change-point; and if
\(d_{\mathcal G}(F_{S_j},F_{S_{j+1}})\ge\vartheta_{\mathcal G}>0\), then
\[
D_{T,m(j)}\ge\frac{\vartheta_{\mathcal G}}{2}.
\]
\end{lemma}
\begin{proof}
Because adjacent grid points are \(w_T\) observations apart, a nearest grid
point satisfies \(r_{T,j}:=|t_{m(j)}-\tau_j|\le w_T/2\). Condition
\(w_T=o(\tau_{\min})\) implies \(2w_T<\tau_{\min}\) eventually, so the
combined region centered at this grid point contains only \(\tau_j\).

If \(t_{m(j)}<\tau_j\), the reference-window distribution is \(F_{S_j}\),
whereas the stride-window distribution is
\[
\frac{r_{T,j}}{w_T}F_{S_j}
+
\left(1-\frac{r_{T,j}}{w_T}\right)F_{S_{j+1}}.
\]
If \(t_{m(j)}\ge\tau_j\), the analogous mixture occurs in the reference
window and the stride-window distribution is \(F_{S_{j+1}}\). In either case,
linearity of integration and positive homogeneity of the IPM give
\[
D_{T,m(j)}
=
\left(1-\frac{r_{T,j}}{w_T}\right)
d_{\mathcal G}(F_{S_j},F_{S_{j+1}})
\ge
\frac{\vartheta_{\mathcal G}}{2}.
\]
\end{proof}

In Lemma \ref{lem:typeII}, we show the consistency of global Type-II
error under the analytical threshold as in Lemma \ref{lem:typeI}.

\begin{lemma}
\label{lem:typeII}
Under the setup of Lemma \ref{lem:typeI}, let the $k$ true change-points
satisfy the assumptions of Lemma \ref{lem:grid-coverage}, with associated
splits $t_{m(j)}$, $j=1,\ldots,k$. Assume the absolute IPM error bound and
its single-change-point extension (Remark \ref{rem:contaminated-window})
hold uniformly over these splits, with common prefactor $C_0>0$. Then $\Pr\bigl(\mathcal E^{II}_{T,w_T}\bigr)\to0$ as $T\to\infty$.
\end{lemma}
\begin{proof}
Let $\delta_T=\tfrac{\vartheta_{\mathcal G}}{2}-b_{T,\max}$ with
$b_{T,\max}=\max_m b_{T,m}$. Since
$b_{T,\max}\le A^*\sqrt{\log T/w_T}=O(T^{-\beta/2}\sqrt{\log T})\to0$,
eventually $\delta_T\in[\vartheta_{\mathcal G}/4,\vartheta_{\mathcal G}/2]$.
By Lemma \ref{lem:grid-coverage},
$D_{T,m(j)}-b_{T,m(j)}\ge\delta_T$, so
$\{T_{t_{m(j)}}\le b_{T,m(j)}\}
\subseteq\{|T_{t_{m(j)}}-D_{T,m(j)}|>\delta_T/2\}$. Then $\inf_{\varepsilon\in K}\widetilde a_{w_T}(\varepsilon)\ge c_K(\log w_T)^{-2}$
for all large $T$, and the uniform error bound with $\varepsilon=\delta_T/2$ and a union bound
over the $k$ splits give
\[
\Pr\bigl(\mathcal E^{II}_{T,w_T}\bigr)
\le
\frac{2kC_0}{\delta_T}
\exp\!\Bigl\{-\tfrac{w_T\delta_T^2}{4}\,
\widetilde a_{w_T}(\delta_T/2)\Bigr\}.
\]
Since eventually $\delta_T/2$ lies in a fixed compact subset of
$(0,\infty)$, the assumed lower bound on $\widetilde a_{w_T}$ yields
\[
\frac{w_T\delta_T^2}{4}\,\widetilde a_{w_T}(\delta_T/2)
\ge
\frac{c_0\vartheta_{\mathcal G}^2}{64}\cdot
\frac{w_T}{(\log w_T)^2}
\to\infty,
\]
while the prefactor $2kC_0/\delta_T$ stays bounded ($k$ fixed,
$\delta_T\ge\vartheta_{\mathcal G}/4$). Hence
$\Pr\bigl(\mathcal E^{II}_{T,w_T}\bigr)\to0$.
\end{proof}

\subsection{Proof of Theorem 3}\label{proof-of-theorem-3}

\begin{proof}
For the window size $w_T$ satisfying $w_T=o(\tau_{\min})$ and thresholds $b_{T,m}$, define the Type-I and Type-II error events
$$
\mathcal E^{I}_{T,w_T}
:=
\bigcup_{m:\,D_{T,m}=0}\bigl\{T_{t_m}>b_{T,m}\bigr\},
\qquad
\mathcal E^{II}_{T,w_T}
:=
\bigcup_{j=1}^{k}\bigl\{T_{t_{m(j)}}\le b_{T,m(j)}\bigr\}.
$$
Here, $D_{T,m}>0$ indicates that the window $(t_m-w_T,t_m+w_T]$ contains a true change-point, and $m(j)$ is the nearest-grid index of $\tau_j$ from Lemma\ \ref{lem:grid-coverage}.
Consider the event $\mathcal Q_T:=(\mathcal E^{I}_{T,w_T}\cup\mathcal E^{II}_{T,w_T})^c$. By Lemmas \ref{lem:typeI} and \ref{lem:typeII} and the union bound,
$$
\Pr(\mathcal Q_T^c)\to0.
$$

Let $\widehat{\mathcal J}_T$ denote the set of rejected grid indices, and partition it into its maximal consecutive components $\mathcal C_{T,1},\dots,\mathcal C_{T,\widehat k}$, where $\widehat k$ is defined as the number of such components. For each $\ell\in\{1,\dots,\widehat k\}$, define the consolidated region
$$
\mathcal I_{T,\ell}:=\bigcup_{m\in\mathcal C_{T,\ell}}(t_m-w_T,\,t_m+w_T].
$$

Then, for each $j\in\{1,\dots,k\}$, define the rejected indices associated with $\tau_j$ by
$$
\mathcal R_{T,j}
:=
\bigl\{m\in\widehat{\mathcal J}_T:\tau_j\in(t_m-w_T,t_m+w_T]\bigr\}.
$$
Since no Type-I error occurs on $\mathcal Q_T$, every $m\in\widehat{\mathcal J}_T$ satisfies $D_{T,m}>0$, and since $2w_T<\tau_{\min}$ for large $T$, its localization region contains exactly one change-point; hence $\widehat{\mathcal J}_T=\bigcup_{j=1}^{k}\mathcal R_{T,j}$, a disjoint union.
The condition $\tau_j\in(t_m-w_T,t_m+w_T]$ is equivalent to $t_m\in[\tau_j-w_T,\tau_j+w_T)$. Since $t_m=mw_T$, this holds if and only if
$$ 
m\in\Bigl[\tfrac{\tau_j}{w_T}-1,\ \tfrac{\tau_j}{w_T}+1\Bigr),
$$
an interval of length $2$, which contains at most two integers;
hence $|\mathcal R_{T,j}|\le2$. Moreover, if $m,m'\in\mathcal R_{T,j}$ with
$m<m'$, then $(m'-m)w_T=t_{m'}-t_m<2w_T$, forcing $m'-m=1$.
Thus each $\mathcal R_{T,j}$ is a singleton or a pair of consecutive
indices, and so lies within a single maximal component. Additionally, the components associated with different change-points are eventually separated, since $m\in\mathcal R_{T,j}$ and $m'\in\mathcal R_{T,j'}$ with $j\ne j'$, then
$$
|t_m-t_{m'}|\ \ge\ |\tau_j-\tau_{j'}|-2w_T\ \ge\ \tau_{\min}-2w_T,
$$
so $|m-m'|\ge \tau_{\min}/w_T-2>1$ for all $T$ large enough, since $w_T=o(\tau_{\min})$. Hence $\mathcal R_{T,j}$ and $\mathcal R_{T,j'}$ cannot belong to the same maximal component. 
Since no Type-II error
occurs, the nearest grid index $m(j)$ satisfies $m(j)\in\widehat{\mathcal J}_T$, and
$|t_{m(j)}-\tau_j|\le w_T/2<w_T$ gives $m(j)\in\mathcal R_{T,j}$, so every
$\mathcal R_{T,j}$ is nonempty. Therefore, $\widehat k=k$ on $\mathcal Q_T$ for all $T$ large enough,
and
$$
\Pr(\widehat k=k)\ \ge\ \Pr(\mathcal Q_T)\longrightarrow1.
$$

Finally, under the event $\mathcal Q_T$, each consolidated region $\mathcal I_{T,j}$ exactly one change-point $\tau_j$, and apply SWAL statistic per region to estimate the exact location of the change-point. By Theorem 2, for
every $\varepsilon>0$,
$$
\Pr\Bigl(\max_{1\le j\le k}|\widehat\tau_j-\tau_j|>\varepsilon w_T\Bigr)
\le
\Pr(\mathcal Q_T^c)
+\sum_{j=1}^{k}
\Pr\bigl(\{|\widehat\tau_j-\tau_j|>\varepsilon w_T\}\cap\mathcal Q_T\bigr)
\to 0,
$$
so $w_T^{-1}\max_{1\le j\le k}|\widehat\tau_j-\tau_j|\xrightarrow{p}0$.
Since $w_T=o(T)$, setting $\rho_T:=w_T/T=o(1)$ yields
$$
\Pr\Bigl(\widehat k=k
\ \text{ and }\
\max_{1\le j\le k}|\widehat\tau_j-\tau_j|\le T\rho_T\Bigr)
\to 1.
$$
\end{proof}

\subsection{Consistency of
Ensemble-SCAN}\label{consistency-of-ensemble-scan}

Ensemble aggregation is used as a stability mechanism rather than as a
separate source of asymptotic validity. The motivation follows the
general ensemble-learning principle that aggregating several competent
base procedures can reduce instability without degrading performance
\citep{Rokach2010, theisen2023ensemblesreallyeffective}. Related
consistency arguments for majority-vote classifiers also show that,
under suitable conditions, majority aggregation can preserve the
consistency of the underlying base learners
\citep{denil2013consistency}. In the present setting, however, the
consistency of Ensemble-SCAN follows directly from the consistency of
the finite collection of base SCAN detectors.

Let \(\mathcal{W} = \{w_1, \ldots, w_d\}\) denote a finite collection of
window sizes, where \(d\) is fixed and does not grow with \(T\). For
each \(w_i \in \mathcal{W}\), let \(\widehat{\mathcal{R}}^{(i)}\) denote
the change-point set estimated by the \(i\)th base SCAN detector, and
let \(\mathcal{R} =\{\tau_1, \ldots, \tau_k\}\) denote the true
change-point set. By Theorem 3, each SCAN detector consistently
estimates both the number and locations of the change-points. That is,
for some \(\rho_T=o(1)\), \[
\Pr\left(
|\widehat{\mathcal{R}}^{(i)}| = k
\ \text{and}\
\max_{1 \le \ell \le k}
|\widehat{\tau}^{(i)}_\ell - \tau_\ell|
\le T\rho_T
\right)
\to 1
\] as \(T \to \infty\), for each \(i = 1, \ldots, d\) in the ensemble.
If the ensemble merging radius \(r_T\) satisfies \(2T\rho_T \le r_T\)
and \(2r_T < \tau_{\min}\) for all sufficiently large \(T\), then, with
probability tending to one, detections associated with the same true
change-point are grouped into the same cluster, while detections
associated with distinct true change-points remain separated. Thus,
\(r_T\) is the theoretical tolerance radius for merging detections.

Since \(d\) is finite, the intersection of the base-detector consistency
events also occurs with probability tending to one. On this event, each
true change-point receives support from all \(d\) SCAN detectors, and no
spurious cluster receives majority support. Therefore, for any fixed
voting threshold \(\nu\in(0,1]\), the ensemble estimator
\(\widehat{\mathcal R}^{\star}\) satisfies \[
\Pr\left(|\widehat{\mathcal R}^{\star}|=k\right)\to 1,
\qquad
\max_{1\le \ell\le k}
\frac{|\widehat{\tau}^{\star}_\ell-\tau_\ell|}{T}
\xrightarrow{p}0.
\] Thus, Ensemble-SCAN preserves the consistency of the base SCAN
change-point detection framework while reducing sensitivity to the
choice of a single window size.

\subsection{Proof of Proposition 2}\label{proof-of-proposition-2}

\begin{proposition}[Window size risk trade-off]
\label{prop:window-risk-tradeoff}
Under Theorems 1 and 2, 
suppose the local error satisfies the uniform stochastic bound $\mathcal{E}_T(w) = O_p (R_T(w))$, where
$$
R_T(w) = \sqrt{\frac{\log w}{w}}+ \frac{w}{T}.
$$
A minimizer $w^*$ of $R_T(w)$ satisfies $w^* \asymp T^{2/3}(\log T)^{1/3}$.
\end{proposition}
\begin{proof}
Under Theorems 1 and 2,
\[
R_T(w)
=
\sqrt{\frac{\log w}{w}}
+
\frac{w}{T}.
\]
Differentiating $R_T(w)$ with respect to $w$ gives
\[
\frac{dR_T(w)}{dw}
=
-\frac{1}{2}
\frac{\log w - 1}{w^{3/2}\sqrt{\log w}}
+
\frac{1}{T}.
\]
At the minimizer $w^*$, the first-order condition $\frac{dR_T(w^*)}{dw}=0$ gives
$$
\frac{1}{T}
=
\frac{1}{2}
\frac{\log w^*-1}{(w^*)^{3/2}\sqrt{\log w^*}}.
$$
As $T\to\infty$, we have $w^*\to\infty$ and therefore $\log w^*-1\asymp\log w^*$. The first-order condition consequently implies
\[
(w^*)^{3/2}\asymp T\sqrt{\log w^*},
\]
or equivalently,
\[
w^*\asymp T^{2/3}(\log w^*)^{1/3}.
\]
Since $\log w^*\asymp\log T$, it follows that $w^*\asymp T^{2/3}(\log T)^{1/3}$. Finally, up to logarithmic factors, the optimal window size satisfies
$w^* \asymp T^{2/3}$.
\end{proof}

\section{Additional Numerical
Implementation}\label{additional-numerical-implementation}

\subsection{Adaptive Threshold
Calibration}\label{adaptive-threshold-calibration}

Algorithm \ref{alg:supp-adaptive-threshold} gives how the adaptive
threshold is calibrated using tapered-block-bootstrap at each local
split.

\begin{algorithm}[!htbp]
\SetAlgoSkip{0pt}
\DontPrintSemicolon
\SetKwInOut{KwIn}{Input}
\SetKwInOut{KwOut}{Output}
\caption{Calibrate the Data-Driven Threshold}
\label{alg:supp-adaptive-threshold}
\KwIn{Reference window $X_{t-w+1:t}$; stride window $X_{t+1:t+w}$; number of bootstrap samples $B$; Bonferroni-corrected significance level $\alpha'=\alpha/M$.}
\KwOut{Bootstrap-calibrated threshold $\widehat b_{T,m}$.}
Estimate the observed IPM discrepancy between the reference and stride windows:
\[
{T}_{t,w} \leftarrow d_{\mathcal{G}}\left(\widehat F_{(w),t},\,\widehat F_{(s),t}\right).
\]
Construct the pooled local sequence:
\(Z_{t,1:2w}\leftarrow(X_{t-w+1},\ldots,X_t,X_{t+1},\ldots,X_{t+w}).\)
\For{$b=1,\ldots,B$}{
Generate a tapered block bootstrap replicate $Z_{t,1:2w}^{*(b)}$ from $Z_{t,1:2w}$\;
Split $Z_{t,1:2w}^{*(b)}$ into the null reference and stride segments $Z_{t,1:w}^{*(b)}$ and $Z_{t,w+1:2w}^{*(b)}$\;
Compute ${T}_{t,w}^{(b)}=d_{\mathcal{G}}\left(\widehat F_{(w),t}^{*(b)},\widehat F_{(s),t}^{*(b)}\right)$\;
}
Estimate the $(1-\alpha')$ empirical quantile of $\{{T}_{t,w}^{(b)}\}_{b=1}^B$:
\[
\widehat b_{T,m}\leftarrow
\inf\left\{x:\frac{1}{B}\sum_{b=1}^B
\mathbf{1}\left({T}_{t,w}^{(b)}\le x\right)\ge1-\alpha'\right\}.
\]
\end{algorithm}

\subsection{Ensemble-SCAN}\label{ensemble-scan}

Algorithm \ref{alg:supp-ensemble-scan} summarizes the complete
aggregation procedure of the proposed ensemble-SCAN method.

\begin{algorithm}[!htbp]
\DontPrintSemicolon
\caption{Ensemble-SCAN with Majority Voting}
\label{alg:supp-ensemble-scan}
\small
\KwIn{Time series $\{X_t\}_{t=1}^T$; window set $\mathcal W=\{w_1,\ldots,w_d\}$; bootstrap replications $B$; significance level $\alpha$; voting threshold $\nu\in(0,1]$.}
\KwOut{Estimated change-point set $\widehat{\mathcal R}^{\star}$ and cluster support proportions.}
Initialize $\widehat{\mathcal R}^{\star}\leftarrow\varnothing$\;
\For{$j=1,\ldots,d$}{
$\widehat{\mathcal R}^{(j)}\leftarrow\textsc{SCAN}\left(\{X_t\}_{t=1}^T,w_j,B,\alpha\right)$\;
}
Pool and sort $\widehat{\mathcal R}\leftarrow\bigcup_{j=1}^{d}\widehat{\mathcal R}^{(j)}$\;
Partition $\widehat{\mathcal R}$ into clusters $\mathcal S=\{\mathcal S_1,\ldots,\mathcal S_L\}$ by joining consecutive detections separated by at most $r_T=\min(\mathcal W)$\;
\ForEach{$\mathcal S_\ell\in\mathcal S$}{
Compute $\mathrm{SP}(\mathcal S_\ell)\leftarrow d^{-1}\sum_{j=1}^{d}\mathbf{1}\!\left\{\widehat{\mathcal R}^{(j)}\cap\mathcal S_\ell\neq\varnothing\right\}$\;
\If{$\mathrm{SP}(\mathcal S_\ell)\ge\nu$}{
Compute $C_{\mathcal S_\ell}(t)\leftarrow\sum_{j=1}^{d}\mathbf{1}\!\left\{t\in\widehat{\mathcal R}^{(j)}\right\}$ for $t\in\mathcal S_\ell$\;
Choose $\widehat\tau(\mathcal S_\ell)\in\arg\max_{t\in\mathcal S_\ell}C_{\mathcal S_\ell}(t)$, resolving ties using the tied location returned by the largest window\;
$\widehat{\mathcal R}^{\star}\leftarrow\widehat{\mathcal R}^{\star}\cup\{\widehat\tau(\mathcal S_\ell)\}$\;
}
}
\Return{$\widehat{\mathcal R}^{\star}$}\;
\end{algorithm}

\subsection{Choice of Tuning
Parameters}\label{choice-of-tuning-parameters}

SCAN does not require a model-specific penalty parameter; instead, its
threshold is data-driven. However, the framework involves several
hyperparameters as follows: the significance level \(\alpha\), number of
bootstrap replications \(B\), ensemble size \(d\), voting threshold
\(\nu\). These control the testing level, Monte Carlo accuracy, ensemble
stability and aggregation respectively. Throughout the simulation study,
we set \(B=400\), \(d=7\), \(\nu=0.5\), and \(w_{\min}=30\). The values
of \(B\) and \(d\) may be increased when additional computational
resources are available for further accuracy. Additionally, practical
guidance for selecting the window-size range and voting threshold is
provided below.

\begin{enumerate}
\item \textbf{Window size: }
The window size governs a trade-off between statistical stability and localization resolution. Small windows provide finer temporal resolution but yield noisier IPM estimates, while larger windows stabilize the empirical discrepancy at the cost of reduced localization accuracy and a greater risk of including more than one change-point in a local comparison.

We select the admissible range $[w_{\min}, w_{\max}]$ as follows. The lower bound $w_{\min}$ is the smallest window size for which the upper bound on the local error probability, derived in Lemmas 1.5--1.6, falls below $1/2$. This threshold is motivated by the competence condition required for majority-vote aggregation. The upper bound $w_{\max}$ preserves the local nature of the scan. If the minimum change-point spacing $\tau_{\min}$ is known, we set $w_{\max} \le \lfloor \tau_{\min}/2 \rfloor$. Otherwise, we use the rate-motivated default $w_{\max} = \lfloor T^{2/3} \rfloor$, up to logarithmic factors, as justified above.

The ensemble window set $\mathcal{W}$ is then constructed by sampling a finite number of window sizes from $[w_{\min}, w_{\max}]$, yielding SCAN detectors that operate at multiple temporal scales while excluding windows that are either too narrow for reliable IPM estimation or too wide for accurate localization.

\item \textbf{Voting threshold: } The voting threshold $\nu\in(0,1]$ controls the minimum support required for a candidate cluster to be retained. Since $\nu$ is unknown in practice, we evaluate a grid of possible values and plot the number of retained change-points against $\nu$. A suitable threshold is chosen near the elbow of this curve, after the main drop in retained detections and before the curve stabilizes; an example is provided in the HASC analysis below. Larger values of $\nu$ yield more conservative detections, while smaller values increase sensitivity to weaker changes.
\end{enumerate}

\subsection{Localization Statistic}\label{localization-statistic}

\begin{figure}[H]
\centering
\includegraphics[width=\textwidth]{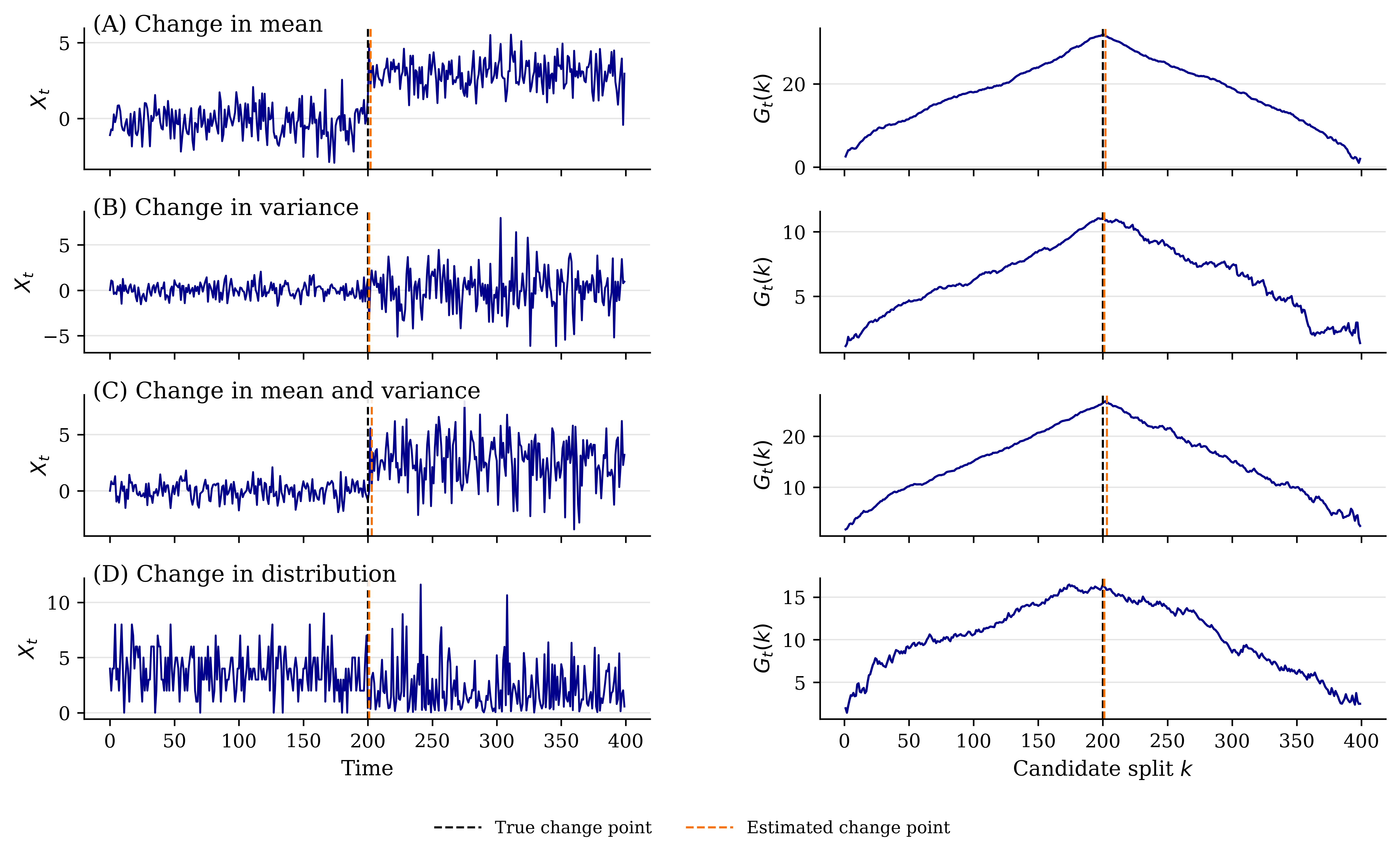}
\caption{Illustration of the proposed localization procedure under four representative single change-point scenarios. Each row corresponds to a different type of structural change: (A) change in mean, \(\mathcal{N}(0,1)\to \mathcal{N}(3,1)\); (B) change in variance, \(\mathcal{N}(0,0.8)\to \mathcal{N}(0,2.2)\); (C) simultaneous change in mean and variance, \(\mathcal{N}(0,0.8)\to \mathcal{N}(2.5,2)\); and (D) change in the overall distribution, \(\mathrm{Poisson}(\lambda=4)\to \mathrm{Exponential}(\theta=2)\).}
\label{fig:localization-stat}
\end{figure}

\section{Additional Simulation
Details}\label{additional-simulation-details}

\subsection{Data Generation}\label{data-generation}

The simulation study allows the number of true change-points, \(k(T)\),
to increase with the series length. This design is related to the
asymptotic framework of \citet{Fryzlewicz2014WBS}, which permits the
number of change-points to diverge with \(T\). Following the
growing-change-point design of
\citet{DingZhouTarokh2017TimeVaryingPrediction}, we set \[
k(T)=\left\lfloor \min\left\{CT^{1/3},\,\frac{T}{C\tau_{\min}}\right\}\right\rfloor,
\] with \(C=2.5\) and minimum change-point spacing \(\tau_{\min}=30\),
except for \(T=500\), for which we use four change-points. This
specification allows \(k(T)\) to grow moderately while preserving the
required minimum spacing between consecutive change-points. The
resulting designs range from four change-points at \(T=500\) to 250
change-points at \(T=10^6\). The Gaussian simulations also evaluate the
empirical behavior of SCAN beyond the compact-support assumptions used
for the formal Wasserstein theory.

\subsection{Benchmarking Methods and Their Parameter
Settings}\label{benchmarking-methods-and-their-parameter-settings}

Binary Segmentation, PELT, and KCP: These methods are implemented using
the \texttt{ruptures} library with the \(\ell_2\) cost function for
changes in the mean and the \texttt{normal} cost function for changes in
both the mean and variance, with a BIC-type penalty parameter.

SBS, WBS: Implemented using the \texttt{sbs} package in \texttt{R}.
These methods were called from Python through the \texttt{rpy2}
interface. The penalty parameter was selected using the BIC criterion.

Functional Pruning Optimal Partitioning (FPOP): Implemented using the
\texttt{fpop} package in \texttt{R}, with the method called from Python
through the \texttt{rpy2} interface. A BIC-type penalty was used to
control the number of estimated change-points.

\section{Additional results}\label{additional-results}

\subsection{Simulation Results}\label{simulation-results}

\begin{figure}[H]
\centering
\includegraphics[width=\textwidth]{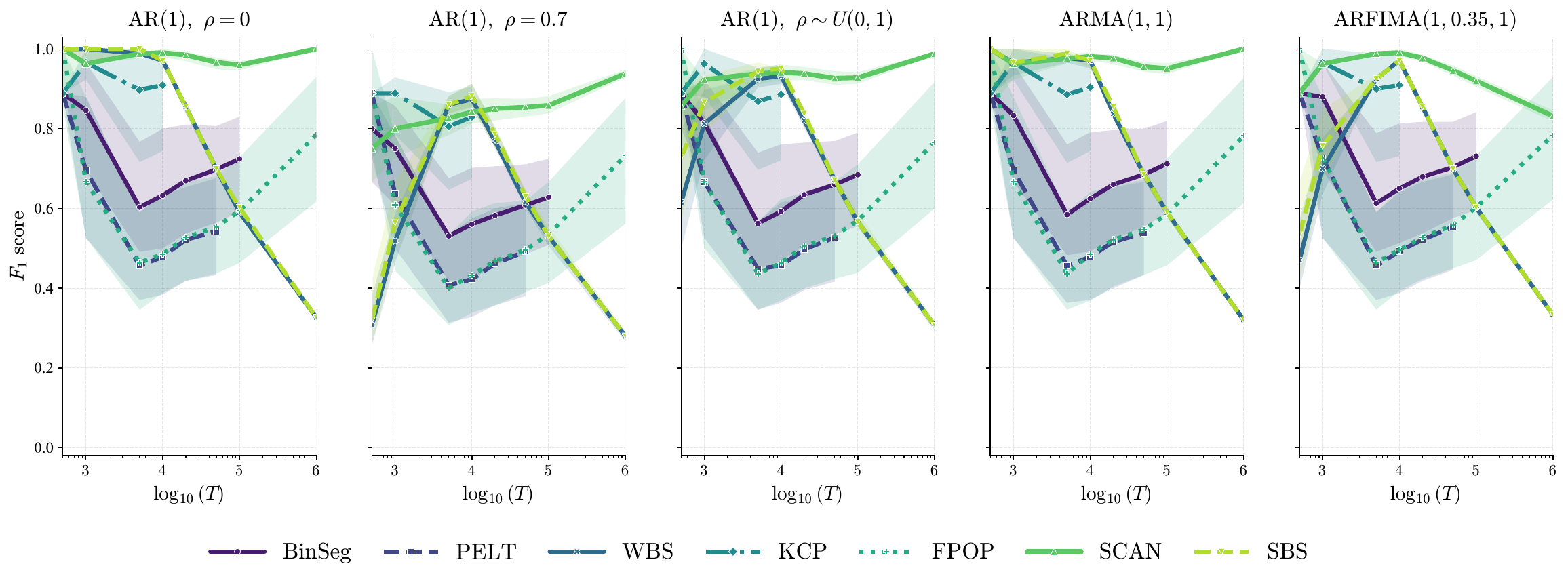}
\caption{Comparison of the $F_1$-score for change-point detection methods under mean changes across series lengths and dependence structures. The central line shows the median $F_1$-score, and the shaded bands represent the interquartile range (IQR).}
\label{fig:f1-score-mean}
\end{figure}

\begin{figure}[H]
\centering
\includegraphics[width=\textwidth]{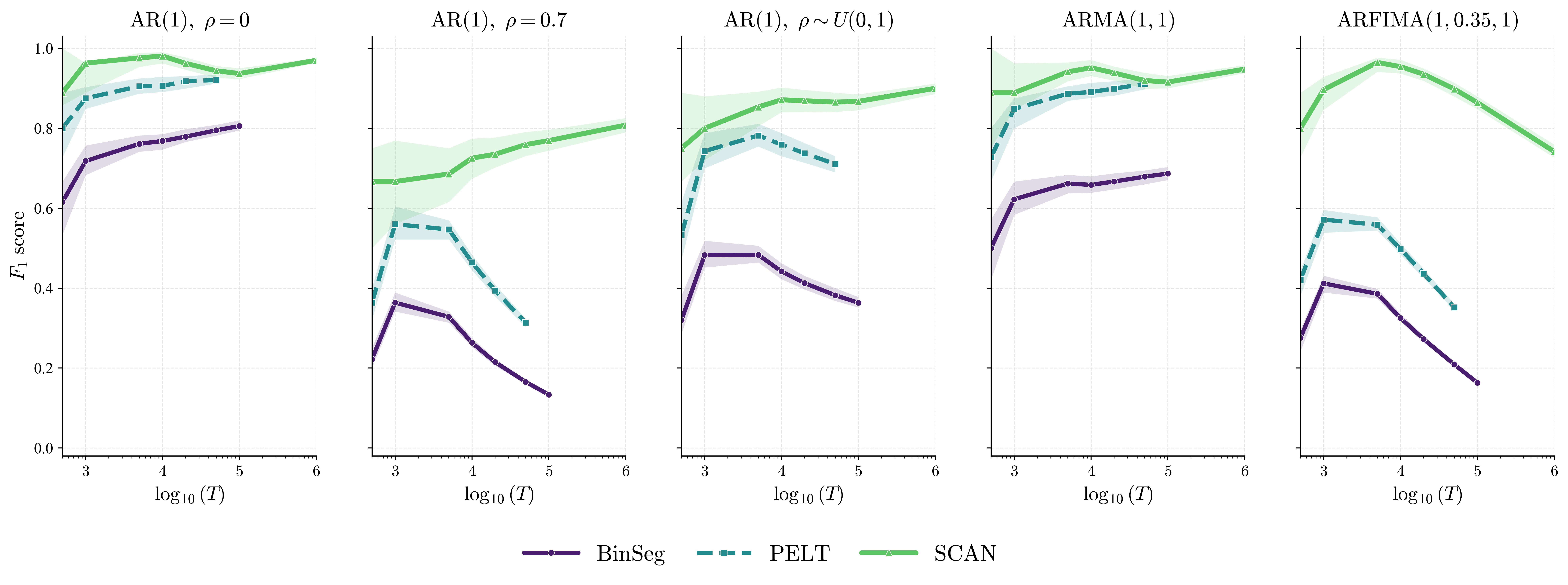}
\caption{Comparison of the $F_1$-score for change-point detection methods under changes in the mean and variance across series lengths and dependence structures.}
\label{fig:f1-score-meanvar}
\end{figure}

\begin{table}

\caption{\label{tbl-covering-f1-comparison}Performance comparison of change-point detection methods using covering metric and $F_1$-score.}

\centering{

\centering

\tiny
\setlength{\tabcolsep}{2.4pt}
\renewcommand{\arraystretch}{1.05}
\resizebox{\linewidth}{!}{%
\begin{tabular}{llcccccccccccccc}
\toprule
\multirow{2}{*}{Model}
& \multirow{2}{*}{Series length}
& \multicolumn{7}{c}{Covering metric}
& \multicolumn{7}{c}{$F_1$-score} \\
\cmidrule(lr){3-9}\cmidrule(lr){10-16}
& & SCAN & BinSeg & PELT & FPOP & WBS & SBS & KCP & SCAN & BinSeg & PELT & FPOP & WBS & SBS & KCP \\
\midrule
\multirow{8}{*}{AR(1), $\rho = 0$} & 500 & 0.971 & 0.973 & 0.973 & \textbf{0.984} & \textbf{0.984} & \textbf{0.984} & \textbf{0.984} & 0.957 & 0.889 & 0.889 & \textbf{1.000} & \textbf{1.000} & \textbf{1.000} & 0.889 \\
 & 1000 & \textbf{0.973} & 0.760 & 0.561 & 0.574 & 0.971 & 0.971 & 0.965 & 0.964 & 0.846 & 0.696 & 0.667 & \textbf{1.000} & \textbf{1.000} & 0.966 \\
 & 5000 & \textbf{0.984} & 0.454 & 0.320 & 0.323 & 0.981 & 0.981 & 0.823 & 0.985 & 0.603 & 0.456 & 0.464 & 0.989 & \textbf{1.000} & 0.897 \\
 & 10000 & \textbf{0.989} & 0.541 & 0.385 & 0.388 & 0.950 & 0.949 & 0.872 & \textbf{0.990} & 0.633 & 0.479 & 0.486 & 0.971 & 0.971 & 0.909 \\
 & 20000 & \textbf{0.983} & 0.620 & 0.462 & 0.465 & 0.825 & 0.825 & - & \textbf{0.980} & 0.670 & 0.522 & 0.527 & 0.855 & 0.855 & - \\
 & 50000 & \textbf{0.977} & 0.691 & 0.539 & 0.541 & 0.694 & 0.696 & - & \textbf{0.962} & 0.698 & 0.543 & 0.554 & 0.699 & 0.699 & - \\
 & 100000 & \textbf{0.979} & 0.745 & - & 0.614 & 0.614 & 0.616 & - & \textbf{0.956} & 0.724 & - & 0.594 & 0.590 & 0.602 & - \\
 & 1000000 & \textbf{0.999} & - & - & 0.771 & 0.289 & 0.289 & - & \textbf{0.999} & - & - & 0.785 & 0.327 & 0.327 & - \\
\midrule
\multirow{8}{*}{AR(1), $\rho = 0.7$} & 500 & 0.799 & 0.917 & 0.942 & \textbf{0.949} & 0.405 & 0.421 & 0.948 & 0.752 & 0.800 & \textbf{0.889} & \textbf{0.889} & 0.308 & 0.320 & \textbf{0.889} \\
 & 1000 & 0.754 & 0.704 & 0.543 & 0.558 & 0.612 & 0.648 & \textbf{0.864} & 0.792 & 0.750 & 0.636 & 0.609 & 0.519 & 0.564 & \textbf{0.889} \\
 & 5000 & 0.767 & 0.452 & 0.314 & 0.318 & \textbf{0.908} & 0.904 & 0.756 & 0.820 & 0.531 & 0.407 & 0.400 & \textbf{0.860} & \textbf{0.860} & 0.806 \\
 & 10000 & 0.825 & 0.527 & 0.381 & 0.386 & \textbf{0.921} & \textbf{0.921} & 0.834 & 0.841 & 0.560 & 0.423 & 0.432 & 0.874 & \textbf{0.882} & 0.830 \\
 & 20000 & \textbf{0.877} & 0.617 & 0.467 & 0.468 & 0.812 & 0.813 & - & \textbf{0.848} & 0.583 & 0.462 & 0.468 & 0.769 & 0.786 & - \\
 & 50000 & \textbf{0.930} & 0.684 & 0.538 & 0.540 & 0.690 & 0.691 & - & \textbf{0.852} & 0.608 & 0.492 & 0.496 & 0.615 & 0.629 & - \\
 & 100000 & \textbf{0.954} & 0.736 & - & 0.609 & 0.612 & 0.613 & - & \textbf{0.859} & 0.628 & - & 0.535 & 0.530 & 0.530 & - \\
 & 1000000 & \textbf{0.996} & - & - & 0.765 & 0.289 & 0.290 & - & \textbf{0.939} & - & - & 0.733 & 0.280 & 0.280 & - \\
\midrule
\multirow{8}{*}{AR(1), $\rho \sim \mathrm{Uniform}$} & 500 & 0.885 & 0.963 & 0.966 & 0.976 & 0.744 & 0.823 & \textbf{0.980} & 0.864 & 0.889 & 0.889 & \textbf{1.000} & 0.615 & 0.727 & 0.889 \\
 & 1000 & 0.864 & 0.741 & 0.552 & 0.567 & 0.868 & 0.899 & \textbf{0.941} & 0.890 & 0.815 & 0.667 & 0.667 & 0.813 & 0.867 & \textbf{0.963} \\
 & 5000 & 0.898 & 0.449 & 0.321 & 0.323 & 0.946 & \textbf{0.955} & 0.793 & 0.929 & 0.563 & 0.448 & 0.436 & 0.925 & \textbf{0.944} & 0.868 \\
 & 10000 & 0.931 & 0.533 & 0.387 & 0.390 & \textbf{0.940} & \textbf{0.940} & 0.857 & 0.941 & 0.593 & 0.457 & 0.464 & 0.932 & \textbf{0.951} & 0.887 \\
 & 20000 & \textbf{0.949} & 0.622 & 0.465 & 0.465 & 0.821 & 0.821 & - & \textbf{0.939} & 0.635 & 0.497 & 0.505 & 0.821 & 0.838 & - \\
 & 50000 & \textbf{0.961} & 0.685 & 0.536 & 0.538 & 0.692 & 0.694 & - & \textbf{0.926} & 0.660 & 0.527 & 0.531 & 0.671 & 0.671 & - \\
 & 100000 & \textbf{0.970} & 0.744 & - & 0.611 & 0.614 & 0.615 & - & \textbf{0.927} & 0.685 & - & 0.570 & 0.566 & 0.566 & - \\
 & 1000000 & \textbf{0.999} & - & - & 0.766 & 0.289 & 0.290 & - & \textbf{0.988} & - & - & 0.766 & 0.307 & 0.307 & - \\
\midrule
\multirow{8}{*}{ARMA(1,1)} & 500 & 0.948 & 0.972 & 0.972 & \textbf{0.984} & 0.980 & 0.980 & \textbf{0.984} & 0.932 & 0.889 & 0.889 & \textbf{1.000} & \textbf{1.000} & \textbf{1.000} & 0.889 \\
 & 1000 & 0.950 & 0.754 & 0.555 & 0.580 & 0.962 & \textbf{0.965} & 0.961 & 0.949 & 0.833 & 0.696 & 0.667 & \textbf{0.966} & \textbf{0.966} & \textbf{0.966} \\
 & 5000 & 0.970 & 0.454 & 0.319 & 0.320 & 0.975 & \textbf{0.978} & 0.821 & 0.976 & 0.585 & 0.456 & 0.436 & 0.977 & \textbf{0.989} & 0.886 \\
 & 10000 & \textbf{0.980} & 0.540 & 0.384 & 0.387 & 0.948 & 0.947 & 0.870 & \textbf{0.983} & 0.625 & 0.479 & 0.486 & 0.971 & 0.971 & 0.904 \\
 & 20000 & \textbf{0.976} & 0.619 & 0.462 & 0.464 & 0.825 & 0.824 & - & \textbf{0.973} & 0.660 & 0.516 & 0.522 & 0.838 & 0.855 & - \\
 & 50000 & \textbf{0.972} & 0.691 & 0.541 & 0.542 & 0.694 & 0.695 & - & \textbf{0.955} & 0.685 & 0.538 & 0.547 & 0.685 & 0.685 & - \\
 & 100000 & \textbf{0.977} & 0.744 & - & 0.614 & 0.614 & 0.615 & - & \textbf{0.950} & 0.712 & - & 0.585 & 0.590 & 0.590 & - \\
 & 1000000 & \textbf{0.999} & - & - & 0.770 & 0.289 & 0.289 & - & \textbf{0.998} & - & - & 0.783 & 0.320 & 0.320 & - \\
\midrule
\multirow{8}{*}{ARFIMA(1, $0.35$, 1)} & 500 & 0.935 & 0.976 & 0.976 & \textbf{0.984} & 0.569 & 0.610 & \textbf{0.984} & 0.920 & 0.889 & 0.889 & \textbf{1.000} & 0.471 & 0.533 & 0.889 \\
 & 1000 & \textbf{0.983} & 0.783 & 0.568 & 0.609 & 0.766 & 0.808 & 0.971 & \textbf{0.966} & 0.880 & 0.727 & 0.727 & 0.700 & 0.757 & \textbf{0.966} \\
 & 5000 & \textbf{0.990} & 0.463 & 0.324 & 0.326 & 0.935 & 0.938 & 0.832 & \textbf{0.987} & 0.613 & 0.456 & 0.464 & 0.925 & 0.925 & 0.900 \\
 & 10000 & \textbf{0.987} & 0.546 & 0.389 & 0.393 & 0.952 & 0.950 & 0.876 & \textbf{0.988} & 0.650 & 0.493 & 0.496 & 0.971 & 0.971 & 0.909 \\
 & 20000 & \textbf{0.972} & 0.627 & 0.466 & 0.468 & 0.827 & 0.825 & - & \textbf{0.975} & 0.680 & 0.522 & 0.527 & 0.855 & 0.855 & - \\
 & 50000 & \textbf{0.945} & 0.695 & 0.540 & 0.539 & 0.695 & 0.695 & - & \textbf{0.945} & 0.703 & 0.554 & 0.558 & 0.699 & 0.699 & - \\
 & 100000 & \textbf{0.919} & 0.748 & - & 0.616 & 0.614 & 0.615 & - & \textbf{0.918} & 0.731 & - & 0.602 & 0.602 & 0.602 & - \\
 & 1000000 & \textbf{0.815} & - & - & 0.767 & 0.289 & 0.290 & - & \textbf{0.832} & - & - & 0.783 & 0.333 & 0.333 & - \\
\bottomrule
\end{tabular}%
}

}

\end{table}%

\subsection{Long-memory Voting-Threshold
Sensitivity}\label{long-memory-voting-threshold-sensitivity}

\begin{figure}[H]
\centering
\includegraphics[width=0.8\textwidth]{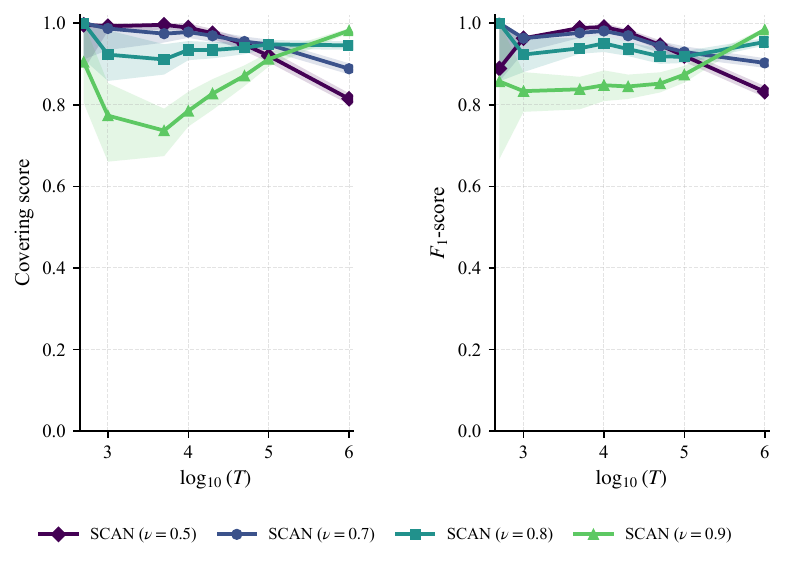}
\caption{Sensitivity of SCAN performance to the ensemble voting threshold in the presence of long-memory.}
\label{fig:arfima-threshold-sensitivity}
\end{figure}

\begin{table}

\caption{\label{tbl-meanvar-covering-f1-comparison}Performance comparison of change-point detection methods under joint mean-variance changes using the covering metric and $F_1$-score.}

\centering{

\centering

\setlength{\tabcolsep}{3pt}
\renewcommand{\arraystretch}{0.95}

\begin{adjustbox}{max width=\linewidth, max totalheight=0.88\textheight, keepaspectratio}
\begin{tabular}{llcccccc}
\toprule
\multirow{2}{*}{Model}
& \multirow{2}{*}{Series length}
& \multicolumn{3}{c}{Covering metric}
& \multicolumn{3}{c}{$F_1$-score} \\
\cmidrule(lr){3-5}\cmidrule(lr){6-8}
& & SCAN & BinSeg & PELT & SCAN & BinSeg & PELT \\
\midrule

\multirow{8}{*}{AR(1), $\rho = 0$} & 500 & \textbf{0.996} & 0.868 & 0.960 & \textbf{1.000} & 0.616 & 0.794 \\
 & 1000 & \textbf{0.985} & 0.880 & 0.942 & \textbf{0.963} & 0.718 & 0.876 \\
 & 5000 & \textbf{0.992} & 0.925 & 0.966 & \textbf{0.988} & 0.759 & 0.905 \\
 & 10000 & \textbf{0.993} & 0.945 & 0.979 & \textbf{0.991} & 0.769 & 0.909 \\
 & 20000 & 0.985 & 0.959 & \textbf{0.987} & \textbf{0.985} & 0.781 & 0.915 \\
 & 50000 & 0.978 & 0.971 & \textbf{0.993} & \textbf{0.967} & 0.795 & 0.922 \\
 & 100000 & \textbf{0.981} & 0.979 & - & \textbf{0.960} & 0.806 & - \\
 & 1000000 & \textbf{1.000} & - & - & \textbf{1.000} & - & - \\
\midrule
\multirow{8}{*}{AR(1), $\rho = 0.7$} & 500 & \textbf{0.815} & 0.392 & 0.513 & \textbf{0.750} & 0.227 & 0.363 \\
 & 1000 & \textbf{0.769} & 0.511 & 0.678 & \textbf{0.800} & 0.366 & 0.565 \\
 & 5000 & \textbf{0.776} & 0.514 & 0.695 & \textbf{0.827} & 0.328 & 0.547 \\
 & 10000 & \textbf{0.829} & 0.469 & 0.635 & \textbf{0.842} & 0.264 & 0.465 \\
 & 20000 & \textbf{0.880} & 0.438 & 0.574 & \textbf{0.850} & 0.215 & 0.394 \\
 & 50000 & \textbf{0.931} & 0.415 & 0.499 & \textbf{0.854} & 0.165 & 0.314 \\
 & 100000 & \textbf{0.955} & 0.407 & - & \textbf{0.858} & 0.134 & - \\
 & 1000000 & \textbf{0.997} & - & - & \textbf{0.938} & - & - \\
\midrule
\multirow{8}{*}{AR(1), $\rho \sim \mathrm{Uniform}$} & 500 & \textbf{0.917} & 0.570 & 0.741 & \textbf{0.857} & 0.336 & 0.555 \\
 & 1000 & \textbf{0.883} & 0.664 & 0.837 & \textbf{0.923} & 0.486 & 0.745 \\
 & 5000 & \textbf{0.903} & 0.703 & 0.882 & \textbf{0.938} & 0.486 & 0.782 \\
 & 10000 & \textbf{0.935} & 0.693 & 0.879 & \textbf{0.941} & 0.443 & 0.759 \\
 & 20000 & \textbf{0.952} & 0.692 & 0.869 & \textbf{0.939} & 0.414 & 0.738 \\
 & 50000 & \textbf{0.962} & 0.696 & 0.854 & \textbf{0.927} & 0.384 & 0.710 \\
 & 100000 & \textbf{0.971} & 0.708 & - & \textbf{0.928} & 0.364 & - \\
 & 1000000 & \textbf{1.000} & - & - & \textbf{0.988} & - & - \\
\midrule
\multirow{8}{*}{ARMA(1,1)} & 500 & \textbf{0.992} & 0.755 & 0.912 & \textbf{1.000} & 0.499 & 0.734 \\
 & 1000 & \textbf{0.975} & 0.805 & 0.920 & \textbf{0.963} & 0.628 & 0.847 \\
 & 5000 & \textbf{0.973} & 0.855 & 0.953 & \textbf{0.976} & 0.662 & 0.887 \\
 & 10000 & \textbf{0.983} & 0.871 & 0.968 & \textbf{0.981} & 0.661 & 0.892 \\
 & 20000 & \textbf{0.978} & 0.886 & 0.977 & \textbf{0.977} & 0.667 & 0.900 \\
 & 50000 & 0.974 & 0.903 & \textbf{0.985} & \textbf{0.956} & 0.678 & 0.910 \\
 & 100000 & \textbf{0.978} & 0.915 & - & \textbf{0.951} & 0.686 & - \\
 & 1000000 & \textbf{1.000} & - & - & \textbf{1.000} & - & - \\
\midrule
\multirow{8}{*}{ARFIMA(1, $0.35$, 1)} & 500 & \textbf{0.996} & 0.473 & 0.619 & \textbf{0.889} & 0.278 & 0.420 \\
 & 1000 & \textbf{0.992} & 0.599 & 0.768 & \textbf{0.963} & 0.411 & 0.570 \\
 & 5000 & \textbf{0.996} & 0.597 & 0.772 & \textbf{0.988} & 0.386 & 0.560 \\
 & 10000 & \textbf{0.989} & 0.533 & 0.696 & \textbf{0.991} & 0.327 & 0.499 \\
 & 20000 & \textbf{0.975} & 0.467 & 0.609 & \textbf{0.977} & 0.273 & 0.437 \\
 & 50000 & \textbf{0.948} & 0.391 & 0.499 & \textbf{0.946} & 0.210 & 0.353 \\
 & 100000 & \textbf{0.920} & 0.334 & - & \textbf{0.920} & 0.164 & - \\
 & 1000000 & \textbf{0.814} & - & - & \textbf{0.832} & - & - \\

\bottomrule
\end{tabular}
\end{adjustbox}

}

\end{table}%

\subsection{Computational Complexity and Parallel
Implementation}\label{computational-complexity-and-parallel-implementation}

For a fixed window size \(w\), SCAN performs \(O(T/w)\) local hypothesis
tests between adjacent reference and stride windows. In the univariate
setting, computing the 1-Wasserstein distance within a window of size
\(w\) has computational cost \(O(w\log w)\), since the observations in
the window must be sorted when constructing the empirical CDFs.
Therefore, with \(B\) tapered block bootstrap replications, the
computational cost for a fixed window size is \(O(BT\log w)\).

For the ensemble version, let \(\mathcal W=\{w_1,\ldots,w_d\}\) denote
the set of window sizes and let \(w_{\max}=\max(\mathcal W)\). Running
all \(d\) SCAN detectors serially has cost

\[\sum_{j=1}^{d} O(BT\log w_j)
\le
O(BdT\log w_{\max}).\]

The number of bootstrap replications \(B\) and the ensemble size \(d\)
are treated as fixed hyperparameters, independent of the series length.
Hence, for fixed \(B\) and \(d\), the computational cost is
approximately linear in \(T\) up to logarithmic factors. When no prior
information is available for selecting the maximum window size, the
implementation uses the rate-motivated default
\(w_{\max}\asymp T^{2/3}\), giving complexity \(O(BdT\log(T^{2/3}))\).

Bootstrap replications and different window sizes can be computed in
parallel as separate tasks. The resulting base detectors are not
statistically independent; they share the same series and may use
overlapping observations, but this does not prevent parallel
computation.

In practice, the \texttt{scan-cpd} implementation leverages multi-core
processing, so the wall-clock time depends on the slowest base-detector
task and the available computational resources. Although the literature
does not provide an explicit optimal choice for the number of SCAN
detectors in the ensemble, increasing this number excessively is not
necessarily beneficial because it increases computational cost. The
ensemble size should therefore be large enough to provide stability
across window sizes, but not increased without bound
\citep{OshiroPerezBaranauskas2012}.

\subsection{Real-world datasets}\label{real-world-datasets}

\textbf{HASC Dataset}

Under standard ensembling theory, a voting threshold of \(0.5\) is a
natural default under the competence framework described in the main
body of the paper. However, this threshold may be overly conservative in
real-world applications. In particular, for changes in variance, the
distributional shift primarily affects the spread rather than the
location of the observations, which may produce a weaker discrepancy
signal than a mean shift. This can reduce the power of the change-point
detection algorithm.

\begin{figure}[H]
\centering
\includegraphics[width=0.65\textwidth]{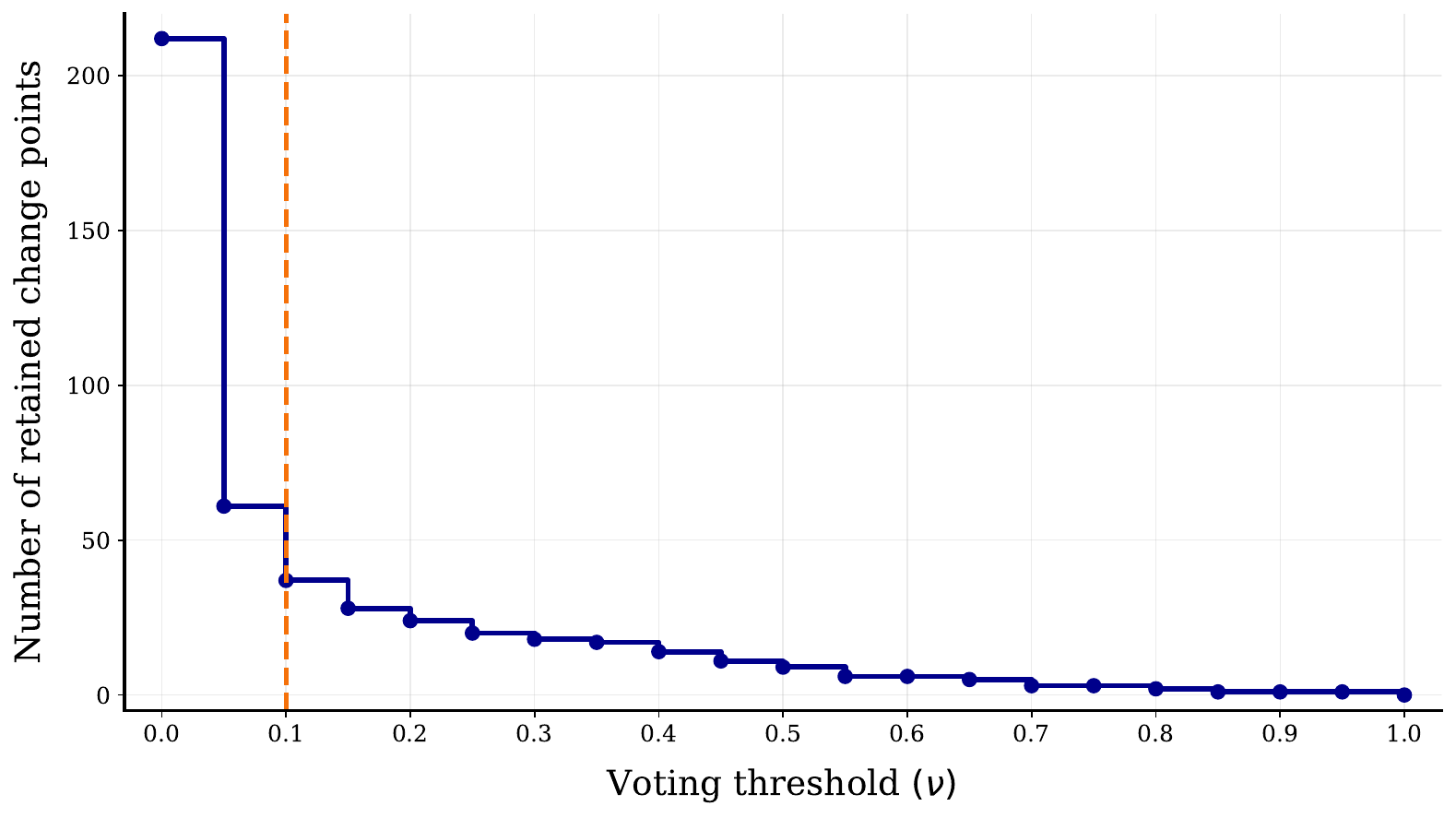}
\caption{Sensitivity of the number of retained change-points to the ensemble voting threshold.}
\label{fig:hasc-elbow}
\end{figure}

For this reason, we replace the default majority-voting threshold of
\(0.5\) with an elbow-type diagnostic plot for selecting the voting
proportion. Figure \ref{fig:hasc-elbow} shows how the voting threshold
was determined for the HASC dataset. This approach provides a
transparent, unsupervised threshold-selection procedure.

The analysis of the HASC dataset used a significance level of
\(\alpha=0.05\), \(B=400\) tapered block bootstrap replications, and an
ensemble of \(d=20\) base SCAN detectors. The window sizes ranged from
\(w_{\min}=30\) to \(w_{\max}=1081\), and the voting threshold was set
to \(\nu=0.1\) based on the elbow shown in Figure \ref{fig:hasc-elbow}.

\textbf{BTC-USD Dataset}

\begin{longtable}[]{@{}
  >{\raggedleft\arraybackslash}p{(\linewidth - 6\tabcolsep) * \real{0.0896}}
  >{\raggedright\arraybackslash}p{(\linewidth - 6\tabcolsep) * \real{0.2687}}
  >{\raggedright\arraybackslash}p{(\linewidth - 6\tabcolsep) * \real{0.2090}}
  >{\raggedright\arraybackslash}p{(\linewidth - 6\tabcolsep) * \real{0.4328}}@{}}
\caption{change-points matched with nearby major Bitcoin-related Reuters
articles, providing possible interpretations of the detected
changes.}\tabularnewline
\toprule\noalign{}
\begin{minipage}[b]{\linewidth}\raggedleft
No.
\end{minipage} & \begin{minipage}[b]{\linewidth}\raggedright
Detected
\end{minipage} & \begin{minipage}[b]{\linewidth}\raggedright
Event date
\end{minipage} & \begin{minipage}[b]{\linewidth}\raggedright
Event summary
\end{minipage} \\
\midrule\noalign{}
\endfirsthead
\toprule\noalign{}
\begin{minipage}[b]{\linewidth}\raggedleft
No.
\end{minipage} & \begin{minipage}[b]{\linewidth}\raggedright
Detected
\end{minipage} & \begin{minipage}[b]{\linewidth}\raggedright
Event date
\end{minipage} & \begin{minipage}[b]{\linewidth}\raggedright
Event summary
\end{minipage} \\
\midrule\noalign{}
\endhead
\bottomrule\noalign{}
\endlastfoot
1 & 2017-02-17 13:00 & 2017-02-23 &
\href{https://www.reuters.com/article/us-global-markets-bitcoin-idUSKBN16221V/}{U.S.
Bitcoin-ETF speculation pushes Bitcoin toward a record high} \\
4 & 2018-03-14 14:00 & 2018-03-14 &
\href{https://www.reuters.com/article/world/uk-google-bans-cryptocurrency-advertising-bitcoin-price-slumps-idUSKCN1GQ0GS/}{Google
bans cryptocurrency advertising} \\
5 & 2018-11-14 22:00 & 2018-11-14 &
\href{https://www.reuters.com/article/technology/bitcoin-drops-to-one-year-low-as-slump-persists-ethereum-down-sharply-idUSKCN1NJ2GS/}{Bitcoin
falls below \$6,000} \\
6 & 2019-04-02 09:00 & 2019-04-02 &
\href{https://www.reuters.com/article/technology/bitcoin-jumps-20-percent-mystery-order-seen-as-catalyst-idUSKCN1RE0JX/}{Large
anonymous order triggers rally} \\
7 & 2019-05-03 07:00 & 2019-05-03 &
\href{https://www.reuters.com/article/amp/idUSKCN1S90QQ/}{Bitcoin jumps
as much as 6 percent to new six-month high} \\
8 & 2019-06-22 04:00 & 2019-06-22 &
\href{https://www.wsj.com/articles/bitcoin-is-back-above-10-000-and-investors-say-this-rally-is-different-11561201454}{Bitcoin
rises above \$10,000 amid renewed cryptocurrency optimism} \\
9 & 2019-09-24 23:00 & 2019-09-24 &
\href{https://www.coindesk.com/markets/2019/09/24/bitcoins-price-slides-1000-in-30-minutes-after-margin-calls-at-bitmex}{Margin
calls trigger a sharp Bitcoin price decline} \\
10 & 2020-01-07 04:00 & 2020-01-06 &
\href{https://www.coindesk.com/markets/2020/01/06/bitcoin-eyes-price-breakout-amid-us-iran-tensions}{U.S.--Iran
tensions support a Bitcoin price breakout} \\
11 & 2020-03-09 09:00 & 2020-03-09 &
\href{https://www.coindesk.com/markets/2020/03/09/market-liquidations-cause-cascade-in-bitcoin-price}{Market
liquidations accelerate Bitcoin's decline} \\
12 & 2020-04-16 13:00 & 2020-04-16 &
\href{https://www.coindesk.com/markets/2020/04/16/bitcoin-price-spikes-above-71k-liquidating-23m-on-bitmex}{Bitcoin
jumps above \$7,100 and triggers BitMEX liquidations} \\
13 & 2020-07-27 21:00 & 2020-07-27 &
\href{https://www.coindesk.com/markets/2020/07/27/market-wrap-bitcoin-blasts-past-10000-ethereum-fees-up-550-in-2020}{High
trading volume pushes Bitcoin toward \$11,000} \\
14 & 2020-11-04 18:00 & 2020-11-04 &
\href{https://www.reuters.com/article/mrkte-kryptowhrungen-idDEL8N2HO5WU/}{U.S.
election uncertainty lifts Bitcoin} \\
15 & 2021-02-09 05:00 & 2021-02-09 &
\href{https://www.reuters.com/world/china/bitcoin-rockets-new-highs-tesla-takes-it-mainstream-2021-02-09/}{Tesla
investment drives Bitcoin to new highs} \\
16 & 2021-05-17 06:00 & 2021-05-17 &
\href{https://www.reuters.com/business/finance/bitcoin-hits-three-month-low-musk-drives-investors-exit-2021-05-17/}{Bitcoin
falls after Musk comments} \\
17 & 2021-10-06 17:00 & 2021-10-06 &
\href{https://www.reuters.com/technology/bitcoin-hits-strongest-level-since-may-2021-10-06/}{Soros
holdings and seasonal rally} \\
18 & 2021-12-04 09:00 & 2021-12-04 &
\href{https://www.reuters.com/technology/bitcoin-extends-downtrend-falls-121-47176-2021-12-04/}{Leveraged
positions liquidated during crash} \\
19 & 2022-04-22 20:00 & 2022-04-22 &
\href{https://www.reuters.com/technology/how-crypto-giant-binance-built-ties-russian-fsb-linked-agency-2022-04-22/}{Binance--Russia
investigation} \\
20 & 2022-06-13 13:00 & 2022-06-13 &
\href{https://www.reuters.com/business/finance/cryptocurrency-market-value-slumps-under-1-trillion-2022-06-13/}{Crypto
market falls below \$1 trillion} \\
21 & 2022-08-26 20:00 & 2022-08-27 &
\href{https://www.reuters.com/markets/us/bitcoin-drops-16-below-20000-2022-08-27/}{Hawkish
Fed speech pressures Bitcoin} \\
22 & 2023-01-12 23:00 & 2023-01-12 &
\href{https://www.reuters.com/markets/currencies/yen-jumps-dollar-tentative-ahead-us-inflation-data-2023-01-12/}{Softer
U.S. inflation supports Bitcoin} \\
23 & 2023-06-21 06:00 & 2023-06-21 &
\href{https://www.reuters.com/technology/bitcoin-eyes-third-straight-day-gains-after-touching-two-month-high-2023-06-21/}{BlackRock
spot-ETF application drives rally} \\
24 & 2023-08-18 02:00 & 2023-08-18 &
\href{https://www.reuters.com/markets/currencies/bitcoin-drops-new-two-month-low-world-markets-sell-off-2023-08-18/}{Global
risk-asset sell-off} \\
25 & 2023-10-20 08:00 & 2023-10-20 &
\href{https://www.reuters.com/technology/bitcoin-tops-30000-first-time-since-august-2023-10-20/}{Technical
rally with no clear catalyst} \\
26 & 2024-02-14 14:00 & 2024-02-14 &
\href{https://www.reuters.com/technology/total-amount-invested-bitcoin-back-over-1-trillion-2024-02-14/}{Spot-ETF
inflows lift Bitcoin market value} \\
27 & 2024-02-27 07:00 & 2024-02-27 &
\href{https://www.reuters.com/technology/bitcoin-breaks-57000-big-buyers-circle-2024-02-27/}{Institutional
demand pushes Bitcoin above \$57,000} \\
28 & 2024-11-06 04:00 & 2024-11-06 &
\href{https://www.reuters.com/technology/bitcoin-leaps-record-high-traders-lean-toward-trump-victory-2024-11-06/}{Bitcoin
leaps to record high as traders lean toward Trump victory} \\
29 & 2025-02-23 13:00 & 2025-02-24 &
\href{https://www.reuters.com/technology/cybersecurity/cryptos-biggest-hacks-heists-after-15-billion-theft-bybit-2025-02-24/}{Crypto's
biggest hacks and heists after \$1.5 billion theft from Bybit} \\
30 & 2025-04-25 12:00 & 2025-04-25 &
\href{https://www.reuters.com/business/finance/swiss-national-bank-chairman-rebuffs-bitcoin-reserve-asset-2025-04-25/}{Swiss
National Bank chairman rebuffs Bitcoin as reserve asset} \\
31 & 2025-05-08 16:00 & 2025-05-09 &
\href{https://www.reuters.com/markets/currencies/bitcoin-tops-100000-trade-deal-optimism-2025-05-08/}{Bitcoin
retakes \$100,000 on global trade deal optimism} \\
32 & 2025-07-09 20:00 & 2025-07-09 &
\href{https://www.reuters.com/world/africa/dollar-gains-against-yen-trumps-trade-war-intensifies-2025-07-09/}{Bitcoin
soars to all-time peak just shy of \$112,000} \\
33 & 2026-01-30 02:00 & 2026-01-30 &
\href{https://www.reuters.com/business/bitcoin-slips-fed-chair-speculation-hits-risky-assets-2026-01-30/}{Bitcoin
slips as Fed chair speculation hits risky assets} \\
\end{longtable}

\renewcommand\refname{References}
\bibliography{bibliography.bib}